\documentclass[pdflatex,sn-mathphys-num]{sn-jnl}

\usepackage{lmodern}
\usepackage{hyperref}
\usepackage{bookmark}

\usepackage{mathrsfs}%

\usepackage{graphicx}%
\usepackage{multirow}%
\usepackage{amsmath,amssymb,amsfonts}%
\usepackage{amsthm}%

\usepackage[title]{appendix}%
\usepackage{xcolor}%
\usepackage{textcomp}%
\usepackage{manyfoot}%
\usepackage{booktabs}%
\usepackage{algorithm}%
\usepackage{algorithmicx}%
\usepackage{algpseudocode}%
\usepackage{listings}%
\usepackage{makecell}
\usepackage{adjustbox}

\usepackage{booktabs}
\usepackage[braket, qm]{qcircuit}
\usepackage{braket}
\usepackage{qcircuit}
\usepackage{longtable}
\usepackage{tikz}
\usetikzlibrary{positioning}

\usepackage{anyfontsize} 

\theoremstyle{thmstyleone}%
\theoremstyle{thmstyletwo}%

\theoremstyle{thmstylethree}%

\begin{document}

\title[Regression of Functions by Quantum Neural Networks Circuits]{Regression of Functions by Quantum Neural Networks Circuits}


\author*[1]{\fnm{Fernando} \sur{M. de Paula Neto}}\email{fernando@cin.ufpe.br}

\author[1]{\fnm{Lucas} \sur{dos Reis Silva}}\email{lrs5@cin.ufpe.br}

\author[1]{\fnm{Paulo S. G.} \sur{de Mattos Neto}}\email{pgsmn@cin.ufpe.br}

\author[2,3]{\fnm{Felipe F.} \sur{Fanchini}}\email{felipe.fanchini@unesp.br}

\affil*[1]{\orgdiv{Center of Informatics}, \orgname{Federal University of Pernambuco - UFPE}, \orgaddress{\street{Av. Jorn. Aníbal Fernandes}, \city{Recife}, \postcode{50.740-560}, \state{Pernambuco}, \country{Brazil}}}

\affil[2]{\orgdiv{Department of Physics and Meteorology}, \orgname{São Paulo State University - UNESP}, \orgaddress{\street{Av. Eng. Luiz Edmundo Carrijo Coube}, \city{Bauru}, \postcode{17033-360}, \state{São Paulo}, \country{Brazil}}}

\affil[3]{\orgname{QuaTI - Quantum Technology \& Information}, \orgaddress{\street{Major José Inácio Street}, \city{São Carlos}, \postcode{13560-161}, \state{São Paulo}, \country{Brazil}}}


\abstract{
The performance of quantum neural network models depends strongly on architectural decisions, including circuit depth, placement of parametrized operations, and data-encoding strategies. Selecting an effective architecture is challenging and closely related to the classical difficulty of choosing suitable neural-network topologies, which is computationally hard.
This work investigates automated quantum-circuit construction for regression tasks and introduces a genetic-algorithm framework that discovers Reduced Regressor QNN architectures. The approach explores depth, parametrized gate configurations, and flexible data re-uploading patterns, formulating the construction of quantum regressors as an optimization process. The discovered circuits are evaluated against seventeen classical regression models on twenty-two nonlinear benchmark functions and four analytical functions. Although classical methods often achieve comparable results, they typically require far more parameters, whereas the evolved quantum models remain compact while providing competitive performance.
We further analyze dataset complexity using twelve structural descriptors and show, across five increasingly challenging meta-learning scenarios, that these measures can reliably predict which quantum architecture will perform best. The results demonstrate perfect or near-perfect predictive accuracy in several scenarios, indicating that complexity metrics offer powerful and compact representations of dataset structure and can effectively guide automated model selection.
Overall, this study provides a principled basis for meta-learning-driven quantum architecture design and advances the understanding of how quantum models behave in regression settings—a topic that has received limited exploration in prior work. These findings pave the way for more systematic and theoretically grounded approaches to quantum regression.
}




\maketitle

\section{Introduction}

Quantum computing promises algorithmic speed-ups for problems that are conjectured to be intractable for classical computation. A landmark example is Shor's polynomial-time quantum algorithm for integer factorization and discrete logarithms \cite{Shor1997}. Beyond algorithmic constructions, experimental demonstrations of \emph{quantum supremacy}---including boson sampling and instantaneous quantum polynomial-time (IQP) circuits---have shown that even shallow quantum devices can generate probability distributions that cannot be efficiently reproduced by classical machines unless the polynomial hierarchy collapses \cite{Du2020Expressive}. These results underscore the intrinsic expressive richness of quantum models and their potential to outperform classical methods in specific computational regimes.

Motivated by these developments, the field of \emph{quantum machine learning} (QML) seeks to combine quantum computation with modern machine-learning methodologies~\cite{biamonte2017quantum}. Quantum learning models exploit high-dimensional Hilbert spaces, controllable entanglement, and quantum interference to construct expressive hypothesis classes that can be difficult to emulate classically~\cite{barreto2024consider}. A prominent line of research has focused on quantum kernel methods~\cite{Havlicek2019}, which map classical data into quantum-enhanced feature spaces where linear classifiers can achieve separations that are provably hard for classical models. In this direction, Liu~\emph{et al.}~\cite{liu2021rigorous} construct a classification task for which a quantum kernel method, implemented via efficiently parameterized unitary circuits and requiring only classical data access, achieves an end-to-end quantum advantage under standard cryptographic assumptions.

More recently, the scope of QML has expanded beyond kernel-based approaches toward more general parametrized quantum models. In particular, quantum neural networks (QNNs) and hybrid quantum--classical architectures have been proposed as flexible learning frameworks capable of achieving large effective dimension and improved trainability compared to classical neural networks~\cite{Abbas2021Power}. These models have been successfully applied to a wide range of learning paradigms~\cite{ding2025quantum,lin2024quantum,onim2025quantum}. 

Beyond classification, QML techniques have also been explored in structured learning settings and real-world applications, such as efficient quantum convolutional neural networks designed under hardware constraints~\cite{roseler2025efficient}, long-term time-series forecasting~\cite{chittoor2024qultsf}, and financial modeling for institutional algorithmic trading~\cite{ciceri2025enhanced}. At the methodological level, recent advances in quantum feature engineering, including digitized counterdiabatic protocols~\cite{hegade2022digitized} and quenched quantum feature maps~\cite{simen2025quenched}, further highlight the growing interest in designing expressive yet trainable quantum embeddings. Collectively, these works illustrate the rapid evolution of QML from theoretically motivated kernel methods toward versatile, application-driven quantum learning architectures.

QNNs can be naturally viewed as instances of \emph{variational quantum circuits} (VQCs)~\cite{peruzzo2014variational}. These hybrid quantum--classical models employ parameterized quantum gates whose trainable weights are optimized to minimize a task-dependent loss function. Several QNN architectures have been proposed in the literature, ranging from early perceptron-inspired models \cite{Tacchino2020Qubit, de2019implementing}, tensor-network embeddings \cite{Huggins2019Towards}, and quantum convolutional networks \cite{cong2019quantum}, to recurrent and reuploading-based designs \cite{perez2020data}. Recently, a study demonstrated the universality of quantum neural networks of the random type \cite{gonon2025universal}, and other study demonstrates that deep data re-uploading QNNs can approximate arbitrary continuous functions, and that the attainable expressive power heavily depends on the structure of the measurement operators and the entangling topology \cite{schuld2008effect}. There are also important analyses in terms of the learning capacity of intelligent quantum models~\cite{wang2024information, coyle2020born, riste2017demonstration}. Some studies have demonstrated the ability of quantum neural networks to solve nonlinear problems in classification tasks~\cite{de2023parametrized,basilewitsch2025quantum, monteiro2021quantum}.

The performance, trainability, and robustness of QNNs depend critically on the chosen variational ansatz. Architectural factors such as gate placement, entanglement patterns, circuit depth, and data-encoding strategies play central roles in determining the expressive capacity and optimization landscape of the model. Indeed, the representational power of parameterized quantum circuits is tightly constrained by the interplay between feature maps and circuit topology \cite{Du2020Expressive, sim2019expressibility}, and suboptimal architectural choices may lead to barren plateaus~\cite{mcclean2018barren}, excessive depth requirements, or undesirable sensitivity to device noise. Consequently, the principled design of QNN architectures—especially those intended for regression—remains a challenging open problem and a key motivation for automated circuit-synthesis approaches.

This has motivated the rapidly growing field of \emph{Quantum Architecture Search} (QAS), which aims to automatically discover quantum circuit structures tailored to specific learning tasks. Approaches based on evolutionary algorithms, reinforcement learning, neural predictors, and differentiable search have shown that automatically synthesized circuits can outperform expert-crafted ansätze while using fewer parameters and shallower depth \cite{Zhang2021QAS,He2024MetaQAS,Martyniuk2024Survey}.



These observations highlight a fundamental open challenge: identifying circuit architectures that are expressive enough for the target learning task while respecting constraints on depth, noise resilience, and hardware feasibility. This challenge is analogous to the classical NP-complete problem of selecting optimal multilayer perceptron topologies. Previous work has already explored optimal constructions of quantum circuits for specific tasks, such as the preparation of quantum states and the synthesis of arbitrary unitaries~\cite{sun2023asymptotically}. A recent study performs systematic comparisons between quantum learning models and classical classifiers \cite{basilewitsch2025quantum}. Consequently, there is a compelling need for automated, data-driven, and hardware-adapted methodologies for synthesizing variational quantum circuits suitable for regression tasks.

In this work, we address this challenge by conducting a systematic comparative analysis between automatically synthesized quantum circuits and three widely used variational ansatz families in the QML literature—\textit{Strongly Entangling Layers}, \textit{Simplified Two Design}, and \textit{Basic Entangler Layers}. We propose a genetic-algorithm-based framework for constructing Reduced Regressor Quantum Neural Network (RRQNN) architectures and treat quantum circuits as design objects whose topology, depth, and data--reuploading patterns must be synthesized rather than manually specified. This approach enables fine-grained exploration of quantum regressors under stringent depth constraints while making their architectural design responsive to the complexity of the underlying function.

Beyond circuit synthesis, we also address the orthogonal yet fundamental question of \emph{model selection for quantum regressors}: given a dataset, which quantum architecture is likely to achieve the best performance? While prior works in Quantum Architecture Search (QAS) have explored architecture optimization using evolutionary search \cite{Zhang2021QAS}, differentiable search \cite{zhang2022differentiable}, reinforcement learning \cite{ostaszewski2021reinforcement}, or meta-trained predictors \cite{He2024MetaQAS}, these methods primarily aim at efficiently discovering a single circuit structure. Similarly, QML research on meta-learning has focused on accelerating training or improving initialization \cite{sotoca2006meta}, but not on predicting the most suitable quantum model for a given regression task. In contrast, our work integrates meta-feature analysis into the QML pipeline and demonstrates, for the first time in quantum regression, that structural descriptors of the dataset can be used to predict—with optimal accuracy in our experiments—which among several competing quantum circuit families (including the GA-generated RRQNNs and the three standard ansätze) yields the best performance. This combination of automated circuit synthesis and meta-learning-based model selection differentiates our approach from existing QAS and QML methodologies.

The paper is structured to progressively introduce the proposed methodology, its experimental validation, and the resulting insights. Section~\ref{sec:qc} provides the necessary background on quantum computing, while Section~\ref{sec:qnn} introduces quantum neural networks, focusing on the concepts required to contextualize the learning models studied in this work. Section~\ref{sec:complexityMetrics} presents the regression complexity measures employed as meta-features, which form the basis of the subsequent meta-learning analysis. The proposed Reduced Regressor Quantum Neural Network (RRQNN) and the genetic-algorithm-based architecture search are detailed in Section~\ref{sec:geneticAlgorithmFORRRQNN}, including the chromosome encoding and the deterministic mapping to quantum circuits. The experimental protocol is described in Section~\ref{sec:experimentalProtocol}. Finally, Section~\ref{sec:resultsDiscussions} discusses the empirical results, statistical analyses, and limitations of the study, while Section~\ref{sec:conclusion} summarizes the main findings and outlines directions for future work.

\section{Quantum Computing: Background and Notation}
\label{sec:qc}

This section provides only the minimum background required for readers from the machine learning community; more comprehensive treatments of quantum computing fundamentals can be found in standard quantum computing textbooks~\cite{nielsen2010quantum}.

\subsection{Quantum Bits}
In quantum information theory, the fundamental unit of data is the quantum bit (qubit). A qubit is formally represented as a normalized vector in the complex Hilbert space $\mathbb{C}^2$, i.e., a two-dimensional complex vector space. Unlike classical bits, which assume deterministic states, a qubit can exist in a coherent superposition of the computational basis states $\ket{0}$ and $\ket{1}$.
Mathematically, any pure qubit state $\ket{\psi}$ can be expressed as a linear combination of the basis vectors, as shown in Equation~\ref{eq:qubit2}:
\begin{equation}
\label{eq:qubit2}
\ket{\psi} = \alpha \ket{0} + \beta \ket{1}, \text{where } \alpha, \beta \in \mathbb{C} \text{ and } |\alpha|^2 + |\beta|^2 = 1.
\end{equation}
The complex amplitudes $\alpha$ and $\beta$ encode probabilistic information about the outcome of a projective measurement in the computational basis. The probability of obtaining $\ket{0}$ is $|\alpha|^2$, while the probability of measuring $\ket{1}$ is $|\beta|^2$. Throughout this manuscript, $i$ denotes the imaginary unit, such that $i^2 = -1$.
Composite quantum systems are described via the tensor product operation, denoted by $\otimes$.
For two qubits $\ket{a}$ and $\ket{b}$, the joint system is given by $\ket{\mathbf{g}} = \ket{a} \otimes \ket{b}$, often denoted succinctly as $\ket{ab}$.
If $\ket{a} = \alpha_1 \ket{0} + \beta_1 \ket{1}$ and $\ket{b} = \alpha_2 \ket{0} + \beta_2 \ket{1}$, then the composite state expands to:

 \begin{equation}
 \ket{ab} = \alpha_1 \alpha_2 \ket{00} + \alpha_1 \beta_2 \ket{01} + \beta_1 \alpha_2 \ket{10} + \beta_1 \beta_2 \ket{11}.
    \label{eq:TensorProduct}
    \end{equation}
    
In the general case of states $\ket{\mathbf{p}} \in \mathbb{C}^{2^n}$ and $\ket{\mathbf{q}} \in \mathbb{C}^{2^m}$, the tensor product $\ket{pq} = \ket{\mathbf{p}} \otimes \ket{\mathbf{q}}$ produces a vector in $\mathbb{C}^{2^{n+m}}$, as detailed in Equation~\ref{eq:TensorProduct2}.

 \begin{equation}
        \left[
    \begin{array}{ll}
    \alpha_1 \\
    \alpha_2 \\
    ... \\
    \alpha_{2^n}
    \end{array}
    \right]
    \otimes
        \left[
    \begin{array}{ll}
    \beta_1 \\
    \beta_2 \\
    ... \\
    \beta_{2^m}
    \end{array}
    \right]
    = \left[
    \begin{array}{ll}
    \alpha_1 \left[
    \begin{array}{ll}
    \beta_1 \\
    \beta_2 \\
    ...\\
    \beta_{2^m}
    \end{array}
    \right]     \\ \\
    \alpha_2 \left[
    \begin{array}{ll}
    \beta_1 \\
    \beta_2 \\
    ...\\
    \beta_{2^m}
    \end{array}
    \right] \\ 
    ...\\
    \alpha_{2^n} \left[
    \begin{array}{ll}
    \beta_1 \\
    \beta_2 \\
    ...\\
    \beta_{2^m}
    \end{array}
    \right]
    \end{array}
    \right]     
    =\left[
    \begin{array}{ll}
    \alpha_1 \beta_1 \\
    \alpha_1 \beta_2 \\
    ...\\
    \alpha_1 \beta_{2^m}\\
    \alpha_2 \beta_1\\
    \alpha_2 \beta_2\\
    ...\\
    \alpha_2 \beta_{2^m}\\
    ... \\
    \alpha_{2^n} \beta_{2^m}
    \end{array}
    \right].
    \label{eq:TensorProduct2}
    \end{equation}
This tensor construction is essential for representing multi-qubit systems and modeling quantum entanglement.
A state $\ket{\psi} \in Q \otimes R$ is entangled if it cannot be factorized as $\ket{\psi} \neq \ket{q} \otimes \ket{r}$ with $\ket{q} \in Q$ and $\ket{r} \in R$.
Entanglement constitutes one of the fundamental non-classical features enabling quantum computational advantage.
\subsection{Quantum Operators}
Quantum evolution is governed by unitary transformations, known as quantum operators or quantum gates.
An operator $\mathbf{U}$ acting on $n$ qubits is a unitary matrix of dimension $2^n \times 2^n$, satisfying $\mathbf{U}^\dagger \mathbf{U} = \mathbf{I}$.
For a single qubit, common elementary gates include the Identity ($\mathbf{I}$), Pauli-X ($\mathbf{X}$), and Hadamard ($\mathbf{H}$) operators, defined as follows:
\begin{equation}
    \label{eq:quantumop1}
    \begin{array}{ll}
    \textbf{I}= \left[
    \begin{array}{ll}
    1 & 0\\
    0 & 1\\
    \end{array}
    \right];
    \begin{array}{l}
    \textbf{I}\ket{0}= \ket{0} \\
    \textbf{I}\ket{1}=\ket{1} 
    \end{array}
    \end{array};
    \begin{array}{ll}
    \textbf{X}= \left[
    \begin{array}{ll}
    0 & 1\\
    1 & 0\\
    \end{array}
    \right];
    \begin{array}{l}
    \textbf{X}\ket{0}= \ket{1} \\
    \textbf{X}\ket{1}=\ket{0} 
    \end{array}
    \end{array}
    \end{equation}
    \begin{equation}
    \label{eq:quantumop2}
    \begin{array}{ll}
    \textbf{H}=\frac{1}{\sqrt{2}} \left[
    \begin{array}{cc}
    1 & 1\\
    1 & -1\\
    \end{array}
    \right];
    &
    \begin{array}{l}
    \textbf{H}\ket{0}= 1/\sqrt{2}(\ket{0}+\ket{1}) \\
    \textbf{H}\ket{1}=1/\sqrt{2}(\ket{0}-\ket{1}) 
    \end{array}
    \end{array}
    \end{equation}
While $\mathbf{I}$ leaves the qubit invariant, $\mathbf{X}$ performs a bit-flip operation analogous to the classical NOT gate.
The Hadamard gate $\mathbf{H}$ induces an equal superposition of $\ket{0}$ and $\ket{1}$, playing a central role in quantum parallelism.
More general single-qubit rotations are represented by the rotation operators $\mathbf{R_x}$, $\mathbf{R_y}$, and $\mathbf{R_z}$:
\begin{equation}
  \mathbf{Rx}(\theta) = \begin{pmatrix}
    cos(\theta/2) & -i \cdot sin(\theta/2)\\ 
    -i \cdot sin(\theta/2) & cos(\theta/2)
  \end{pmatrix},
  \label{eq:Rx}
\end{equation}

\begin{equation}
  \mathbf{Ry}(\theta) = \begin{pmatrix}
    cos(\theta/2) & -sin(\theta/2)\\ 
    sin(\theta/2) & cos(\theta/2)
  \end{pmatrix},
  \label{eq:Ry}
\end{equation}

\begin{equation}
  \mathbf{Rz}(\theta) = \begin{pmatrix}
    e^{-i \cdot \theta/2} & 0\\ 
    0 & e^{i \cdot \theta/2}
  \end{pmatrix},
  \label{eq:Rz}
\end{equation}

\begin{equation}
  \mathbf{U}(\theta, \beta, \gamma) = \begin{pmatrix}
    cos(\theta/2) & -e^{i \cdot \gamma} \cdot sin(\theta)\\ 
    e^{i \cdot \beta} sin(\theta) & e^{i \cdot (\beta + \gamma)} cos(\theta)
  \end{pmatrix}.
  \label{eq:U}
\end{equation}
Any arbitrary single-qubit unitary operator can be decomposed as a product of these rotation gates,
$\mathbf{U} = \mathbf{R_z}(\alpha)\mathbf{R_y}(\beta)\mathbf{R_z}(\gamma)$, or equivalently expressed in the general parametric form shown in Equation~\ref{eq:U2}:
\begin{equation}
  \mathbf{U}(\theta, \beta, \gamma) = \begin{pmatrix}
    cos(\theta/2) & -e^{i \cdot \gamma} \cdot sin(\theta)\\ 
    e^{i \cdot \beta} sin(\theta) & e^{i \cdot (\beta + \gamma)} cos(\theta)
  \end{pmatrix}.
  \label{eq:U2}
\end{equation}
Multi-qubit interactions are typically introduced via controlled operations.
The controlled-NOT (CNOT) gate, for instance, acts on two qubits—a control and a target—and applies $\mathbf{X}$ to the target if and only if the control qubit is in the $\ket{1}$ state:
\begin{equation}
\label{eq:CNOTworking}
\begin{array}{ll}
\mathbf{CNOT} = 
\begin{bmatrix}
 1 & 0 & 0 & 0\\
 0 & 1 & 0 & 0\\
 0 & 0 & 0 & 1\\
 0 & 0 & 1 & 0\\
\end{bmatrix};
&
\begin{array}{l}
\mathbf{CNOT}\ket{00}= \ket{00} \\
\mathbf{CNOT}\ket{01}= \ket{01} \\
\mathbf{CNOT}\ket{10}= \ket{11} \\
\mathbf{CNOT}\ket{11}= \ket{10} 
\end{array}
\end{array}
\end{equation}
This construction generalizes to controlled-$\mathbf{U}$ gates, where any unitary transformation $\mathbf{U}$ can be applied conditionally on the state of one or more control qubits.
The set ${\mathbf{H}, \mathbf{R_z}, \mathbf{CNOT}}$ forms a universal gate set capable of implementing any quantum computation \cite{nielsen2010quantum}.
\subsection{Quantum Measurement}
Quantum measurement constitutes a non-unitary and irreversible process that projects a superposed state into one of the basis states of the measurement operator.
Given a single-qubit state $\ket{\psi} = \alpha\ket{0} + \beta\ket{1}$, a projective measurement in the computational basis collapses the state to $\ket{0}$ with probability $|\alpha|^2$ or to $\ket{1}$ with probability $|\beta|^2$.
For a general multi-qubit state $\ket{\mathbf{g}}$, the probability of observing a configuration $\ket{\mathbf{\phi}}$ is given by $|\langle \mathbf{\phi} | \mathbf{g} \rangle|^2$.
The expectation value of an observable $\mathbf{Z}$ in the state $\ket{\psi}$ is defined as:
\begin{equation}
\braket{Z} = \bra{\psi} \mathbf{Z} \ket{\psi},
\quad
\mathbf{Z} =
\begin{bmatrix}
 1 & 0 \\
 0 & -1 \\ 
\end{bmatrix},
\end{equation}
yielding real outcomes in the interval $[-1, 1]$.
Measurement plays a crucial role in extracting classical information from quantum systems and forms the interface between quantum and classical computation layers.
\subsection{Quantum Circuits}
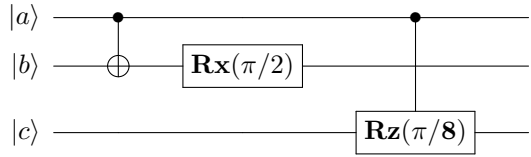
\begin{figure}[ht]
\[
\Qcircuit @C=2em @R=0.9em {
\lstick{\ket{a}} & \ctrl{1} & \qw & \ctrl{2} & \qw\\
\lstick{\ket{b}} & \targ & \gate{\mathbf{Rx}(\pi/2)} & \qw & \qw \\
\lstick{\ket{c}} & \qw & \qw & \gate{\mathbf{Rz(\pi/8)}} & \qw
}
\]
\caption{An example of quantum circuit with one $\mathbf{CNOT}$ operator and two rotation gates, $\mathbf{Rx}$ and $\mathbf{Rz}$, one of them being controlled (controlled rotation).}
\label{fig:cnot}
\end{figure}

Quantum algorithms are typically modeled as quantum circuits, which provide a graphical abstraction of the sequence of quantum operations applied to a register of qubits.
In this representation, qubits are depicted as horizontal wires, and gates correspond to unitary transformations applied along these wires, with computation proceeding from left to right.
An example of a circuit containing a $\mathbf{CNOT}$ gate, a controlled $\mathbf{R_z}$ rotation, and a single-qubit $\mathbf{R_x}$ rotation is illustrated in Figure~\ref{fig:cnot}.

\section{Quantum Neural Networks}
\label{sec:qnn}
\begin{figure}[ht]
    \centering
\[ \Qcircuit @C=0.71em @R=.47em {
&                                 &     &\mbox{$n$ times} \\
&                                 &     & \\
& \multigate{5}{\text{Input Embedding($\boldsymbol{\chi}$)}} & \qw & \multigate{5}{\text{Parametric Layer($\boldsymbol{\theta}$)}} & \qw \\
& \ghost{\text{Input Embedding($\boldsymbol{\chi}$)}}        & \qw & \ghost{\text{Parametric Layer($\boldsymbol{\theta}$)}} & \qw \\
& \ghost{\text{Input Embedding($\boldsymbol{\chi}$)}}        & \qw & \ghost{\text{Parametric Layer($\boldsymbol{\theta}$)}} & \qw \\
& \ghost{\text{Input Embedding($\boldsymbol{\chi}$)}}        & \qw & \ghost{\text{Parametric Layer($\boldsymbol{\theta}$)}} &\qw \\
& \ghost{\text{Input Embedding($\boldsymbol{\chi}$)}}        & \qw & \ghost{\text{Parametric Layer($\boldsymbol{\theta}$)}} &\qw  \\
& \ghost{\text{Input Embedding($\boldsymbol{\chi}$)}}        & \qw & \ghost{\text{Parametric Layer($\boldsymbol{\theta}$)}} &\qw  
 \gategroup{3}{4}{8}{4}{1em}{--}
} \]
    \caption{Traditional generic architecture of a parametric quantum circuit where the Parametric Layer can be repeated $n$ times.}
    \label{fig:genericPQC}
\end{figure}
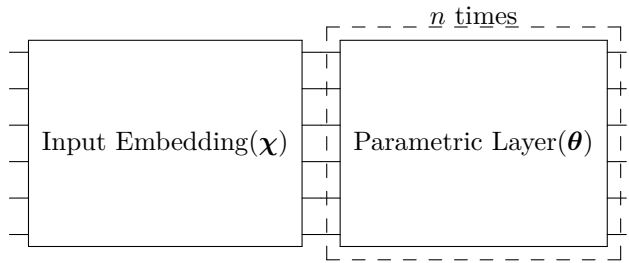

As stated in \cite{schuld2021machine}, modern quantum neural networks combine variational quantum circuits with classical optimization techniques, allowing for efficient training of small quantum circuits. The variational quantum circuit, also known as the parametric quantum circuit (PQC), was introduced in \cite{peruzzo2014variational} as a strategy to construct quantum circuits with a limited number of gates, or equivalently, low circuit depth. A PQC typically consists of two main stages: an information encoding stage and a parameterized operation stage.
Figure~\ref{fig:genericPQC} illustrates a general representation of a PQC, in which a layer dedicated to loading input data, called Input Embedding, as well as one (or $n$) layer for loading circuit parameters, called Parametric Layer. 
Quantum circuits with tunable parameters have demonstrated their capability to solve complex real-world problems \cite{de2023parametrized, monteiro2021quantum, grant2018hierarchical, de2019implementing}. As discussed in \cite{schuld2021effect}, the expressive power of a parameterized circuit—namely, the number of functions it can approximate—increases as the number of repeated layers grows. Furthermore, the results in \cite{ballarin2023entanglement} show that the entanglement difference between layers tends to converge as the number of parameterized layers increases, highlighting a form of structural saturation in deep PQCs.

\subsection{Encoding Classical Data}

There are several strategies for \textbf{encoding classical data} into quantum states \cite{lloyd2020quantum}. 
A common strategy for loading a classical input vector 
\(
\mathbf{x} = (x_1, \dots, x_d) \in \mathbb{R}^d
\)
into a variational quantum circuit is to encode each component \(x_j\) 
into the rotation angle of a single-qubit gate. More precisely, one 
selects a rotation operator \(R_\alpha(\cdot)\), with 
\(\alpha \in \{x,y,z\}\), and applies these gates \emph{in parallel} 
to \(d\) qubits, each receiving the information of one coordinate of 
\(\mathbf{x}\).

A single-qubit rotation around axis \(\alpha\) is defined as
\begin{equation}
    R_\alpha(\theta)
    =
    \exp\!\left(-\,\frac{i\theta}{2}\,\sigma_\alpha\right),
    \qquad
    \alpha \in \{x,y,z\},
\end{equation}
where \(\sigma_\alpha\) denotes the corresponding Pauli operator.
Thus, a general phase-embedding unitary takes the form
\begin{equation}
    U_{\mathrm{embed}}(\mathbf{x})
    =
    \bigotimes_{j=1}^{d}
    R_\alpha\!\big( \phi_j(x_j) \big),
\end{equation}
where \(\phi_j : \mathbb{R} \rightarrow \mathbb{R}\) is typically a linear function,
such as
\begin{equation}
    \phi_j(x_j) = \omega_j x_j,
\end{equation}
with \(\omega_j\) being a frequency or scaling hyperparameter.

The resulting quantum state after embedding is
\begin{equation}
    |\psi(\mathbf{x})\rangle
    =
    U_{\mathrm{embed}}(\mathbf{x})
    \,|0\rangle^{\otimes d}
    =
    \left(
        \bigotimes_{j=1}^{d}
        R_\alpha\!\big( \phi_j(x_j) \big)
    \right)
    |0\rangle^{\otimes d}.
\end{equation}

When encoding through \(Z\)-axis rotations, we have explicitly
\begin{equation}
    R_z(\theta)
    =
    \exp\!\left(-\,\frac{i\theta}{2} Z\right)
    =
    \begin{pmatrix}
        e^{-i\theta/2} & 0 \\[4pt]
        0 & e^{i\theta/2}
    \end{pmatrix},
\end{equation}
and therefore the corresponding phase embedding is
\begin{equation}
    U_{\mathrm{embed}}^{(Z)}(\mathbf{x})
    =
    \bigotimes_{j=1}^{d}
    R_z\!\big( \omega_j x_j \big).
\end{equation}

This type of encoding through parallel single-qubit rotations is widely used in variational quantum neural networks and data reuploading models due to its simplicity and because it introduces controlled phase contributions that are later processed by trainable variational blocks. In practice, phase rotations such as \(R_z\) are most meaningful once the quantum states have been brought into superposition—typically via the application of Hadamard gates—since global or relative phases become observable only in the presence of interference. Under this condition, \(R_z\)-based embeddings effectively modulate relative phases across computational basis components, which are subsequently transformed into measurable amplitudes by entangling and variational layers. A detailed discussion of such encoding strategies in regression-oriented quantum neural networks can be found in Panadero~\emph{et al.}~\cite{panadero2024}, where the authors show how phase embeddings contribute to the Fourier expressive structure of QNNs.

There are some works that demonstrate the effect of repeatedly loading data into the circuit. This routine is called \textit{reuploading}. In \cite{perez2020data}, a reuploading approach shows the universality of the proposed quantum classifier. In \cite{schuld2021effect}, it is possible to verify that variational quantum circuits with reuploading can be represented as Fourier series, where the more repetitions of data applied to variational quantum circuits, the closer the circuit function is to its true value.

\subsection{Parametric Layers}

\begin{figure}[ht]
\centering
\includegraphics[scale=0.5]{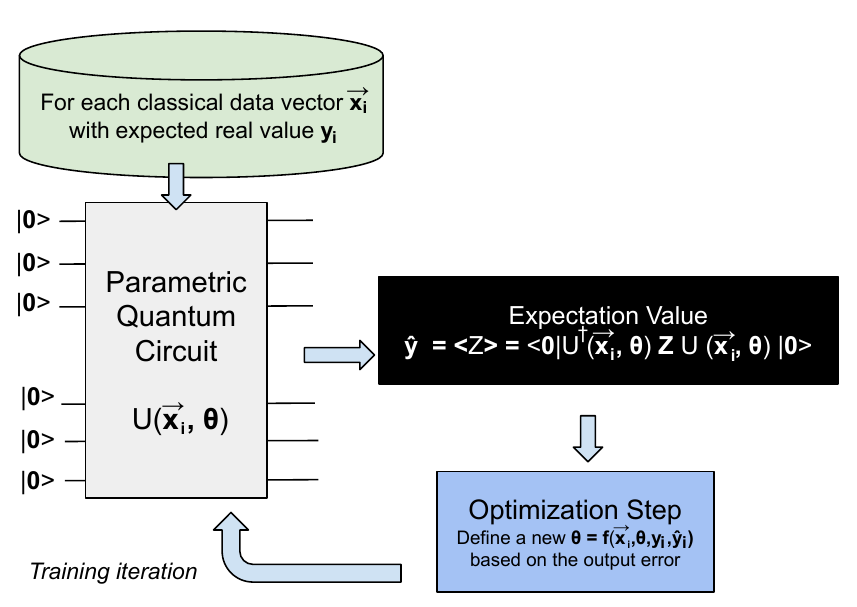}
\caption{Learning process of a parametric quantum circuit. The classical optimizer iteratively updates the parameters $\boldsymbol{\theta}$ based on the feedback from quantum measurements.}
\label{fig:learningPQC}
\end{figure}

In the context of parametric quantum circuits, any composition of quantum gates containing tunable rotation parameters can serve as a valid architecture.
To train such circuits, it is essential to define a loss function that quantifies the prediction error for a given input and parameter configuration.
This loss function guides a classical optimization algorithm, which iteratively updates the circuit parameters to minimize the loss value, thereby improving the model’s predictive performance.
Figure~\ref{fig:learningPQC} illustrates a generic overview of the parameter optimization process in a variational quantum circuit.

\subsubsection{Strongly Entangling Layers}
\label{sec:stronglyEntang}

Among the most widely used variational templates in quantum machine learning is the \textit{Strongly Entangling Layers} architecture, which provides an expressive yet hardware-efficient structure for parametrized quantum circuits~\cite{schuld2020circuit}. Each layer consists of two components arranged sequentially: a block of single-qubit rotations followed by a pattern of fixed, non-parametric two-qubit entangling gates. The parameter tensor supplied to the template has shape $(L, M, 3)$, where $L$ denotes the number of layers, $M$ the number of qubits, and the final dimension corresponds to the three rotation angles applied to each qubit in every layer. Thus, each layer implements three learnable single-qubit rotations on all $M$ wires, contributing a total of $3M$ trainable parameters. The circuit template is drawn in \ref{fig:StronglyEntanglingLayers}.

Following the local rotations, the layer applies a set of entangling operations designed to generate strong, globally distributed correlations across the register. These entanglers are instantiated through a fixed two-qubit gate---typically a CNOT, although other non-parametric two-qubit operation may be used---and are arranged according to a cyclic connectivity rule governed by the range hyperparameter $r$. For a given qubit $i$, the template couples it with qubit $(i+r) \bmod M$, ensuring that every qubit interacts with a partner displaced by $r$ positions along the register. By selecting different values of $r$ across successive layers, the architecture sweeps through complementary coupling patterns, thereby enhancing the propagation of information and the circuit's effective expressive power. If a single-qubit system is supplied, the template automatically omits entanglers, reducing to a purely local parametrized layer.

This construction yields a compact yet highly expressive variational block, capable of generating substantial entanglement with depth linear in the number of qubits. The design is inspired by the circuit-centric classifier \textit{et al.}~\cite{schuld2020circuit}, and has become a standard choice in QML applications due to its balanced trade-off between expressivity, trainability, and compatibility with gradient-based optimization methods. Its layered structure, consisting of parametrized rotations interleaved with deterministic entanglers, makes it well suited for supervised learning, quantum neural network models, and general variational tasks where strong entanglement and efficient parameterization are required.

\newcommand{\Rbox}[3]{R(\alpha_{#1}^{#3},\,\beta_{#1}^{#3},\,\gamma_{#1}^{#3})}

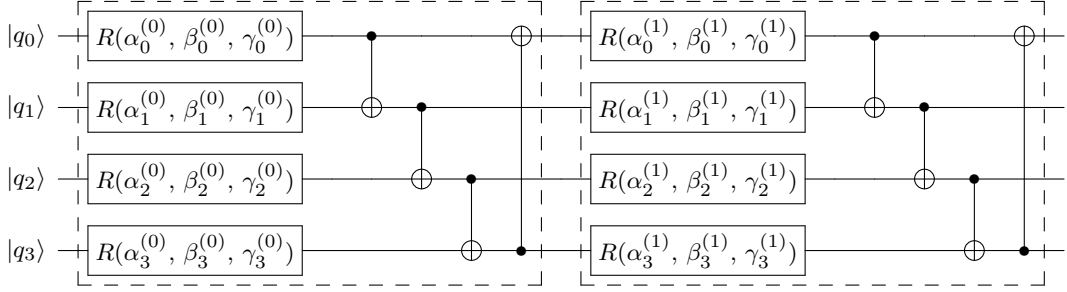
\begin{figure}[ht]
    \centering
    \small
\Qcircuit @C=1.2em @R=.9em {
\lstick{\ket{q_0}} &
  \gate{\Rbox{0}{ }{(0)}} & \qw & \ctrl{1} & \qw      & \qw      & \targ  & \qw & 
  \gate{\Rbox{0}{ }{(1)}} & \qw & \ctrl{1} & \qw      & \qw      & \targ  & \qw  \\
\lstick{\ket{q_1}} &
  \gate{\Rbox{1}{ }{(0)}} & \qw & \targ    & \ctrl{1} & \qw      & \qw    & \qw &
  \gate{\Rbox{1}{ }{(1)}} & \qw & \targ    & \ctrl{1} & \qw      & \qw    & \qw \\
\lstick{\ket{q_2}} &
  \gate{\Rbox{2}{ }{(0)}} & \qw & \qw      & \targ    & \ctrl{1} & \qw    & \qw &
  \gate{\Rbox{2}{ }{(1)}} & \qw & \qw      & \targ    & \ctrl{1} & \qw    & \qw  \\
\lstick{\ket{q_3}} &
  \gate{\Rbox{3}{ }{(0)}} & \qw & \qw      & \qw      & \targ    & \ctrl{-3} & \qw &
  \gate{\Rbox{3}{ }{(1)}} & \qw & \qw      & \qw      & \targ    & \ctrl{-3} & \qw 
  \gategroup{1}{2}{4}{7}{.8em}{--}
  \gategroup{1}{9}{4}{14}{.8em}{--}
}
\caption{Strongly Entangling Layers (PennyLane implementation).
Each layer applies three-parameter rotations $R(\alpha,\beta,\gamma)$ on every qubit, followed by CNOTs with periodic pattern: in layer $l$ with range $r=l \bmod M$, each qubit $i$ controls qubit $(i+r) \bmod M$. The weight tensor has shape $(L, M, 3)$~\cite{bergholm2022pennylane}.}
    \label{fig:StronglyEntanglingLayers}
\end{figure}

\subsubsection{Basic Entangler Layer}
\label{sec:basicEntang}
\newcommand{\R}[2]{R(\theta_{#1}^{(#2)})}

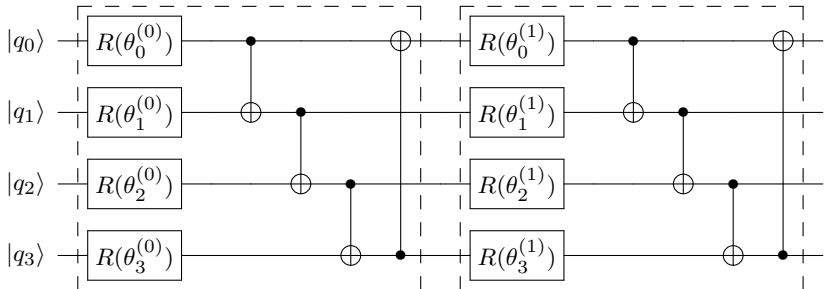
\begin{figure}[ht]
    \centering
    \small
\Qcircuit @C=1.2em @R=.9em {
\lstick{\ket{q_0}} &
  \gate{\R{0}{0}} & \qw & \ctrl{1} & \qw      & \qw      & \targ  & \qw & 
  \gate{\R{0}{1}} & \qw & \ctrl{1} & \qw      & \qw      & \targ  & \qw  \\
\lstick{\ket{q_1}} &
  \gate{\R{1}{0}} & \qw & \targ    & \ctrl{1} & \qw      & \qw    & \qw &
  \gate{\R{1}{1}} & \qw & \targ    & \ctrl{1} & \qw      & \qw    & \qw \\
\lstick{\ket{q_2}} &
  \gate{\R{2}{0}} & \qw & \qw      & \targ    & \ctrl{1} & \qw    & \qw &
  \gate{\R{2}{1}} & \qw & \qw      & \targ    & \ctrl{1} & \qw    & \qw  \\
\lstick{\ket{q_3}} &
  \gate{\R{3}{0}} & \qw & \qw      & \qw      & \targ    & \ctrl{-3} & \qw &
  \gate{\R{3}{1}} & \qw & \qw      & \qw      & \targ    & \ctrl{-3} & \qw 
  \gategroup{1}{2}{4}{7}{.8em}{--}
  \gategroup{1}{9}{4}{14}{.8em}{--}
}
    \caption{Basic Entangler Layers (PennyLane implementation). Each layer applies single-parameter rotations (default $R_X$) on all qubits, followed by a closed ring of CNOT gates connecting neighboring qubits with periodic boundary conditions. The weight tensor has shape $(L, M)$ where $L$ is the number of layers and $M$ the number of qubits; for two qubits, the periodic connection is omitted to avoid repeated entanglement~\cite{bergholm2022pennylane}.}
    \label{fig:BasicEntanglerLayer}
\end{figure}

Another commonly employed variational template in quantum machine learning is the \textit{Basic Entangler Layers} architecture, which offers a simple yet effective mechanism for generating entanglement across a quantum register. Each layer of this template consists of a block of single-qubit rotations followed by a closed ring of fixed, non-parametric two-qubit entangling gates. The trainable parameters are supplied through a tensor of shape $(L, M)$, where $L$ denotes the number of layers and $M$ is the number of qubits acted upon by the template. Each element of this tensor specifies the angle of a one-parameter rotation applied to a corresponding qubit; unless otherwise specified through the \texttt{rotation} argument, these rotations default to $R_X$ gates. Consequently, each layer contributes exactly $M$ trainable parameters. The circuit template is drawn in \ref{fig:BasicEntanglerLayer}.

Following the local rotations, the template applies a sequence of two-qubit entanglers arranged in a ring topology. In its standard configuration, each qubit $i$ is coupled to its immediate neighbor $(i+1) \bmod M$ via a CNOT gate, thereby forming a closed chain that wraps around the register. This cyclic structure ensures that correlations can propagate across the entire system, enabling the circuit to capture global features of the data with minimal depth. When the template operates on only two qubits, it follows the established convention of applying a single CNOT per layer, thereby avoiding repeated use of the same entangler on an identical pair of wires. If a single qubit is provided, the template naturally reduces to a purely local parametrized layer, since no entanglement can be generated.

The \textit{Basic Entangler Layers} template is widely used due to its simplicity, hardware friendliness, and compatibility with gradient-based variational optimization. Its fixed entangling pattern minimizes architectural complexity while still enabling the formation of sufficiently expressive quantum states for many learning tasks. The combination of lightweight parameterization and an efficient entanglement structure makes this template a standard baseline for benchmarking variational quantum circuits and serves as a natural point of comparison for more expressive or automatically synthesized architectures.

\subsubsection{Simplified Two Design}
\label{sec:simp2design}

The \textit{Simplified Two Design} architecture is a variational template inspired by the 2-design constructions analyzed by Cerezo \emph{et al.}~\cite{cerezo2021cost}, proposed originally to study trainability and the emergence of barren plateaus in quantum optimization landscapes. In random quantum circuit theory, a unitary 2-design is an ensemble of unitaries whose first and second statistical moments match those of the Haar measure, thereby providing an efficient surrogate for sampling random unitaries. While the present template does not constitute a strict unitary 2-design---since it is not built from a universal set of two-qubit gates---it retains several of the structural properties associated with 2-designs and has been demonstrated to induce expressive quantum states suitable for the study of variational optimization behavior. The circuit template is drawn in \ref{fig:SimplifiedTwoDesign}.

The construction begins with an initial layer of single-qubit $R_Y$ rotations, controlled by a vector of parameters of dimension $M$, where $M$ denotes the number of qubits. This initialization is followed by $L$ main layers, each designed to alternate entanglement patterns in a manner reminiscent of brickwork-style random circuits. Each layer is composed of two parts: an ``even'' block and an ``odd'' block. The even block applies controlled-$Z$ (CZ) entanglers to qubit pairs $(0,1)$, $(2,3)$, and so forth, followed by a pair of parametrized $R_Y$ rotations applied independently to each qubit in the entangled pair. The odd block then shifts this pattern by one qubit, applying CZ gates to pairs $(1,2)$, $(3,4)$, and so on, again followed by the corresponding local rotations. Through this alternating pattern, every qubit interacts with both of its neighbors across successive layers, promoting effective information scrambling and a diverse entanglement structure.

The parameterization of the template reflects this layered organization. The initial layer weights form a vector of shape $(M)$, while the weights for the main body of the circuit are organized in a tensor of shape $(L, M-1, 2)$, corresponding to $M-1$ pairs of rotation angles per layer. As a result, each layer applies $2(M-1)$ trainable single-qubit rotations interleaved with deterministic CZ gates. This structure preserves a relatively shallow depth while enabling the circuit to approximate, in practice, several of the expressivity properties expected from true 2-design ensembles.

Overall, the \textit{Simplified Two Design} offers a balance between architectural simplicity and expressive richness, providing a template that is both theoretically motivated and practically useful in variational quantum algorithms. Its layered CZ--$R_Y$ structure makes it well suited for analyzing optimization landscapes, benchmarking variational models, and comparing the trainability of different ansatz families under controlled entanglement and parameterization patterns.

\newcommand{\RY}[1]{\ensuremath{R_Y(#1)}}
\newcommand{\RX}[1]{\ensuremath{R_X(#1)}}
\newcommand{\RZ}[1]{\ensuremath{R_Z(#1)}}

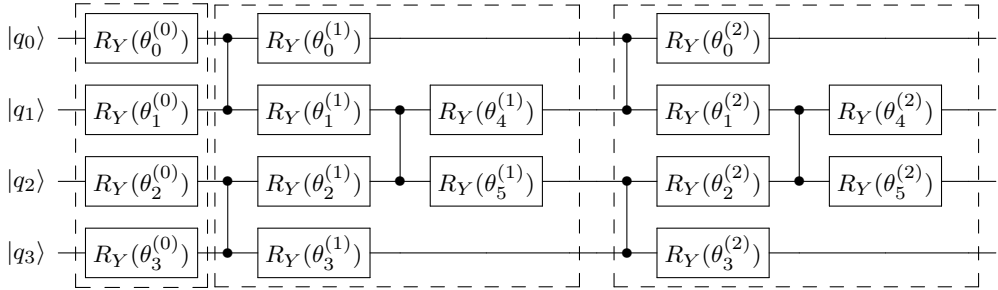
\begin{figure}[ht]
    \centering
    \small
\Qcircuit @C=1.1em @R=.9em {
\lstick{\ket{q_0}} &
  \gate{\RY{\theta_{0}^{(0)}}} & \ctrl{1} & \gate{\RY{\theta_{0}^{(1)}}} & \qw & \qw & \qw & \qw &
  \ctrl{1} & \gate{\RY{\theta_{0}^{(2)}}} & \qw & \qw & \qw & \qw & \\
\lstick{\ket{q_1}} &
  \gate{\RY{\theta_{1}^{(0)}}} & \ctrl{0} & \gate{\RY{\theta_{1}^{(1)}}} & \ctrl{1} & \gate{\RY{\theta_{4}^{(1)}}} & \qw & \qw &
  \ctrl{0} & \gate{\RY{\theta_{1}^{(2)}}} & \ctrl{1} & \gate{\RY{\theta_{4}^{(2)}}} & \qw & \qw &\\
\lstick{\ket{q_2}} &
  \gate{\RY{\theta_{2}^{(0)}}} & \ctrl{1} & \gate{\RY{\theta_{2}^{(1)}}} & \ctrl{0} & \gate{\RY{\theta_{5}^{(1)}}} & \qw & \qw &
  \ctrl{1} & \gate{\RY{\theta_{2}^{(2)}}} & \ctrl{0} & \gate{\RY{\theta_{5}^{(2)}}} & \qw & \qw &\\
\lstick{\ket{q_3}} &
  \gate{\RY{\theta_{3}^{(0)}}} & \ctrl{0} &
  \gate{\RY{\theta_{3}^{(1)}}} & \qw & \qw & \qw & \qw &
  \ctrl{0} & \gate{\RY{\theta_{3}^{(2)}}} & \qw & \qw & \qw & \qw 
  \gategroup{1}{2}{4}{2}{.8em}{--}   
  \gategroup{1}{3}{4}{7}{.8em,2.5em}{--}   
  \gategroup{1}{9}{4}{13}{.8em,2.5em}{--}  
}
    \caption{Simplified Two-Design architecture (PennyLane implementation). The circuit consists of an initial layer of Pauli-Y rotations on all qubits, followed by two processing units (dashed boxes). Each unit implements a CZ entangler (alternating control positions between units) followed by two RY rotations per qubit. The weight tensor has shape $(L, M-1, 2)$ where $L$ is the number of layers and $M$ is the number of qubits~\cite{bergholm2022pennylane}.}
\label{fig:SimplifiedTwoDesign}
\end{figure}

\section{Complexity Measures for Regression}
\label{sec:complexityMetrics}

Metalearning investigates how to use metaknowledge to adapt machine-learning and data-mining processes more efficiently. Although many algorithms can produce good models, choosing the most appropriate one still requires a systematic strategy. Metalearning offers this strategy by enabling systems to improve their performance through accumulated experience~\cite{brazdil2022metalearning}. 
In \cite{lorena2018data}, it was introduced and evaluated novel regression complexity measures that effectively characterize the functional structure of regression datasets and serve as informative meta-features, enabling accurate prediction of function types, tuning of support vector regressor hyperparameters, and forecasting of regressor performance, often matching or surpassing classical meta-features in meta-learning tasks.
This section introduces that collection of numerical indicators designed to quantify the intrinsic difficulty of regression tasks. These measures examine the learning problem from distinct viewpoints and are organized into four major groups:

\begin{itemize}
    \item \textbf{Feature--output association}: characterizes how individual predictors relate to the target variable;
    \item \textbf{Linearity indicators}: evaluate whether linear functions are adequate to approximate the underlying data-generating process;
    \item \textbf{Smoothness criteria}: assess how smoothly the target varies with respect to the input features;
    \item \textbf{Geometric, topological, and density descriptors}: capture how samples are spatially arranged in the input--output domain.
\end{itemize}

Let $\mathbf{X}$ denote the data matrix composed of $n$ observations described by $d$ predictors. The $i$-th instance is expressed as $\mathbf{x}_i \in \mathbb{R}^{d}$, whereas the $j$-th feature corresponds to the column vector $\mathbf{x}_{j}$. The response vector is given by $\mathbf{y} = (y_1,\dots,y_n)^\top$, where each $y_i \in \mathbb{R}$.

Several metrics rely on properties of a multiple linear regression model of the form
\[
f(\mathbf{x}) = \beta_{0} + \beta_{1} x_{1} + \dots + \beta_{d} x_{d} + \varepsilon,
\]
where $\beta_{k}$ are the regression coefficients and $\varepsilon$ denotes the residual error.

\subsection{Feature--Output Association}

Measures in this category quantify the strength of dependence between each predictor and the target variable. High dependence implies that the regression function can be approximated using simpler models. These metrics treat each predictor independently—an intentional simplification that allows a coarse yet informative assessment of the factors contributing to task difficulty.

\subsubsection{Maximum Correlation (C1)}

For each feature $\mathbf{x}_{j}$, we compute the absolute Spearman rank correlation with $\mathbf{y}$, denoted by $|\rho(\mathbf{x}_{j}, \mathbf{y})|$. Since the correlation domain is $[-1,1]$, large values (positive or negative) reflect strong monotonic associations. The measure $C1$ is defined as
\[
C1 = \max_{j=1,\dots,d} |\rho(\mathbf{x}_{j},\mathbf{y})|.
\]
Spearman correlation is employed because it is nonparametric and does not assume any specific form for the predictor–target relationship. Computing $C1$ requires $O(d\,n\log n)$ operations. Larger values indicate simpler problems, as at least one predictor strongly aligns with the response.

\subsubsection{Average Correlation (C2)}

Instead of focusing on the best predictor, $C2$ summarizes the overall predictor--target dependency by averaging the absolute Spearman correlations:
\[
C2 = \frac{1}{d} \sum_{j=1}^{d} |\rho(\mathbf{x}_{j},\mathbf{y})|.
\]
Its computational complexity is also $O(d\,n\log n)$. High values imply that many features contribute meaningful monotonic information regarding the target.

\subsubsection{Individual Feature Efficiency (C3)}

This measure evaluates how many samples must be progressively discarded from a dataset for a predictor to reach a high correlation threshold with the target. For every feature $\mathbf{x}_j$, we remove the minimum number of samples necessary so that $|\rho(\mathbf{x}_j, \mathbf{y})| > 0.9$. Let $n_j$ denote the count of discarded samples. Then,
\[
C3 = \min_{j=1,\dots,d} \frac{n_j}{n}.
\]
A naive procedure would require cubic time in $n$, but an optimized version using ranking updates yields a worst-case complexity of $O(d\,n^{2})$. Low values indicate simpler problems because a strong monotonic relationship emerges without removing many points.

\subsubsection{Collective Feature Efficiency (C4)}

This metric evaluates how effectively the full set of predictors, taken sequentially, can explain the data through simple linear fits. Starting from the most correlated feature, points whose absolute residuals satisfy $|\varepsilon_i| \leq 0.1$ are removed. The process repeats with the next most correlated feature applied to the remaining subset. If $T_{\ell}$ denotes the final data subset after $\ell$ iterations, the measure is
\[
C4 = \frac{\#\{ \mathbf{x}_i \in T_{\ell} \mid |\varepsilon_i| > 0.1 \}}{|T_{\ell}|}.
\]
The computational cost is $O(d^{2} + d\,n\log n)$. Higher values indicate more complex datasets, as many points fail to align with simple linear explanations even when features are used sequentially.

\subsection{Linearity Indicators}

These metrics quantify the adequacy of a linear model to describe the dataset. When linear regression yields small residuals, the problem is considered inherently simpler.

\subsubsection{Mean Absolute Residual (L1)}

The first linearity measure is the mean absolute deviation of residuals obtained from a multiple linear regression:
\[
L1 = \frac{1}{n} \sum_{i=1}^{n} |\varepsilon_i|.
\]
Lower values denote problems that are well approximated by linear functions. The overall computational complexity is $O(n d^{2})$ due to the regression estimation.

\subsubsection{Residual Variance (L2)}

The second measure computes the mean squared residual:
\[
L2 = \frac{1}{n} \sum_{i=1}^{n} \varepsilon_i^{2}.
\]
Smaller values again suggest that the target can be described with a linear model. This measure also requires $O(n d^{2})$ time.

\subsection{Smoothness Criteria}

Smoothness indicators evaluate how gradually the target variable changes with respect to the input. Problems where nearby samples exhibit similar outputs tend to be easier for regression algorithms.

\subsubsection{Output Variation Along a Minimum Spanning Tree (S1)}

Following the spirit of MST-based measures in earlier literature, we construct a Minimum Spanning Tree on the input space, where vertices correspond to data points and edge weights represent Euclidean distances. Let $MST$ denote the set of adjacent pairs $(i,j)$. The measure is defined as
\[
S1 = \frac{1}{n} \sum_{(i,j)\in MST} |y_i - y_j|.
\]
Low values imply that nearby points in the input space also have similar outputs. Constructing the MST requires $O(n^{2} d)$ for distance computation plus $O(n^{2})$ for Prim's algorithm.

\subsubsection{Input Variation Across Ordered Outputs (S2)}

After sorting samples by their target values, we compute the Euclidean distance between successive samples:
\[
S2 = \frac{1}{n}\sum_{i=2}^{n} \| \mathbf{x}_i - \mathbf{x}_{i-1} \|_{2}.
\]
Low values indicate that samples with similar outputs lie close in the input space. The computation requires $O(n(d + \log n))$ operations.

\subsubsection{Nearest-Neighbor Regression Error (S3)}

This measure computes the leave-one-out mean squared error of a 1-nearest-neighbor regressor:
\[
S3 = \frac{1}{n} \sum_{i=1}^{n} \big( NN(\mathbf{x}_i) - y_i \big)^{2},
\]
where $NN(\mathbf{x}_i)$ is the prediction given by the nearest neighbor of $\mathbf{x}_i$ in the input space. High values correspond to irregular or sparse regions in the dataset. The asymptotic complexity is $O(d\,n^{2})$.

\subsection{Geometric, Topological, and Density Descriptors}

These measures capture how samples populate the input space and how the regression function behaves under synthetic perturbations or geometric transformations.

\subsubsection{Nonlinearity of Linear Regression (L3)}

Adapted from earlier nonlinearity measures for classification, this metric evaluates how a linear regressor performs on artificially generated points. Pairs of samples with adjacent target values are linearly interpolated to form new test points $(\mathbf{x}_i', y_i')$. A linear model is fitted to the original data, and its mean squared error on the interpolated set is computed as
\[
L3 = \frac{1}{\ell} \sum_{i=1}^{\ell} \big( f(\mathbf{x}_i') - y_i' \big)^{2},
\]
where $\ell = n - 1$. Low values indicate that the linear model behaves consistently in regions between observed samples. The complexity is $O(n(d^{2} + \log n))$.

\subsubsection{Nonlinearity of Nearest-Neighbor Regression (S4)}

Using the same interpolated points as $L3$, this measure replaces the linear regressor with a 1-nearest-neighbor predictor:
\[
S4 = \frac{1}{\ell} \sum_{i=1}^{\ell} \big( NN(\mathbf{x}_i') - y_i' \big)^{2}.
\]
Datasets requiring highly nonlinear behaviors yield higher values. Using a KD-tree implementation results in an asymptotic complexity of $O(n\,d\log n)$.

\subsubsection{Average Sample Density per Dimension (T2)}

This simple indicator evaluates data sparsity by computing the average sample count per feature dimension:
\[
T2 = \frac{n}{d}.
\]
Low values correspond to sparse or high-dimensional datasets, which typically impose additional complexity on regression algorithms. Its complexity is linear in $n$ and $d$, i.e $O(n+d)$.

\section{Genetic Algorithm Search Proposed Approach: Reduced Regressor Quantum Neural Network}
\label{sec:geneticAlgorithmFORRRQNN}

In this work, we introduce a Reduced Regressor Quantum Neural Network (RRQNN) whose variational architecture is automatically discovered through a Genetic Algorithm (GA).  
Each candidate quantum circuit is encoded as an integer vector
\begin{equation}
    \mathbf{\Lambda_i} = [ \Lambda_1, \Lambda_2, \ldots, \Lambda_{3n_{\text{layers}}} ],
    \label{eq:lambda_vector}
\end{equation}
whose total size is exactly $3n_{\text{gates}}$, where $n_{gates}$ is the maximum quantity of gates in the circuit.
This full vector represents the complete structural characterization of the quantum circuit discovered by the GA.

The chromosome is partitioned into three equal segments:
{\footnotesize 
\begin{equation}
    \mathbf{\Lambda_i}
    =
\underbrace{[\Lambda_1,\ldots,\Lambda_{n_{\text{gates}}}]}_{\text{Gate IDs}}
    \;\Vert\;
    \underbrace{[\Lambda_{n_{\text{gates}}+1},\ldots,\Lambda_{2n_{\text{gates}}}]}_{\text{Control Qubits}}
    \;\Vert\;
    \underbrace{[\Lambda_{2n_{\text{gates}}+1},\ldots,\Lambda_{3n_{\text{gates}}}]}_{\text{Target Qubits}}.
    \label{eq:lambda_partition}
\end{equation}
}
Only the first segment, containing the Gate IDs, determines the type of operation applied at each layer.  
The second and third segments specify, respectively, the control and target qubits associated with the operations.  
These indices are meaningful only when the circuit has two or more qubits; when $n_{\text{qubits}} = 1$, controlled operations collapse into their single-qubit counterparts and the control/target entries are ignored.

For single-qubit architectures, each gate identifier satisfies $\Lambda_j \in \{0,\ldots,6\}$, while for two-qubit architectures the search space expands to $\Lambda_j \in \{0,\ldots,13\}$, enabling the encoding of controlled operations, where 6 and 13 indicate the index of the operators that will be mapped in the quantum circuit (explained in detail in Section \ref{sec:mappingAG}).

The RRQNN receives as input a real-valued sample
\begin{equation}
    \mathbf{x} = \{ x_1, x_2, \ldots, x_{\upsilon} \},
\end{equation}
where $\upsilon$ corresponds to the dimensionality of the training set, and it contains a vector of trainable parameters
\begin{equation}
    \mathbf{\theta} = \{ \theta_1, \theta_2, \ldots, \theta_{n_{gates}} \},
\end{equation}

\subsection{Genetic Algorithm Formulation}

The genetic algorithm was implemented using the PyGAD Python library with default operator configurations. The algorithmic flow begins by instantiating a population of 20 randomly generated individuals, where each individual represents a distinct quantum circuit architecture. Each individual is subsequently evaluated and assigned a fitness score.

The fitness function is defined as $\text{Loss} = -1 \times R^2$, where lower values indicate superior performance. Fitness scores are computed based on the training performance of each circuit, as evaluated by the OPTAX optimizer on the training dataset.
Following the evaluation phase, we employ a steady-state selection strategy combined with single-point crossover. The four best-performing individuals are identified and retained as elites. Two parents are randomly selected from this elite set to generate offspring through single-point crossover, where a random gene index is chosen as the crossover point and genetic material is exchanged between parents at this location.

During offspring generation, a mutation operator is applied where 10\% of the genome is randomly modified, with mutated genes assigned new randomly selected values drawn from the predefined gene space. The mutation operation has a per-individual probability of 20\%.

\begin{figure}[ht]
    \centering
\includegraphics[width=1.\linewidth]{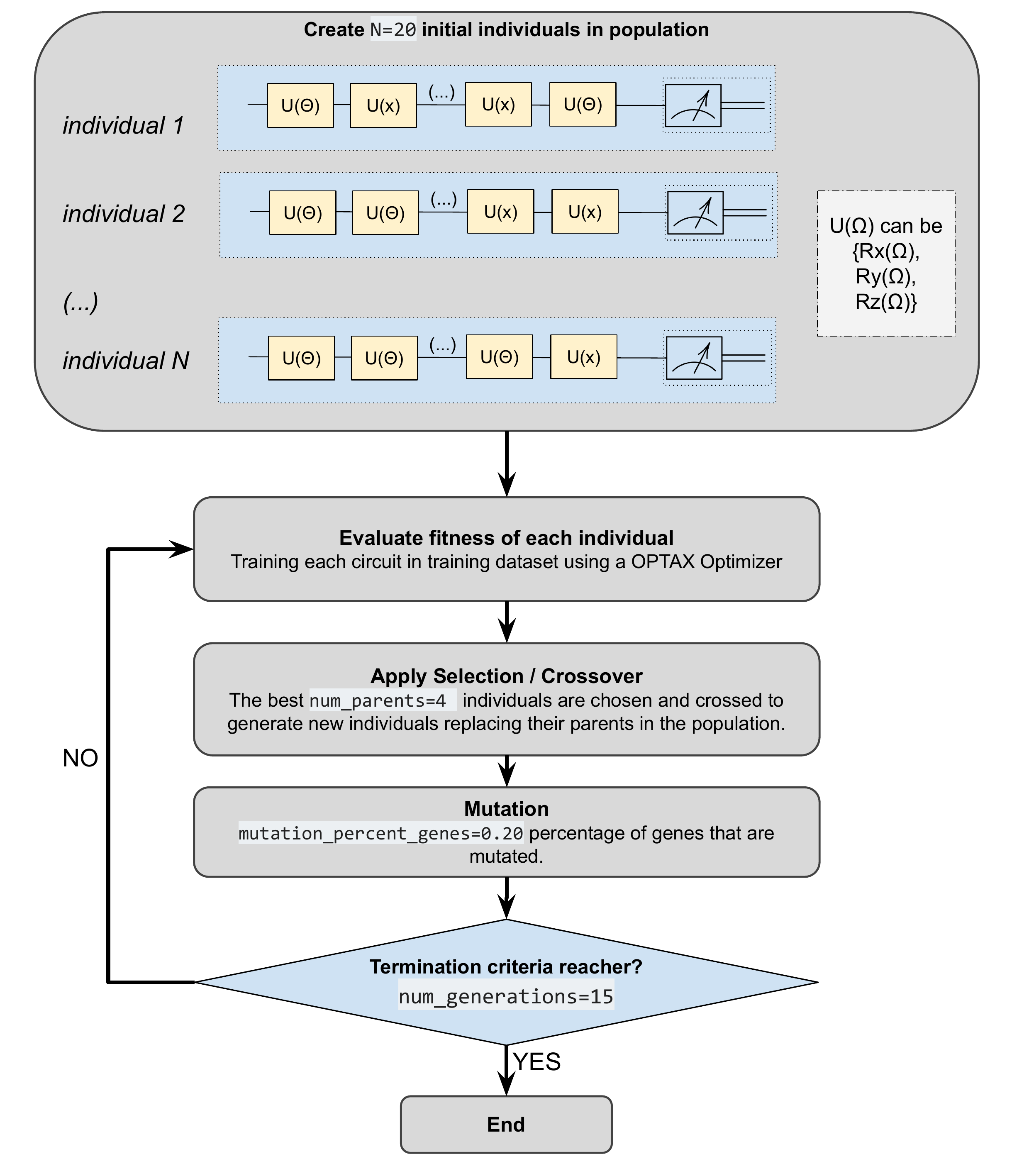}
    \caption{Execution flow of the genetic algorithm generating reduced quantum neural networks with multiple input loading.}
    \label{fig:geneticAlgoFlow}
\end{figure}

After the new population is generated, fitness evaluation is repeated for all individuals. This iterative process continues until the termination criterion is satisfied. In our formulation, the algorithm evolves for 15 generations, after which the best individual in the population is selected as the final optimized quantum circuit. The complete algorithmic flow is illustrated in Fig.~\ref{fig:geneticAlgoFlow}.

\subsection{Deterministic Mapping from Chromosome to Quantum Circuit}
\label{sec:mappingAG}

Given the chromosome 
\(
\mathbf{\Lambda_i} \in \mathbb{Z}^{3n_{\text{layers}}}
\),
the construction of the quantum circuit is performed deterministically and sequentially.  
For a circuit composed of \(n_{\text{layers}}\) gate positions, the chromosome is partitioned as follows as described in Equation \ref{eq:lambda_partition}.

Thus, each layer \(i\) of the circuit is fully specified by the triplet:
\begin{equation}
    g_i = \Lambda_i,\quad
    c_i = \Lambda_{n_{\text{layers}}+i} \bmod n_{\text{qubits}},\quad
    t_i = \Lambda_{2n_{\text{layers}}+i} \bmod n_{\text{qubits}}.
\end{equation}

The value \(g_i\) selects the gate type; \(c_i\) and \(t_i\) determine the qubits involved in the operation.  
For multi-qubit circuits, this encoding supports both single-qubit and controlled operations.  
For single-qubit circuits (\(n_{\text{qubits}} = 1\)), control and target indices collapse trivially to the same qubit, and controlled gates automatically reduce to their single-qubit versions.  
This uniform representation allows the GA to explore architectures across any number of qubits while maintaining a fixed and consistent chromosome structure.

During circuit construction, gates that require trainable parameters or input-dependent angles simply draw their corresponding values in the order they appear in the parameter vector and in the input feature vector. In this way, the sequential progression of the chromosome naturally determines the order in which trainable weights and feature angles are consumed, ensuring that each gate receives the correct argument without the need for explicitly tracking additional indices or counters.

The mapping rules applied to each layer are:

\begin{itemize}
    \item If $g_i = 0$, no operation is applied.
    \item If $g_i = 1$ or $(g_i = 5 \wedge c_i = t_i)$, apply $RX(\theta_{k}, \text{wires}=t_i)$ .
    \item If $g_i = 2$ or $(g_i = 6 \wedge c_i = t_i)$, apply $RY(\theta_{k}, \text{wires}=t_i)$.
    \item If $g_i = 3$ or $(g_i = 7 \wedge c_i = t_i)$, apply $RZ(\theta_{k}, \text{wires}=t_i)$.
    \item If $g_i = 4$ and $c_i \neq t_i$, apply $CNOT([c_i,t_i])$.
    \item If $g_i \in \{5,6,7\}$ and $c_i \neq t_i$, apply:
    \begin{align}
        g_i=5 &: CRX(\theta_k, [c_i,t_i]),\\
        g_i=6 &: CRY(\theta_k, [c_i,t_i]),\\
        g_i=7 &: CRZ(\theta_k, [c_i,t_i]).
    \end{align}
    \item If $g_i = 8$ or $(g_i = 11 \wedge c_i = t_i)$, apply $RX(x_{m},\text{wires}=t_i)$.
    \item If $g_i = 9$ or $(g_i = 12 \wedge c_i = t_i)$, apply $RY(x_{m},\text{wires}=t_i)$.
    \item If $g_i = 10$ or $(g_i = 13 \wedge c_i = t_i)$, apply $RZ(x_{m},\text{wires}=t_i)$.
    \item If $g_i \in \{11,12,13\}$ and $c_i \neq t_i$, apply:
    \begin{align}
        g_i=11 &: CRX(x_m,[c_i,t_i]),\\
        g_i=12 &: CRY(x_m,[c_i,t_i]),\\
        g_i=13 &: CRZ(x_m,[c_i,t_i]).
    \end{align}
\end{itemize}

Here, \(x_m\) denotes the feature angle and \(\theta_k\) the trainable parameter associated with the operation.

\begin{figure}[ht]
    \centering
\Qcircuit @C=1.2em @R=.9em {
\lstick{\ket{0}} &
  \gate{\RY{\theta_1}} & \gate{\RX{x_1}} & \gate{ \RX{\theta_2} }  & \gate{\RZ{x_2}}  & \gate{\RY{x_3}}%
}
    \caption{Quantum circuit that is encoded by the chromosome vector [2,\,4,\,1,\,6,\,5] as explained in Section ~\ref{sec:geneticAlgorithmFORRRQNN}.}
    \label{fig:exampleQC_GeneticAlgo}
\end{figure}
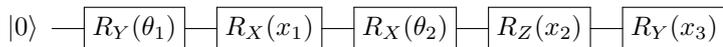

An example of representative encoded individual $[2,\,4,\,1,\,6,\,5]$ produces the quantum circuit shown in Figure~\ref{fig:exampleQC_GeneticAlgo}.

\section{Experimental Protocol}
\label{sec:experimentalProtocol}

This section describes the experimental methodology employed to evaluate and compare a broad spectrum of quantum and classical regression models under controlled and reproducible conditions. All experiments were conducted on a suite of 22 three-dimensional benchmark functions, comprising unimodal and multimodal landscapes of varying structural complexity and nonlinear behavior. Each benchmark dataset consisted of \(900\) samples, generated uniformly over the input domain. For every dataset, \(70\%\) of the samples were used for training, while the \emph{entire dataset} (\(100\%\)) was employed to evaluate each model’s ability to reproduce the full underlying function, thus providing a comprehensive assessment of generalization performance and functional fidelity. For every model and every function, a total of ten independent runs were executed.

All experiments were conducted on a workstation equipped with 16\,GB of RAM and a 12th~Generation Intel\textsuperscript{\textregistered} Core\texttrademark{} i7--12700H processor with 20 logical cores, and a 512\,GB solid-state drive. The operating system was Ubuntu~22.04~LTS. The software stack used in the experiments included Python~3.11.5, \texttt{scikit-learn}~1.6.1 for classical regression models implementation, PennyLane~0.42.3 for quantum circuit simulation, and \texttt{jax}~0.6.2 together with \texttt{optax}~0.2.6 for differentiable programming and gradient-based optimization. The genetic algorithm experiments were implemented using the \texttt{pygad} library (version~2.5.0). The \texttt{problexity} library (version~0.5.11) \cite{komorniczak2023problexity} was used to generate the complexity metrics of the regression functions. This configuration ensured a stable and reproducible environment for the execution, optimization, and benchmarking of all quantum and classical regression models.

\subsection{Quantum Models}

A total of $44$ quantum regression models were evaluated. These models fall into two major categories: (i) fixed QNN ansatz architectures widely adopted in the literature, and (ii) architectures automatically constructed by a Genetic Algorithm (GA). All quantum models were trained using the Adam optimizer with a learning rate of $0.05$, implemented via \texttt{optax} library and leveraging the parallelization capabilities of \texttt{JAX}.

\subsubsection{Fixed QNNs Ansatz Families}

We considered three well-established $\text{QNN}$ as type of variational quantum circuit templates (explained in Sections \ref{sec:stronglyEntang}, \ref{sec:simp2design}, and \ref{sec:basicEntang}), i.e. $\text{QNN} \in \{ \textbf{StronglyEntanglingLayers}, \textbf{SimplifiedTwoDesign}, \textbf{BasicEntanglerLayers}\}$.
Each template was instantiated with different depths, corresponding to the following $L$ numbers of layers, where $L \in \{1, 2, 3, 4, 5, 6, 7, 8, 9, 10, 20, 40, 60\}$. Given the three types of circuits and the twelve possible depths, this results in 36 distinct quantum models. We named each one of these models as $\text{QNN-}L$. Before each layer of the quantum ansatz, the input features are loaded in parallel using $R_z$ rotation gates. That is, for an ansatz with $L$ layers, the input is re-encoded $L$ times, once before each layer. This repeated data re-uploading strategy significantly enhances the expressive capacity and supports the universal approximation properties of these models~\cite{schuld2008effect}.

\subsubsection{Genetic-Algorithm–Generated Quantum Circuits named as Reduced Regressor Quantum Neural Networks}

Beyond fixed architectures, we explored model discovery through an evolutionary design process. A Genetic Algorithm (GA) was executed ten times independently, with each run allowed to construct circuits acting on either one or two qubits. To control model complexity, the GA was constrained by upper bounds on the number of optimizable parameters. The generated RRQNN circuits were limited to $P_{max}$ quantum gates, where $P_{max} \in \{120, 60, 40, 20\}$, and $q$ for both 1-qubit and 2-qubit configurations. This yields 8 \text{RRQNNs} configurations, named each one as $\text{RRQNN-}{P_{max}}\text{-}{q}$.


\subsection{Classical Regression Models - Baselines}

To contextualize the performance of quantum models, we employed a diverse suite of $17$ classical regression baselines spanning distance-based, tree-based, kernel-based, and neural networks models. 
Each model and its settings are detailed in Table ~\ref{table:ClassicalRegressors}. The parameters not mentioned are the defaults in the \texttt{sklearn} library.
The classical regressor models were executed ten times on each dataset, using data partitions identical to the quantum models. 

The effective number of trainable parameters in each classical model reflects the intrinsic model capacity and serves as a baseline for discussing the expressive efficiency of quantum architectures. The parameter-counting procedure was systematically defined for each model family considered in this study, following the structural properties of their learning mechanisms.
For K-Nearest Neighbors (KNN) regressors, it was considered that KNN does not learn parametric representations; instead, they store the training samples and make predictions through distance-based interpolation. Consequently, their number of trainable parameters is defined as zero, since no optimization is performed over a parametric function space.

For decision-tree regressors, the model complexity is directly tied to the structure of the learned tree. Each internal or terminal node corresponds to a decision rule or leaf prediction, respectively. Therefore, the total number of trainable parameters is defined as the number of nodes in the learned tree, which captures the full set of data-dependent decisions encoded by the model.
Random Forests aggregate multiple decision trees. Thus, their total number of trainable parameters is computed as the sum of the node counts of all individual trees in the ensemble. This definition reflects the distributed capacity of the model, as each tree contributes an independent substructure to the overall decision process.

For neural-network-based regressors, the total number of trainable parameters comprises all weight matrices and bias vectors across layers. Specifically, the number of parameters is obtained by summing the sizes (number of scalar entries) of all learned weight tensors and all corresponding bias vectors. This yields a comprehensive measure of the network's expressive capacity, fully characterizing its parametric footprint.

The parameter count in SVR depends on the kernel employed. For linear kernels, the learned weight vector provides a direct parameterization of the decision function, and the parameter count is given by the total number of coefficients. In contrast, nonlinear kernels such as RBF, polynomial, or sigmoid do not maintain explicit weight vectors. Instead, their learned representation is defined through the support vectors, whose number and dimensionality jointly determine the effective parameter count. The total number of parameters is thus given by the size of the support-vector matrix.


\begin{table}[ht]
\footnotesize
\centering
\caption{Classical Regressors used in the experiments.}
\begin{tabular}{lr}
\hline
Model acronym                    & Model name and configuration \\ \hline
DT                               & Decision Tree Regressor, default configuration\\
RF                               & Random Forest Regressor, default configuration\\
MLP100-ReLU & MLP, one hidden layer with 100 neurons, ReLU function \\
MLP500-ReLU & MLP, one hidden layer with 500 neurons, ReLU function \\
MLP100-100-ReLU & MLP, two hidden layers, 100 neurons, ReLU function \\
MLP500-500-ReLU & MLP, two hidden layers, 500 neurons, ReLU function \\
MLP100-Id & MLP, one hidden layer with 100 neurons, identity function \\
MLP500-Id & MLP, one hidden layer with 500 neurons, identity function \\
MLP100-100-Id & MLP, two hidden layers, 100 neurons, identity function \\
MLP500-500-Id & MLP, two hidden layers, 500 neurons, identity function \\
SVR-RBF                          & Support Vector Regressor, C = 1.0, Kernel = RBF, Gamma = Scale\\
SVR-linear                       & Support Vector Regressor, C = 1.0, Kernel = Linear, Gamma = Scale\\ 
SVR-sigmoid                      & Support Vector Regressor, C = 1.0, Kernel = Sigmoid, Gamma = Scale\\
SVR-poly                         & Support Vector Regression, C = 1.0, Kernel = Poly, Gamma = Scale\\
knn2, knn3, knn4                 & Regression based on k-nearest neighbors, k = 2\\ 
knn3 & Regression based on k-nearest neighbors, k = 3 \\
knn4 & Regression based on k-nearest neighbors, k = 4\\
\hline
\end{tabular}
\label{table:ClassicalRegressors}
\end{table}

\subsection{Training and Evaluation Procedure}

For each of the 22 benchmark regression functions, the full dataset of $900$ points was randomly split into $70\%$ for training and $30\%$ for testing. 
Only the training portion was used to fit the models. However, to enable direct comparison across all methods, all reported performance metrics—$R^2$ score and Root-Mean-Square Error (RMSE)—were computed on the \emph{entire} dataset of 900 points. This protocol was repeated for ten independent random seeds, producing statistically meaningful performance estimates for every model~\cite{hastie2009elements}.

The $R^2$ score is defined as:
\begin{equation}
R^2 = 1 - \frac{\sum_{i=1}^{N} (y_i - \hat{y}_i)^2}{\sum_{i=1}^{N} (y_i - \bar{y})^2},
\end{equation}
where $y_i$ are the observed outputs, $\hat{y}_i$ are the predicted values, and $\bar{y}$ is the sample mean of the target variable.

The RMSE metric is computed as:
\begin{equation}
\mathrm{RMSE} = \sqrt{\frac{1}{N} \sum_{i=1}^{N} (y_i - \hat{y}_i)^2}.
\end{equation}

Both metrics were averaged over ten repetitions of the training–testing cycle for each model and each benchmark function.



\subsection{Proof of concept - 1-dimensional functions}
\label{sec:proofOfConcept}

To assess whether a Genetic Algorithm (GA) is capable of discovering high-quality Regressor-Ready Quantum Neural Networks (RRQNNs) in low-dimensional settings, we conducted a proof-of-concept experiment involving four nonlinear one-dimensional functions, described in Equations \ref{eq:fx1}, \ref{eq:fx2}, \ref{eq:fx3}, and \ref{eq:fx4}. The objective of this evaluation was twofold: (i) to verify whether the GA can assemble expressive quantum circuits from a restricted set of parameterized gates, and (ii) to compare the resulting RRQNN regressors against a diverse collection of classical models under identical training and testing conditions. The results obtained for all functions are summarized in Table~\ref{tab:resultsConceptProof}.

\begin{equation}
    f_1^{1D}(x) = x^2,
    \label{eq:fx1}
\end{equation}

\begin{equation}
    f_2^{1D}(x) = x^3,
    \label{eq:fx2}
\end{equation}

\begin{equation}
    f_3^{1D}(x) = 2x^4-1,
    \label{eq:fx3}
\end{equation}

\begin{equation}
    f_4^{1D}(x) = \frac{0.9}{1+e^{-10x}}.
    \label{eq:fx4}
\end{equation}

The GA was allowed to generate RRQNN architectures with varying circuit sizes, here expressed as the total number of parameterized quantum gates (10, 15, 20,  and 25). These RRQNNs were evaluated alongside standard classical baselines, including $k$-nearest neighbors (KNN), Support Vector Regressors (SVR), Random Forests (RF), Decision Trees (DT), and multilayer perceptrons (MLP). For each model, we report the $R^2$ score, the RMSE on the all dataset, providing a clear picture of the trade-off between model performance and complexity.

\subsubsection*{Discussion of Proof of Concept Results}

The results in Table~\ref{tab:resultsConceptProof} reveal several notable patterns.
First, across all four functions, RRQNNs with 20--25 gates consistently achieve the best or near-best performance among all tested models. For instance, on $f_{1}^{1D}$, the RRQNN with 25 gates attains $R^2 = 0.999$ with an RMSE of only $0.0092$, outperforming all classical baselines, including Random Forests and Decision Trees. This trend persists in $f_{2}^{1D}$, $f_{3}^{1D}$, and $f_{4}^{1D}$, where RRQNN-25 again exhibits top performance, achieving optimal accuracy ($R^2 = 1.000$) on $f_{4}^{1D}$ with only 25 parameters.

Second, classical models show a broader range of behaviors. While Random Forests and Decision Trees are competitive on most tasks, they do so at a much higher parameter cost; for example, RF uses more than 3600 parameters on $f_{1}^{1D}$, compared to just 25 for RRQNN-25. Simpler classical models such as KNN demonstrate reasonable predictive performance but cannot match the precision of the quantum circuits on more complex targets. Support Vector Regression shows highly heterogeneous behavior: although SVR-poly achieves moderate accuracy on $f_{2}^{1D}$, all kernel variants exhibit numerical instability or degenerate solutions for the other functions, reflected by $R^2 = -\infty$.

Third, multilayer perceptrons (MLPs) perform surprisingly poorly in this setting, often yielding extreme negative $R^2$ values, particularly for $f_{1}^{1D}$ and $f_{3}^{1D}$. This indicates severe overfitting or optimization failure, despite their large parameter count (from 301 up to more than 10,000 trainable weights). By contrast, RRQNNs succeed with only a few dozen parameters, highlighting their favorable inductive bias for structured, low-dimensional nonlinear problems.

Overall, the proof of concept demonstrates that the Genetic Algorithm is indeed capable of discovering expressive and parameter-efficient RRQNN architectures for 1D regression. These automatically generated quantum models do not merely match classical regressors; in most cases, they surpass them—while requiring drastically fewer parameters - which was expected as demonstrated in \cite{huang2022quantum}. The clear performance gap between RRQNNs and classical baselines, especially on the more intricate functions, confirms the viability of evolutionary search as a tool for quantum circuit design in regression tasks.

\begin{figure}[ht]
\centering
  \begin{tabular}{@{}cccc@{}}
    \includegraphics[width=.23\textwidth]{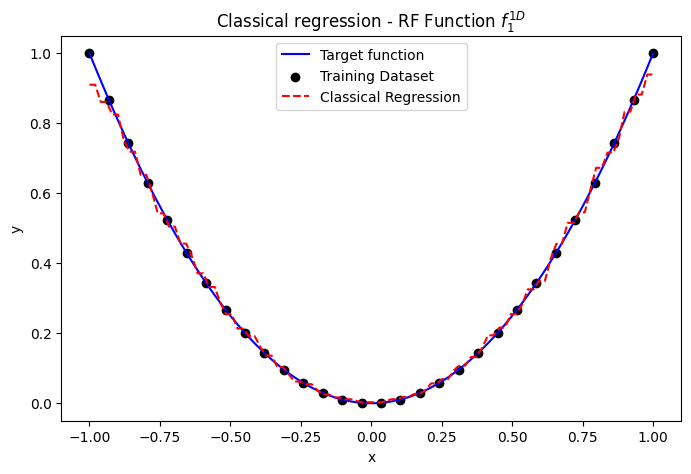} &
    \includegraphics[width=.23\textwidth]{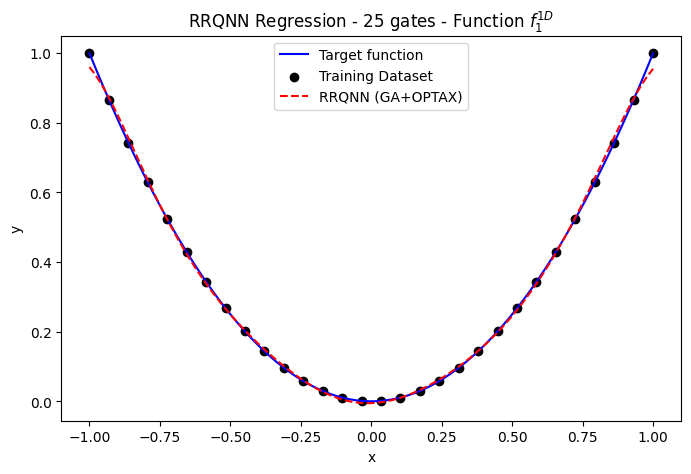} &
    \includegraphics[width=.23\textwidth]{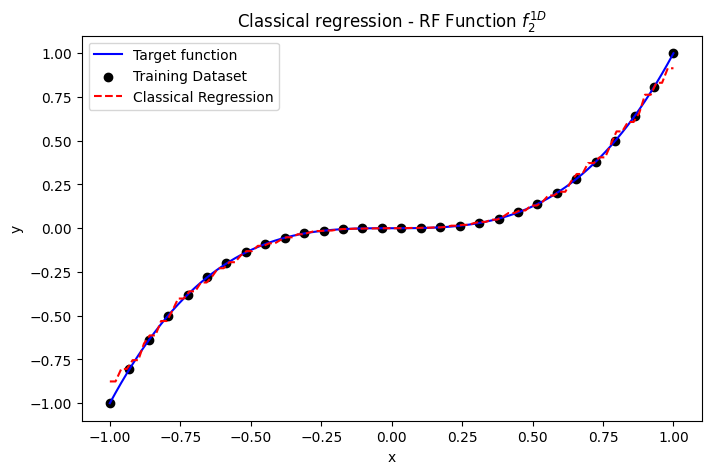} &
    \includegraphics[width=.23\textwidth]{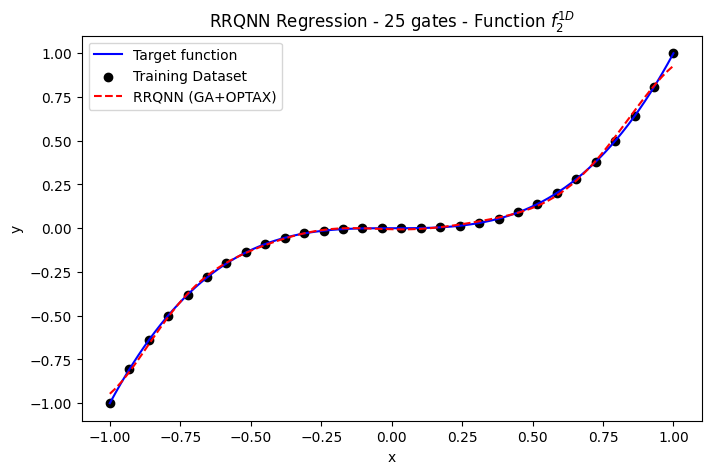}   \\
    \includegraphics[width=.23\textwidth]{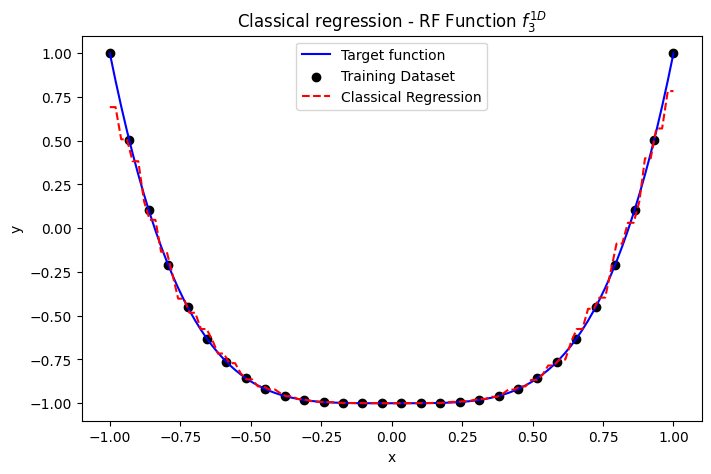} &
    \includegraphics[width=.23\textwidth]{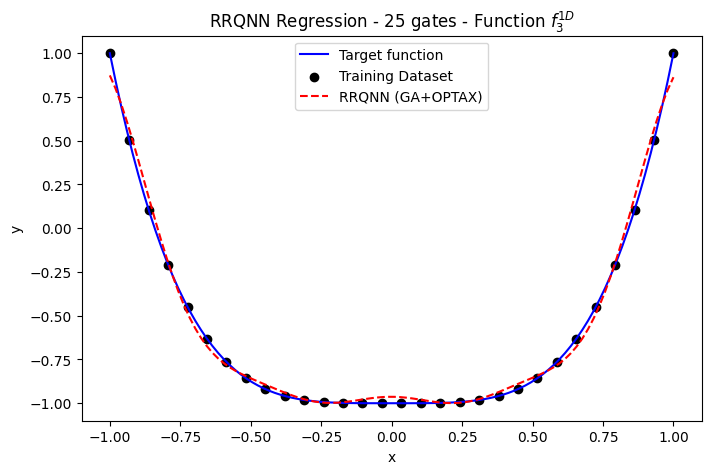} &
    \includegraphics[width=.23\textwidth]{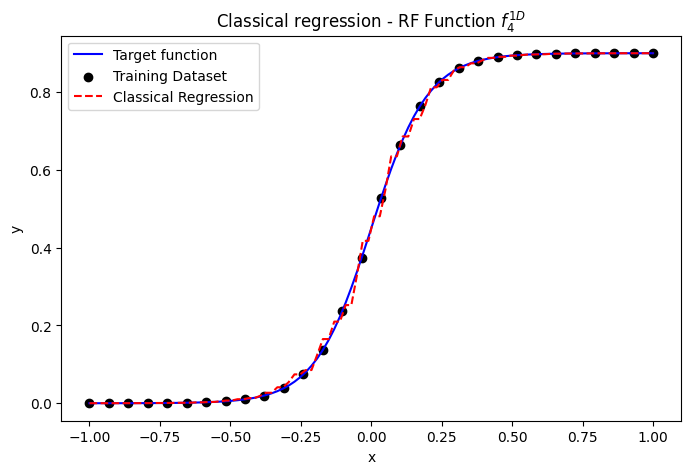} &
    \includegraphics[width=.23\textwidth]{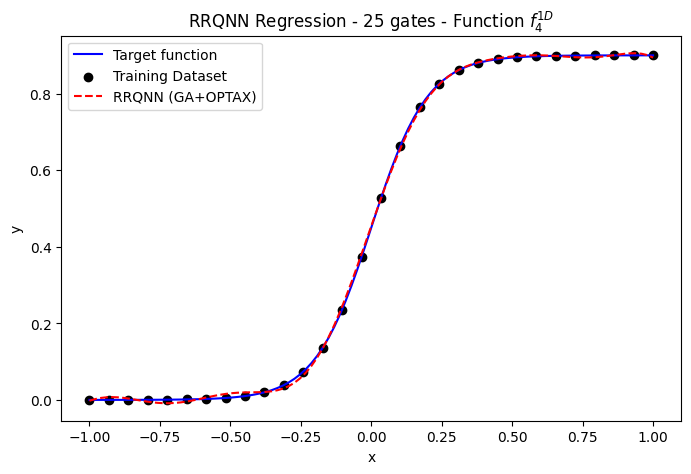} 
  \end{tabular}
  \caption{Results of the best regression models for the four one-dimensional functions defined in Equations~\ref{eq:fx1}--\ref{eq:fx4}, as described in the proof-of-concept protocol (Section~\ref{sec:proofOfConcept}), comparing classical regression models with the proposed RRQNN quantum model.
}
  \label{fig:resultsConceptProof}
\end{figure}

The best-performing RRQNN models identified in the proof-of-concept study were subsequently used to visualize the reconstructed regression functions. These visualizations, presented in Figure~\ref{fig:resultsConceptProof}, illustrate the ability of the automatically discovered quantum circuits to closely approximate the underlying nonlinear target functions across all four one-dimensional benchmarks. The plots clearly show that the RRQNN architectures not only capture the global shape of each function but also reproduce fine–grained variations with high fidelity, corroborating the quantitative superiority reported in Table~\ref{tab:resultsConceptProof}. 

\subsection{Benchmark Suite of Nonlinear Regression Functions}

In order to thoroughly evaluate the performance and generalization capacity of the proposed regression framework, a benchmark suite composed of 22 nonlinear functions was employed. This suite includes the ten functions from the CEC~2020 benchmark and an additional set of thirteen classical nonlinear functions broadly adopted in the evolutionary computation and global optimization literature \cite{plevris2022collection}. Together, these functions cover a wide range of properties—such as modality, separability, conditioning, ruggedness, and the presence of deceptive or composite landscape features—ensuring a rigorous experimental evaluation.



The IEEE Congress on Evolutionary Computation (CEC) benchmark series is one of the most influential standardized problem sets used for testing optimization and learning algorithms. Specifically, the CEC~2020 benchmark \cite{cec2019} contains ten functions designed to provide diverse challenges, including unimodal, multimodal, hybrid, and composite formulations. These functions incorporate transformations such as shifting, rotation, noise components, and hierarchical composition, thereby simulating realistic high-dimensional optimization scenarios.

The ten CEC~2020 functions used in the present work correspond to:
Unimodal Sphere Function,
High Conditioned Elliptic function,
    Bent Cigar Function,
    Discus Function,,
    Sum of Different Powers 2 Function,
    Rosenbrock Function,
    Expanded Schaffer's F6 Function,
    Ackley Function,
    Rastrigin Function,
    Weierstrass Function.
These functions are widely recognized for their diversity and controlled complexity, and they have become a standard tool for benchmarking advanced optimization algorithms. Their inclusion ensures compatibility with contemporary results in the literature and allows direct comparison with state-of-the-art techniques.

As mentioned previously, to complement the CEC~2020 benchmark, thirteen additional functions were selected from the classical optimization literature. 
They vary significantly in terms of modality, separability, and conditioning, offering a broader spectrum of complexity beyond the CEC set.These functions constitute a comprehensive nonlinear testbed, including heavy-tailed landscapes (Schwefel), recursively fractal structures (Katsuura), hybrid formulations combining multiple landscape types (Grie--Rosen), and smooth low-curvature problems (Zakharov, Lévy). Their inclusion ensures a diversified and challenging environment for evaluating regression models, particularly those relying on evolutionary, stochastic, or variational optimization components.

In Table \ref{tab:detailsBenchmark}, all functions are detailed and numbered to facilitate their referencing throughout this work.
This combined dataset of 22 nonlinear regression functions forms a robust and widely recognized benchmark suite. It provides an extensive and challenging environment for assessing model performance under a wide range of nonlinear behaviors and landscape difficulties.

\begin{table*}[ht]
\tiny
\centering
\caption{Details of benchmark functions. The function types are categorized as follows: 
U = unimodal; 
M = multimodal; 
EM = extreme multimodal; 
D = differentiable; 
ND = non-differentiable; 
NS = nonseparable; 
NC = non-convex; 
NCt = non-continuous.}
\label{tab:detailsBenchmark}
\begin{adjustbox}{angle=90}
\begin{tabular}{l|l|l|l|l}
\hline
Function name &
  No. &
  Equation &
  \begin{tabular}[c]{@{}l@{}}Domain\\ Interval\end{tabular} &
  Type \\ \hline
Sphere function &
  1 &
  $  f(x) = \sum_{i=1}^{d}x_i^2$ &
  $\left[-5,5\right]^2$ &
  \begin{tabular}[c]{@{}l@{}}U, \\ D\end{tabular} \\ \hline
Ellipsoid &
  2 &
  $ f(x) = \sum_{i=1}^{d}i \cdot x_i^{2}$ &
  $\left[-5,5\right]^2$ &
  \begin{tabular}[c]{@{}l@{}}U,\\ D\end{tabular} \\ \hline
Bent Cigar &
  3 &
  $f(x) = x_1^{2} + 10^{6}\sum_{i=2}^{d}x_i^{2}$ &
  $\left[-5,5\right]^2$ &
  \begin{tabular}[c]{@{}l@{}}U,\\ NS\end{tabular} \\ \hline
Discus &
  4 &
  $f(x) = 10^{6} \cdot x_1^{2} + \sum_{i=2}^{d}x_i^{2}$ &
  $\left[-5,5\right]^2$ &
  U \\ \hline
Different Powers &
  5 &
  $f(x) = \sum_{i=1}^{d}|x|^{i+1}$ &
  $\left[-5,5\right]^2$ &
  U \\ \hline
Rosenbrock &
  6 &
  $f(x) = \sum_{i=1}^{d-1} \left ( 100\left( x_{i+1} - x_i^2\right)^2 + \left( x_i + 1\right)^2 \right)$ &
  $\left[-5,5\right]^2$ &
  M \\ \hline
Schaffer F7 &
  7 &
  \begin{tabular}[c]{@{}l@{}}$f(x) = \left( \frac{1}{d-1}\sum_{i=1}^{d-1} \left( \sqrt{s_i} + \sqrt{s_i} \cdot sin^{2}\left( 50\cdot s_i^{1/5}\right) \right) \right)^2$\\ $\textnormal{ where }  s_i=\sqrt{x_i^{2} + x_{ i+1}^{2}}$\end{tabular} &
  $\left[-5,5\right]^2$ &
  \begin{tabular}[c]{@{}l@{}}M,\\ NS\end{tabular} \\ \hline
Ackley &
  8 &
  $f(x) = -20\cdot exp\left(-0.2\cdot \sqrt{\frac{1}{d}\cdot \sum_{i=1}^{d}x_i^2} \right) - \sqrt{\frac{1}{d}\sum_{i=1}^{d} cos\left( 2\cdot \pi \cdot x_i\right)} + e + 20$ &
  $\left[-5,5\right]^2$ &
  \begin{tabular}[c]{@{}l@{}}M,\\ NC\end{tabular} \\ \hline
Rastrigin &
  9 &
  $f(x) = \sum_{i=1}^{d}\left( x_i^2 - 10\cdot cos\left( 2\cdot \pi \cdot x_i\right) \right)+10 \cdot d$ &
  $\left[-5,5\right]^2$ &
  M \\ \hline
Weierstrass &
  10 &
  \begin{tabular}[c]{@{}l@{}}$f(x) = \sum_{i=1}^{d} \left( \sum_{k=0}^{k_{\max}} \left( a^{k} \cdot \cos\left( 2\cdot \pi \cdot b^{k}\cdot (x_i + 0.5) \right) \right)\right)-d \cdot \sum_{k=0}^{k_{\max}} \left( a^{k} \cdot \cos(\pi \cdot b^{k}) \right)$\\ $\textnormal{ where }a = 0.5, b = 3,k_{\max} = 20$\end{tabular} &
  $\left[-5,5\right]^2$ &
  \begin{tabular}[c]{@{}l@{}}M,\\ ND\end{tabular} \\ \hline
Griewank &
  11 &
  $f(x) = \frac{1}{4000}\cdot \sum_{i=1}^{d}x_i^{2} - \prod_{i=1}^{d}cos\left( \frac{x_i}{\sqrt{i}}\right)+1$ &
  $\left[-10,10\right]^2$ &
  M \\ \hline
Schwefel &
  12 &
  $f(x) = - \sum_{i=1}^{d}x_i \cdot sin\left( \sqrt{|x_i|} \right) + 418.9828872724337 \cdot d$ &
  $\left[-500,500\right]^2$ &
  M \\ \hline
Katsuura &
  13 &
  $f(\mathbf{x})=\frac{10}{d^{2}}\left[\prod_{i=1}^{d}\left(1+i \cdot \sum_{j=1}^{32}\frac{\left| 2^{j} \cdot x_i-\left\lfloor 2^{j} \cdot x_i + \tfrac{1}{2} \right\rfloor\right|}{2^{j}}\right)^{\frac{10}{d^{1.2}}}\right]-\frac{10}{d^{2}}$ &
  $\left[-500,500\right]^2$ &
  EM \\ \hline
Griewank–Rosenbrock &
  14 &
  $f(\mathbf{x}) =\sum_{i=1}^{d}\left[\frac{\left(100\,(x_i^{2}-x_{i+1})^{2} + (x_i - 1)^2\right)^{\!2}}{4000}-\cos\!\left(100\,(x_i^{2}-x_{i+1})^{2} + (x_i - 1)^2\right)+1\right]$ &
  $\left[-300,300\right]^2$ &
  M \\ \hline
Expanded Schaffer 6 &
  15 &
  $f(x) = \sum_{i=1}^{D-1} \bigl( g(x_i, x_{i+1}) \bigr)+g(x_D, x_1)\textnormal{ where }g(x,y)=0.5+\frac{\sin^{2}\left( \sqrt{x^{2} + y^{2}} \right) - 0.5}{\left( 1 + 0.001 \cdot (x^{2}+y^{2}) \right)^{2}}$ &
  $\left[-10,10\right]^2$ &
  \begin{tabular}[c]{@{}l@{}}M,\\ NS\end{tabular} \\ \hline
Step-Rastrigin &
  16 &
  \begin{tabular}[c]{@{}l@{}}$f(x)=\sum_{i=1}^D \left(z_i^{\,2} - 10\cos(2\pi z_i) + 10 \right)$\\ $\textnormal{ where }z_i = \widetilde{z_i}, \qquad\widetilde{z}_i =\begin{cases}\left\lfloor \frac{5.12}{100}\cdot z_i + 0.5 \right\rfloor, & |z_i| > 0.5, \\[4pt]\frac{5.12}{100} \cdot z_i, & \text{otherwise}\end{cases}$\end{tabular} &
  $\left[-30,30\right]^2$ &
  \begin{tabular}[c]{@{}l@{}}M,\\ NCt\\ non-D\end{tabular} \\ \hline
HappyCat &
  17 &
  $f(\mathbf{x})=\left|\sum_{i=1}^{d} x_i^{2} - d\right|^{1/4}+\frac{0.5 \cdot \sum_{i=1}^{d} x_i^{2}+\sum_{i=1}^{d} x_i}{d}+ 0.5$ &
  $\left[-50,50\right]^2$ &
  M \\ \hline
HGBat &
  18 &
  $f(\mathbf{x})=\left|\left( \sum_{i=1}^{d} x_i^{2} \right)^{2}-\left( \sum_{i=1}^{d} x_i \right)^{2}\right|^{1/2}+\frac{0.5 \cdot \sum_{i=1}^{d} x_i^{2}+\sum_{i=1}^{d} x_i}{d}+ 0.5$ &
  $\left[-30,30\right]^2$ &
  M \\ \hline
\begin{tabular}[c]{@{}l@{}}Different Powers\\ Modified\end{tabular} &
  19 &
  $f(\mathbf{x})=\left(\sum_{i=1}^{d}\left| z_i \right|^{\, 2 + \frac{4(i-1)}{d-1}}\right)^{1/2}$ &
  $\left[-5,5\right]^2$ &
  \begin{tabular}[c]{@{}l@{}}U,\\ NS\end{tabular} \\ \hline
Zakharov &
  20 &
  $f_(\mathbf{x})=\sum_{i=1}^{d} x_i^{2}+\left( \sum_{i=1}^{d} 0.5 \cdot i \cdot x_i \right)^{2}+\left( \sum_{i=1}^{d} 0.5 \cdot i \cdot x_i \right)^{4}.$ &
  $\left[-5,5\right]^2$ &
  U \\ \hline
Lévy &
  21 &
  \begin{tabular}[c]{@{}l@{}}$f(\mathbf{x})=\sin^{2}(\pi w_1)+\sum_{i=1}^{d-1}(w_i - 1)^{2}\,\bigl[\,1 + 10\sin^{2}(\pi w_i + 1)\,\bigr]+(w_d - 1)^{2}\,\bigl[\,1 + \sin^{2}(2\pi w_d)\,\bigr]$\\ $\textnormal{ where }w_i = 1 + \frac{x_i - 1}{4}, \text{for } i = 1,\ldots,d.$\end{tabular} &
  $\left[-500,500\right]^2$ &
  M \\ \hline
Dixon–Price &
  22 &
  $f(\mathbf{x})=(x_1 - 1)^2+\sum_{i=2}^{d}i \cdot \left[\, (2 x_i)^{2} - x_{i-1} \,\right]^{2}.$ &
  $\left[-10^3,10^3\right]^2$ &
  \begin{tabular}[c]{@{}l@{}}U,\\ NS\end{tabular} \\ \hline
\end{tabular}
\end{adjustbox}
\end{table*}

\subsection{Scalability and Experimental Scope}
In this experimental study, quantum models are primarily analyzed and compared in terms of architectural complexity, quantified by the number of trainable parameters and quantum operators composing each circuit. This abstraction enables a systematic comparison across a large number of quantum and classical models while keeping the experimental protocol tractable and reproducible.

Experiments involving noisy quantum simulators or real quantum hardware were intentionally not included in the present evaluation. Given the density and breadth of the experimental campaign---which encompasses multiple circuit families, depths, benchmark functions, and repeated runs---the analysis was conducted under noiseless simulation in order to isolate architectural effects and training behavior. A more detailed investigation of noise effects, hardware constraints, and error mitigation strategies is therefore left for future work, where a reduced set of representative architectures can be examined in greater depth on actual quantum processors platforms~\cite{preskill2018quantum}.

Future studies may also extend the analysis to quantum neural networks with a larger number of qubits, explore alternative architectural paradigms beyond the ans\"atze considered here, and investigate different training strategies, including gradient-free optimization, adaptive learning schemes, and noise-aware training methods. Such extensions could enable a more thorough assessment of the scalability, robustness, and practical viability of automated quantum model design in real-world quantum computing settings.

\section{Results and Discussions}
\label{sec:resultsDiscussions}

We outline here the structure of the analyses conducted in this section. Our results begin with a comparison between quantum and classical regression models (Section~\ref{sec:analysisR2_RMSE}). This design allows us to evaluate average performance, variability, and parameter efficiency, while also determining—via statistical significance testing—whether classical models that surpass quantum ones in mean performance are in fact \emph{significantly} better, or whether quantum models remain statistically equivalent despite using far fewer parameters (Section~\ref{sec:statisticalComparison}).

We then investigate whether the genetic algorithm responsible for generating the Reduced Regressor Quantum Neural Networks (RRQNNs) exhibits systematic architectural tendencies. By applying clustering techniques to the circuits produced across all benchmarks, we explore the emergence of recurring design motifs and potential structural regularities that may reflect an implicit evolutionary preference for certain quantum topologies (Section~\ref{sec:similarityRRQNN_clustering}).

Next, we analyze how the performance of the three quantum model families—\emph{StronglyEntanglingLayers}, \emph{SimplifiedTwoDesign}, \emph{Basic} - varies as circuit depth increases. This examination reveals important insights into the relationship between expressivity, generalization, and the number of layers, clarifying whether deeper architectures consistently yield better regression performance (Section~\ref{sec:analysisQNNBySize}).

Finally, we report results from a meta-learning study that evaluates whether dataset-level complexity measures can predict the optimal quantum model class, the best RRQNN architecture, and its associated hyperparameters. As shown in this section, the meta-learning pipeline demonstrates remarkably strong predictive ability, offering a compelling path toward automated selection of quantum models for regression tasks (Section~\ref{sec:RQQNN_metalearning}).

\subsection{Analysis of \texorpdfstring{$R^2$}{R2} score and RMSE}
\label{sec:analysisR2_RMSE}

The average $R^2$-Score and RMSE results for each model run on each of the functions can be found in Tables~\ref{tab:resultCEC2020F1}-\ref{tab:resultCEC2020F22}. The first group of functions, from $f_1$ to $f_6$, consists of smooth and locally regular functions. These functions generally benefit from the simpler classical models, especially tree-based methods like RF and DT, which yield high $R^2$ scores close to 1, demonstrating near-optimal fit with minimal error. Classical models such as $k$-Nearest Neighbors also perform well, achieving $R^2$ values exceeding 0.995, despite having no trainable parameters. However, quantum models also perform well on these functions. For example, the StronglyEntanglingLayers-20/40/60 models, with depths of 20 to 60 layers, achieve $R^2$ values around 0.995 to 0.997, which is very close to the performance of classical models. Similarly, the SimplifiedTwoDesign (STD) and BasicEntanglerLayers (BEL) models also deliver comparable results, albeit with slightly fewer parameters than their classical counterparts.

The RRQNN models show remarkable efficiency, especially for those with larger parameter budgets. For instance, the RRQNN-120-2q model, which uses only about 80 parameters, reaches $R^2$ scores around 0.996-0.997, which is impressive given its minimal expressivity. However, models with tighter parameter budgets, such as RRQNN-10-1q, exhibit a reduction in performance, but still maintain an $R^2$ greater than 0.90. These results demonstrate that quantum models, particularly resource-restricted networks, can match or even exceed classical performance with far fewer parameters, marking a significant advantage for quantum machine learning approaches.

Moving to Function $f_7$, we observe a sharp transition in performance. This function introduces nonlinearity and structural complexity that confounds even the most robust classical models. Classical models such as RF still perform relatively well, but their $R^2$ values drop significantly compared to the previous functions, with RF scoring around 0.85. Quantum models face similar challenges, with SEL-40/60 and STD-40/60 yielding $R^2$ values around 0.65-0.71. The RRQNN models, especially those with fewer parameters, struggle to achieve higher accuracy, with performance collapsing to near-random predictions as the parameter budget is reduced further. This function serves as an example where both classical and quantum models face challenges, suggesting that higher expressivity or a different model architecture may be required to handle such complex, irregular landscapes.

For Functions $f_8$ to $f_{13}$, which feature moderate nonlinearity and some degree of multi-modality, classical models once again excel. Random Forest and other ensemble models maintain strong performance, with $R^2$ values consistently above 0.95. Quantum models, however, begin to close the gap, particularly with deeper SEL and STD circuits. The SEL models with 40 to 60 layers reach $R^2$ values between 0.90 and 0.97, which is comparable to classical models in this range. The RRQNN models, particularly those with larger budgets (e.g., RRQNN-120), also perform excellently, demonstrating their ability to capture complex patterns with fewer parameters than classical models.

In particular, Function $f_{11}$ stands out. While classical decision tree models fail to capture the underlying complexity of the function, the SEL and STD models achieve competitive performance, outperforming their classical counterparts. This indicates that quantum models, particularly those with deep entangling layers, are better suited for capturing non-axis-aligned feature relationships that simpler classical models struggle with.

Function $f_{14}$ marks a significant milestone in this study, as it is the first function where quantum models outperform all classical baselines. The StronglyEntanglingLayers-60 model achieves an $R^2$ of approximately 0.987, while the RF model lags behind at $R^2$ of 0.93. This result highlights a key advantage of quantum models: their ability to model complex, non-linear relationships that classical models cannot easily capture, especially when the function exhibits entangling or high-dimensional characteristics.

Functions $f_{15}$ to $f_{18}$ represent a transition to more challenging landscapes with higher-dimensional interactions and local irregularities. In these cases, classical models remain strong, with Random Forests and Decision Trees continuing to outperform most quantum models, achieving near-optimal $R^2$ scores. However, quantum models, particularly the deeper SEL and STD circuits, remain competitive, reaching $R^2$ values around 0.95-0.99, while still demonstrating the ability to model these complex interactions with fewer parameters. The RRQNNs, once again, show their strength in terms of parameter efficiency, achieving performance comparable to deep quantum circuits while using a fraction of the number of parameters.

In the final group of functions, from $f_{19}$ to $f_{22}$, we observe the effects of highly structured regimes. These functions introduce significant complexity, making them particularly difficult for all models. Classical ensembles, including RF and MLPs, perform well, but their performance drops off at higher levels of complexity, with $R^2$ values ranging from 0.90 to 0.99. Quantum models, particularly SEL and STD circuits, show a similar trend, with performance ranging from 0.90 to 0.97. RRQNNs, particularly those with larger parameter budgets, maintain solid performance in this regime, although their accuracy begins to degrade as the complexity of the functions increases. The compact RRQNNs, such as RRQNN-10-1q, still achieve an $R^2$ of around 0.96, but they fall short on the most complex functions.

$R^2$-Score values less than -100 were represented by \textit{inf} (infinity) in the respective results tables.

\begin{figure}[ht]
\centering
  \begin{tabular}{@{}cc@{}}
    \includegraphics[width=.45\textwidth]{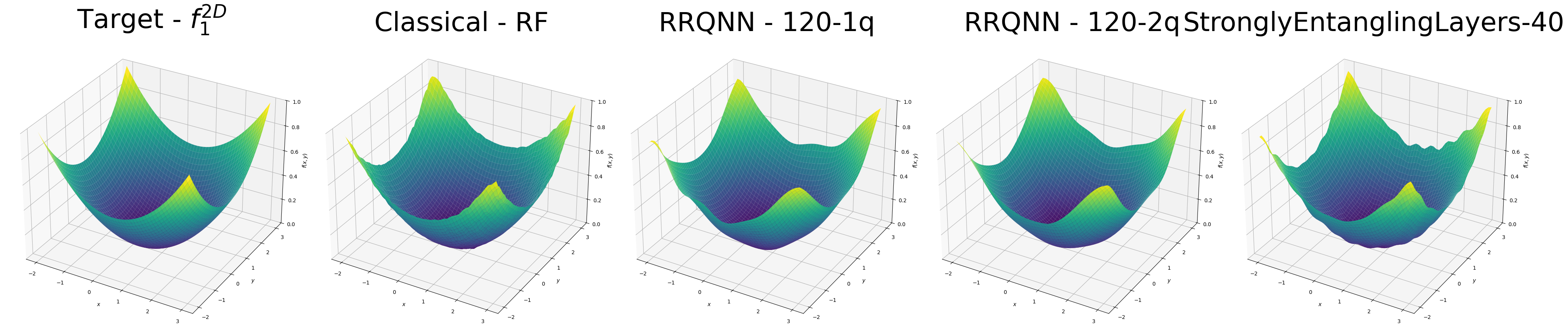} &
    \includegraphics[width=.45\textwidth]{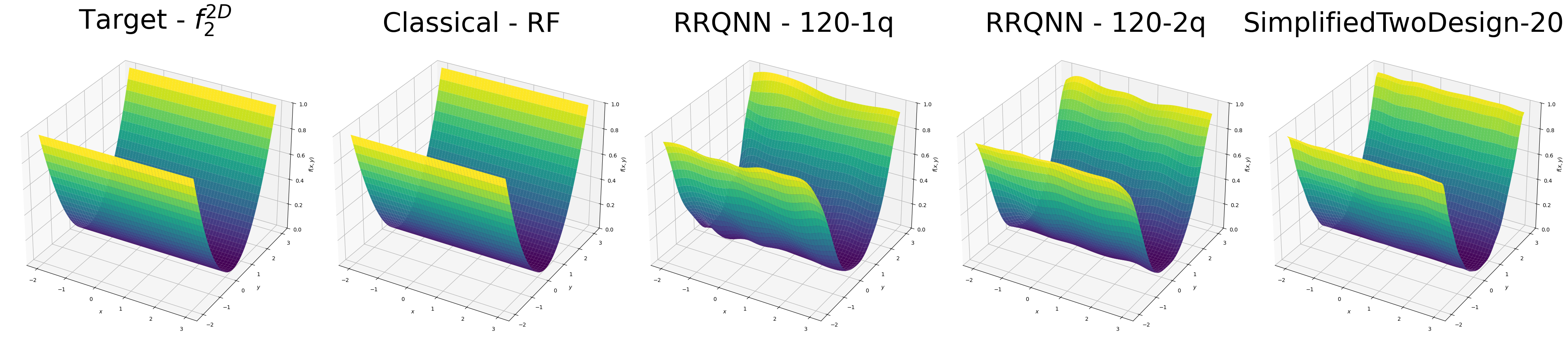} \\
    
    \includegraphics[width=.45\textwidth]{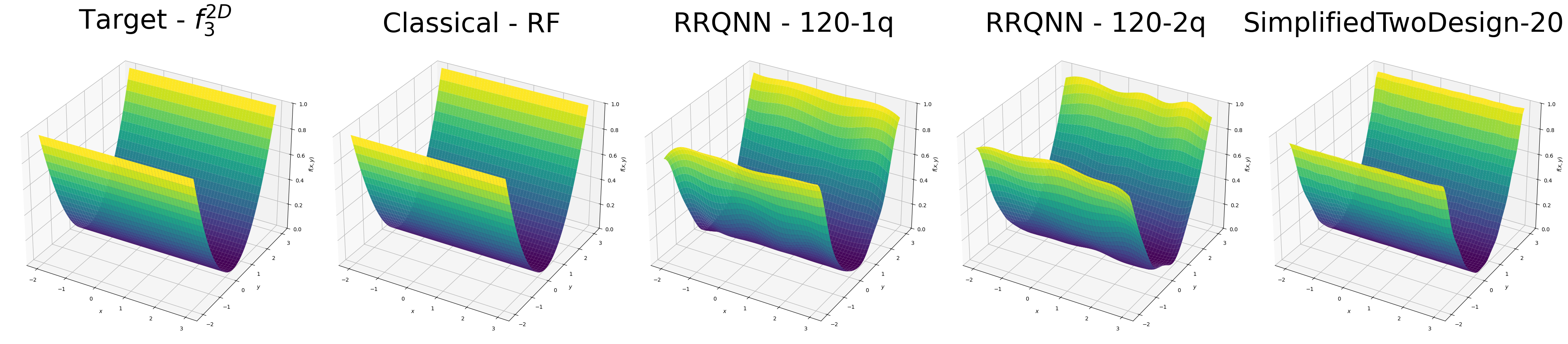} &
    \includegraphics[width=.45\textwidth]{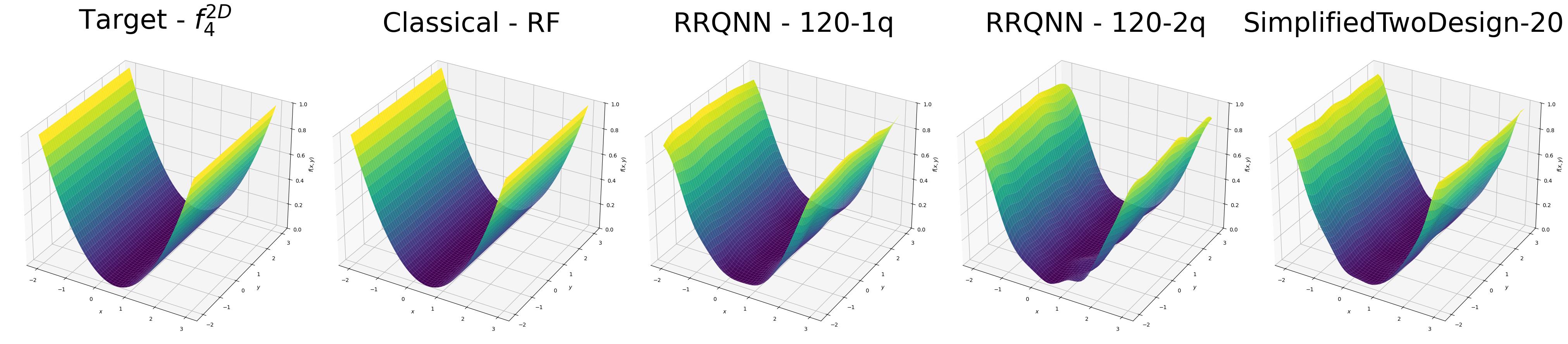} \\
    
    \includegraphics[width=.45\textwidth]{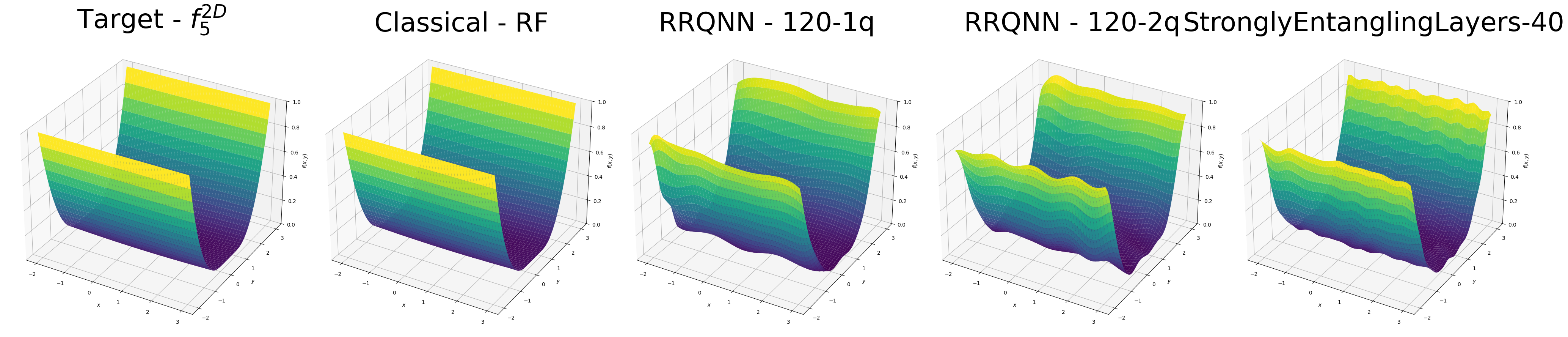} &
    \includegraphics[width=.45\textwidth]{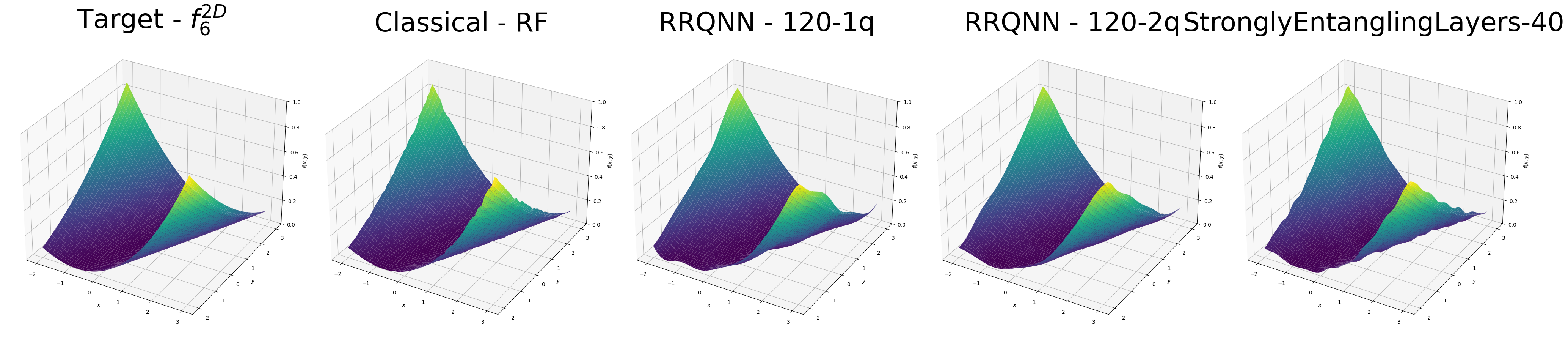} \\
    
    \includegraphics[width=.45\textwidth]{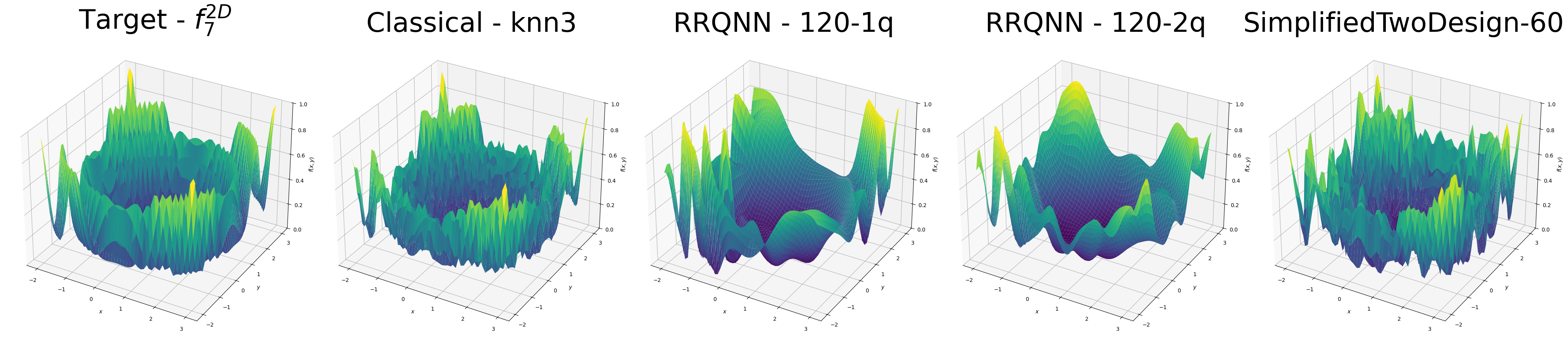} &
    \includegraphics[width=.45\textwidth]{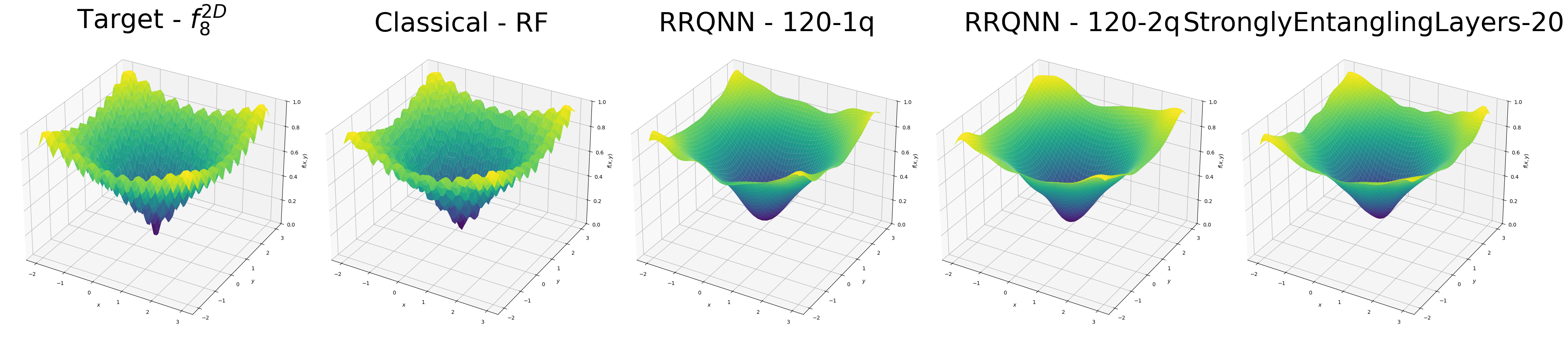} \\
    
    \includegraphics[width=.45\textwidth]{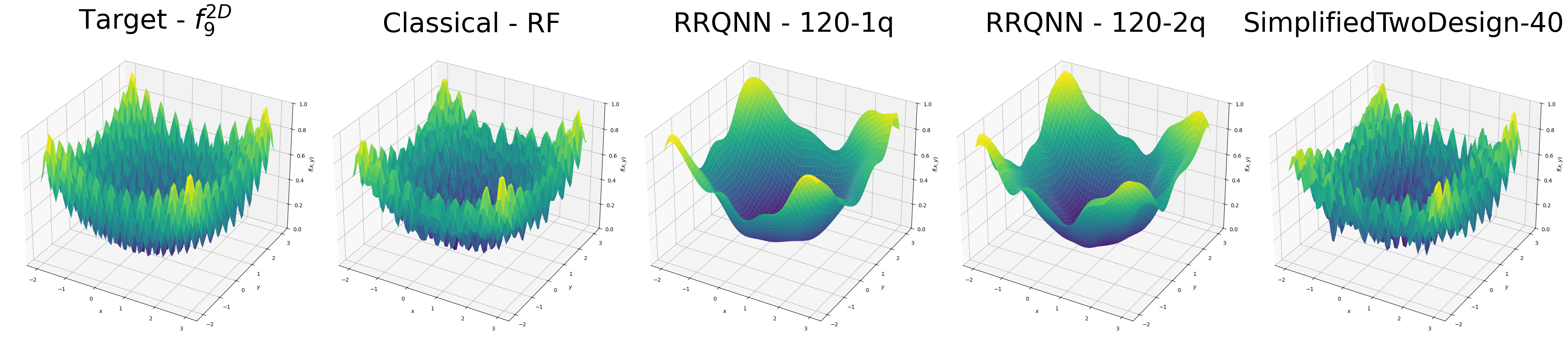} &
    \includegraphics[width=.45\textwidth]{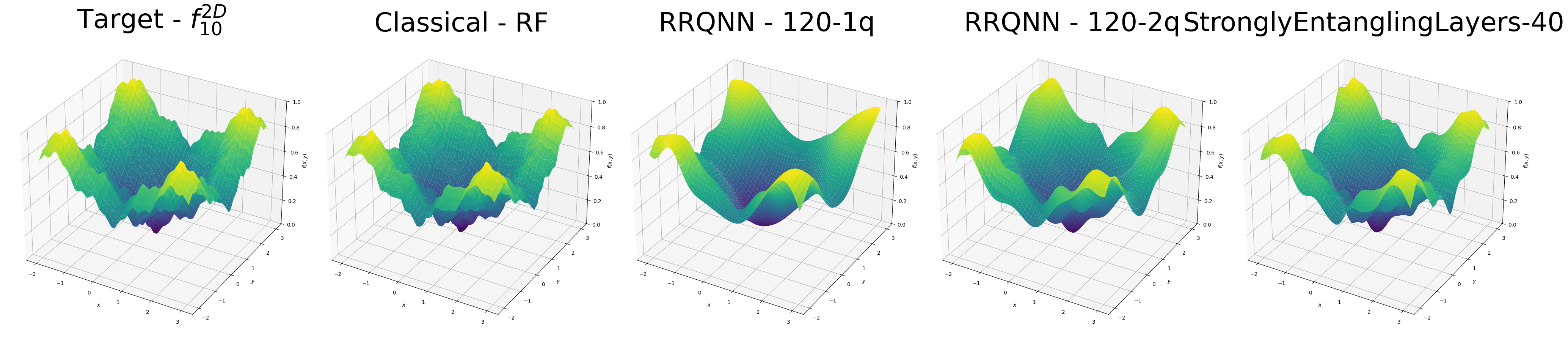} \\
    
    \includegraphics[width=.45\textwidth]{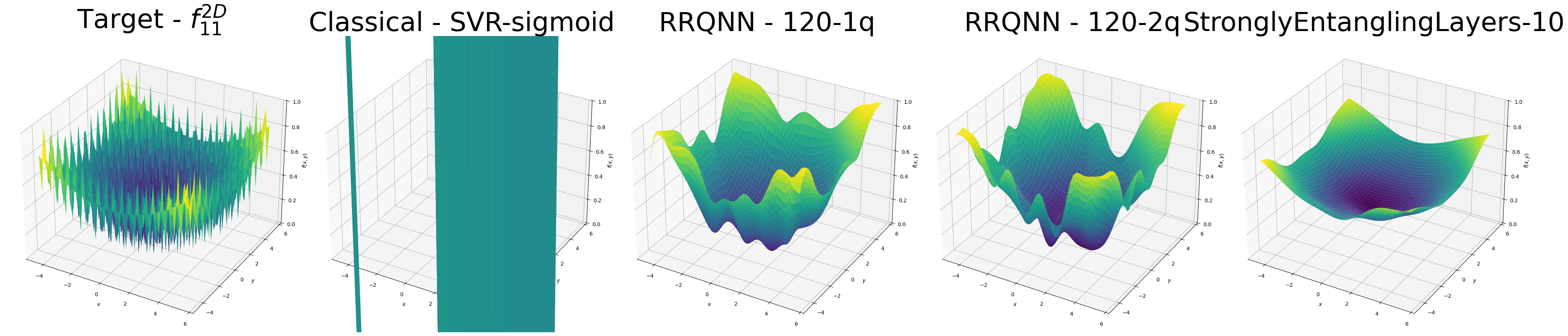} &
    \includegraphics[width=.45\textwidth]{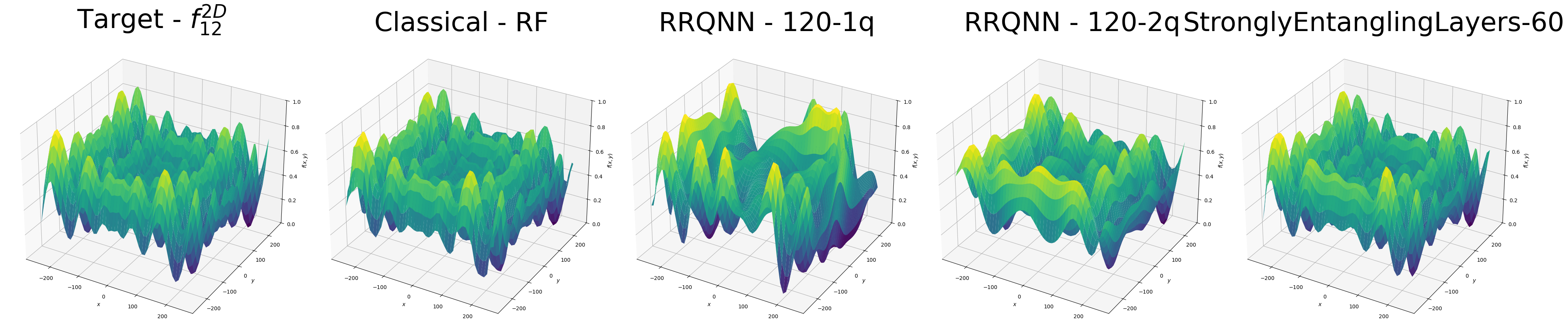} \\
    
    \includegraphics[width=.45\textwidth]{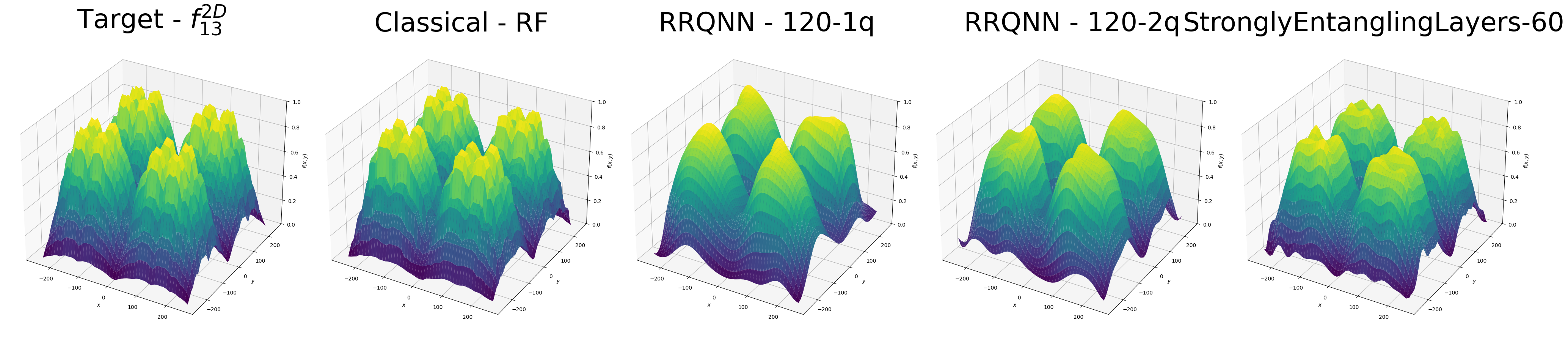} &
    \includegraphics[width=.45\textwidth]{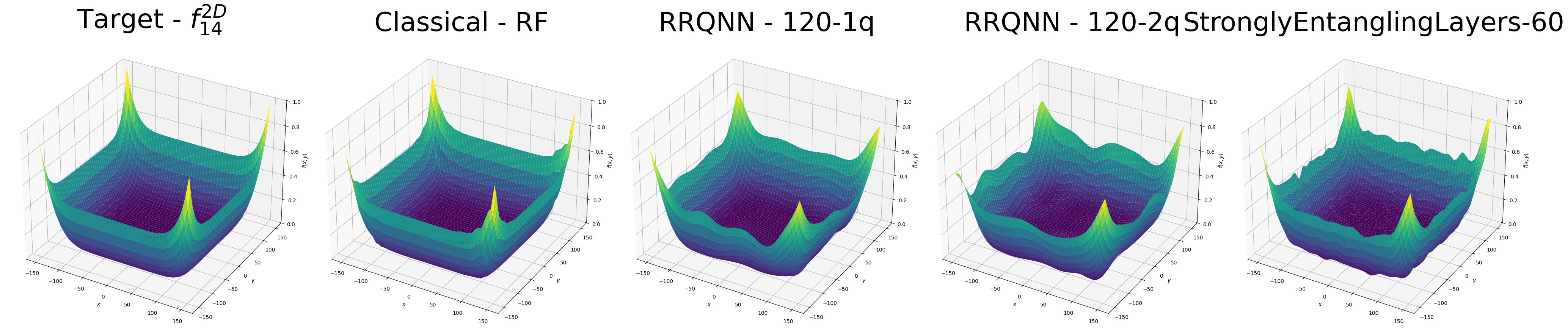} \\
    
    \includegraphics[width=.45\textwidth]{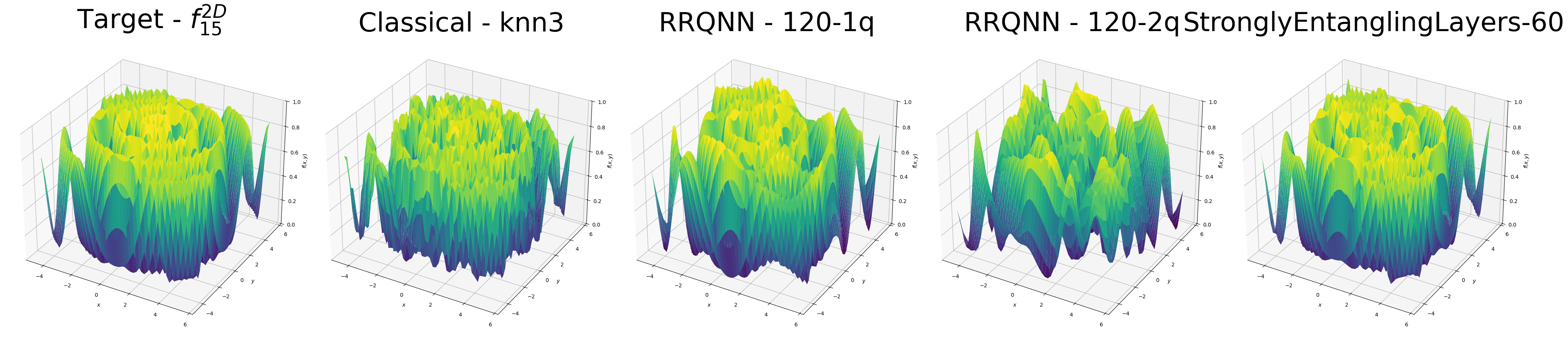} &
    \includegraphics[width=.45\textwidth]{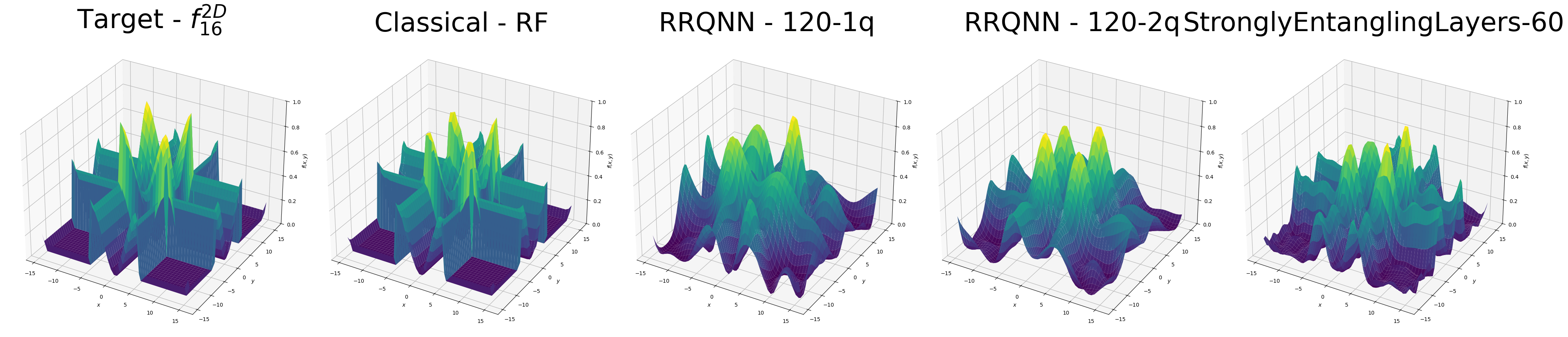} \\
    
    \includegraphics[width=.45\textwidth]{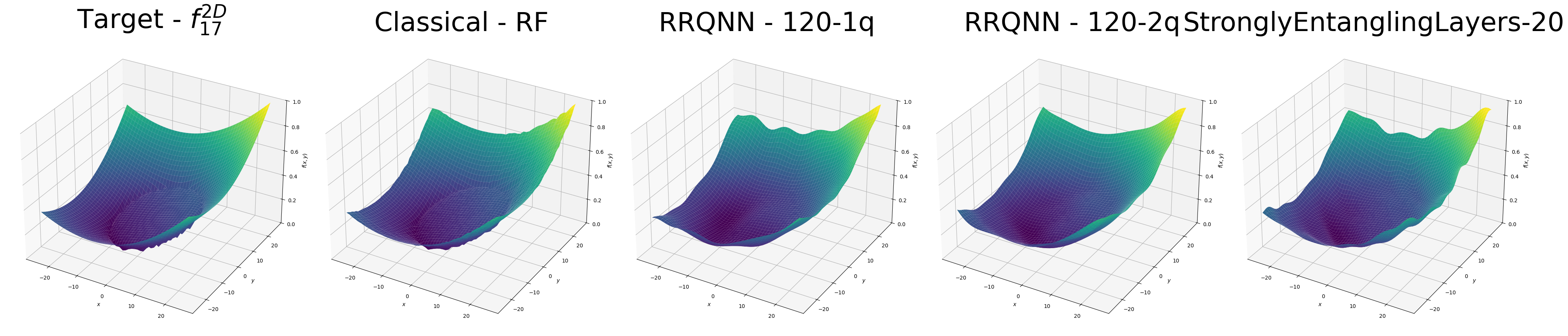} &
    \includegraphics[width=.45\textwidth]{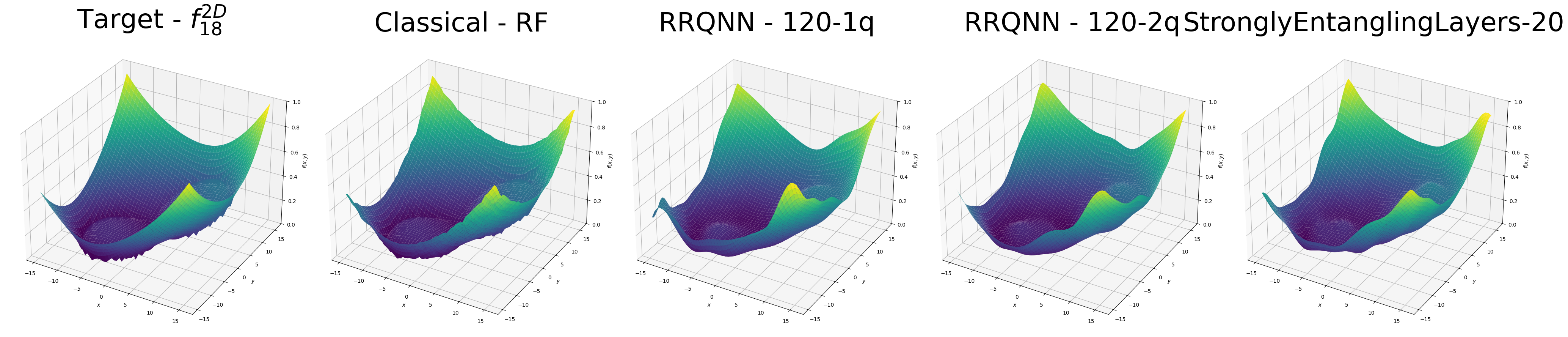} \\
    
    \includegraphics[width=.45\textwidth]{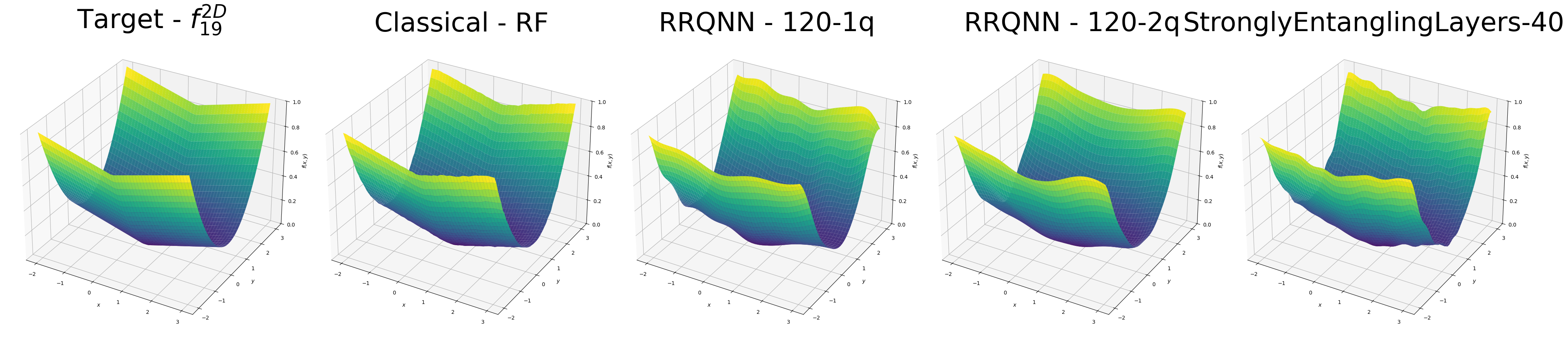} &
    \includegraphics[width=.45\textwidth]{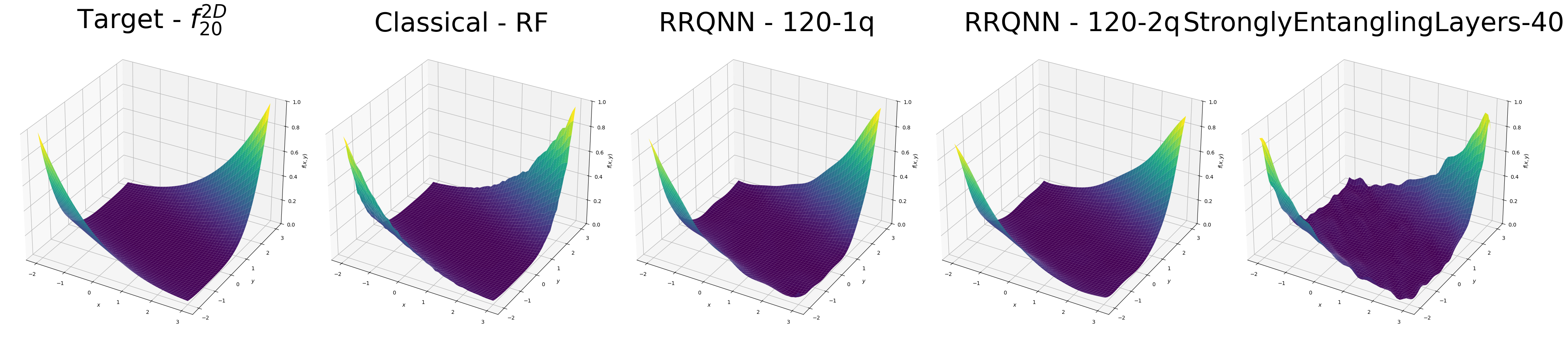} \\
    
    \includegraphics[width=.45\textwidth]{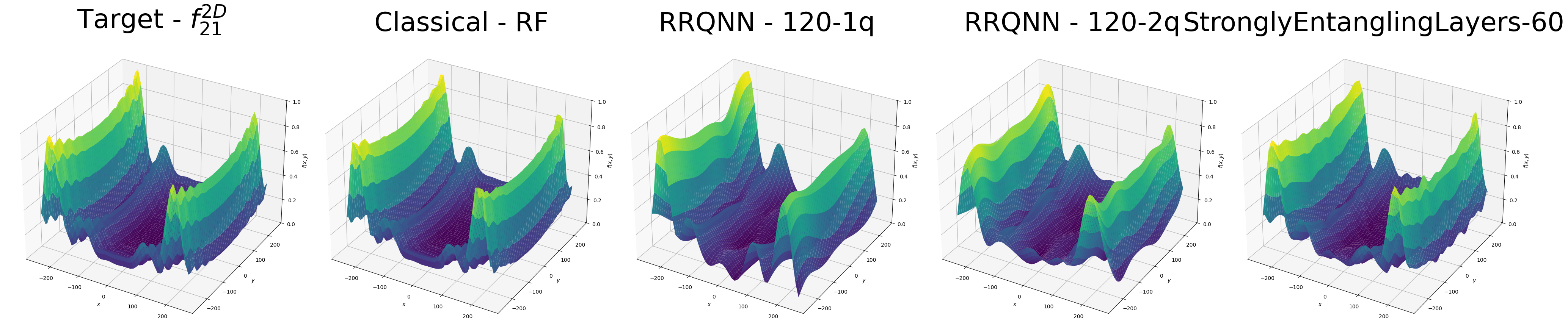} &
    \includegraphics[width=.45\textwidth]{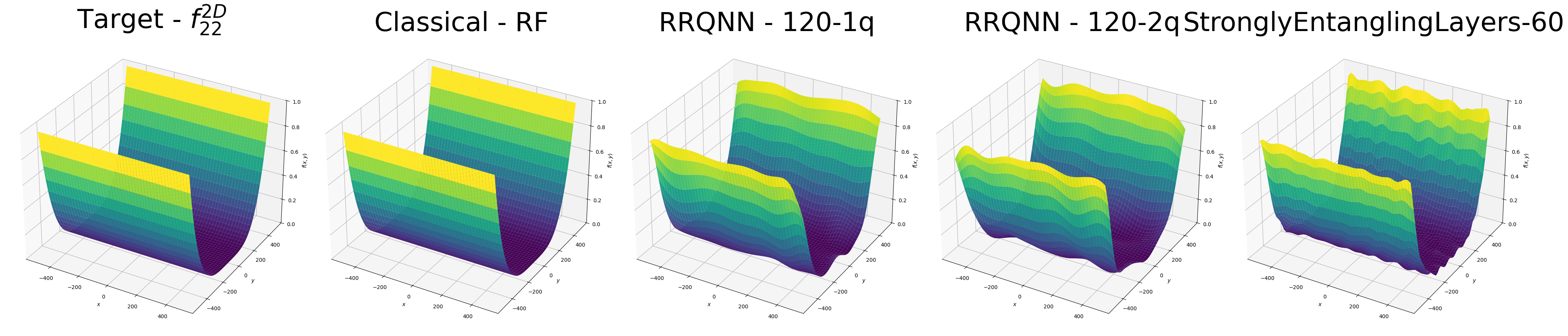} \\
    
  \end{tabular}
  \caption{Results of the best regression of 3D functions of the benchmark dataset by classical models, QNNs and the RRQNNs models.}
  \label{fig:resultsPlots}
\end{figure}

\begin{figure}[ht]
    \centering
    \includegraphics[width=1.\linewidth]{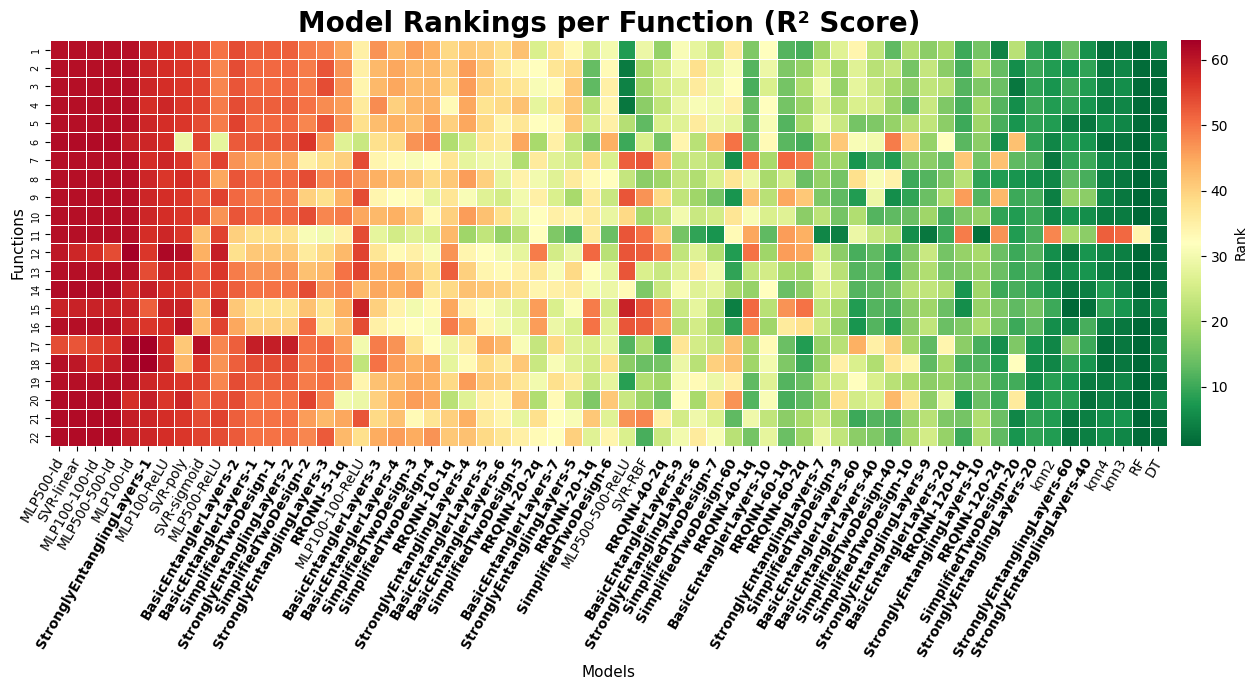}
    \caption{Heatmap of model rankings based on the $R^2$ score across the 22 benchmark functions. Each cell represents the rank of a given model for a specific function, with lower ranks indicating better performance. The color scale ranges from red (higher rank, poorer performance) to green (lower rank, better performance). Models are ordered from left to right according to their average rank across all functions, and quantum models are highlighted by bold labels on the x-axis.}
    \label{fig:model_r2_matrix}
\end{figure}

Figure~\ref{fig:model_r2_matrix} provides a global overview of the relative performance of all quantum and classical models across the benchmark suite. The heatmap reveals a clear stratification of models when aggregated over functions, with classical tree-based and instance-based regressors consistently achieving lower ranks, while several quantum architectures occupy intermediate positions in the ranking.

Notably, deeper instances of the \emph{StronglyEntanglingLayers} and selected \emph{RRQNN} configurations exhibit competitive performance on a subset of functions, as reflected by localized regions of low rank values. At the same time, the variability observed across rows highlights that no single model dominates uniformly over all regression tasks, reinforcing the importance of function-dependent and data-driven model selection strategies.

Table~\ref{tab:averageRankinALL} presents a comprehensive ranking aggregation of all quantum and classical models across the 22 benchmark functions, computed via a rank-based methodology applied to mean $R^2$ scores. Within each function, models were assigned ranks based on their predictive performance, with rank 1 denoting the best-performing model. These per-function ranks were subsequently aggregated across all optimization landscapes to compute the mean rank and standard deviation of ranks for each model. The mean rank serves as a single-valued summary of overall model quality, while the standard deviation of ranks quantifies consistency of performance across heterogeneous problem domains. Models appearing in earlier positions demonstrate superior aggregate performance, whereas those with lower standard deviations exhibit more robust and generalizable behavior independent of function-specific characteristics. This rank-based aggregation is particularly valuable for comparative assessment across functions with disparate inherent difficulty, as ranking ensures equal weighting of each function regardless of problem hardness, thereby providing an unbiased evaluation of model performance across diverse regression tasks.

As expected, several classical models---notably decision trees, random forests, and nearest-neighbor regressors---occupy the top positions in the overall ranking, reflecting their strong and robust performance on a wide range of benchmark functions. These models benefit from mature training procedures and flexible inductive biases that are well suited to the characteristics of the considered datasets. Nevertheless, it is noteworthy that a subset of quantum models, particularly instances of the \emph{StronglyEntanglingLayers} and \emph{SimplifiedTwoDesign} ans\"atze with moderate to large depths, achieve competitive average rankings, in some cases outperforming several classical baselines such as support vector regressors and multilayer perceptrons.
An additional information from Table~\ref{tab:model_ranking_combined} is that the top ten positions in the average ranking are evenly split between classical and quantum models. Specifically, five of the ten best-ranked models correspond to classical regressors, while the remaining five are quantum neural network architectures. This balance indicates that, although classical methods remain highly competitive, appropriately designed quantum models can achieve comparable overall performance when evaluated across a diverse set of regression functions.

The ranking also reveals substantial variability among quantum architectures, emphasizing the importance of architectural choices in quantum regression. Deeper instances of the \emph{StronglyEntanglingLayers} ansatz consistently rank higher than their shallow counterparts, suggesting that increased circuit depth plays a central role in capturing complex functional relationships. 
Similarly, the genetic-algorithm-generated RRQNN models are distributed across a wide range of ranks; however, the best-performing instances consistently correspond to architectures with larger parameter budgets and two-qubit configurations. This observation suggests that increased representational capacity and multi-qubit interactions play an important role in the effectiveness of evolved quantum regressors, while also reinforcing the motivation for automated architecture selection and meta-learning strategies.

From a broader perspective, these results suggest that quantum neural networks, when appropriately structured, can attain performance levels comparable to well-established classical models in regression tasks, albeit with a larger variance across architectures. While the average ranking does not imply systematic superiority of quantum models over classical ones, it supports the claim that quantum regressors constitute a viable and competitive modeling alternative within a heterogeneous model pool. This observation further motivates the use of data-driven model selection mechanisms, as discussed in subsequent sections, to identify those quantum architectures that are best suited to a given regression problem.

\begin{table}[ht]
\centering
\tiny
\caption{Average model rankings (mean $\pm$ standard deviation) computed from $R^2$ scores across the 22 benchmark regression functions. Lower ranks indicate better overall performance.}
\label{tab:model_ranking_combined}

\setlength{\tabcolsep}{1.5pt}
\begin{tabular}{@{}rll @{\hspace{1em}} rll @{\hspace{1em}} rll@{}}
\toprule
Rank & Model & Mean $\pm$ Std & Rank & Model & Mean $\pm$ Std & Rank & Model & Mean $\pm$ Std \\
\midrule
1 & \scalebox{0.8}{DT} & 2.50 $\pm$ 1.25 & 
22 & \textbf{\scalebox{0.8}{RRQNN-60-1q}} & 23.80 $\pm$ 14.16 & 
43 & \textbf{\scalebox{0.8}{SimplifiedTwoDesign-3}} & 39.70 $\pm$ 6.17 \\
2 & \scalebox{0.8}{RF} & 2.70 $\pm$ 7.00 & 
23 & \textbf{\scalebox{0.8}{BasicEntanglerLayers-10}} & 26.05 $\pm$ 5.64 &
44 & \textbf{\scalebox{0.8}{BasicEntanglerLayers-4}} & 40.48 $\pm$ 5.99 \\
3 & \scalebox{0.8}{knn3} & 6.34 $\pm$ 10.06 & 
24 & \textbf{\scalebox{0.8}{RRQNN-40-1q}} & 26.18 $\pm$ 14.68 & 
45 & \textbf{\scalebox{0.8}{BasicEntanglerLayers-3}} & 41.34 $\pm$ 5.42 \\
4 & \scalebox{0.8}{knn4} & 6.59 $\pm$ 10.30 & 
25 & \textbf{\scalebox{0.8}{SimplifiedTwoDesign-60}} & 26.50 $\pm$ 14.65 &
46 & \scalebox{0.8}{MLP100-100-ReLU} & 42.20 $\pm$ 11.20 \\
5 & \textbf{\scalebox{0.8}{StronglyEntanglingLayers-40}} & 7.41 $\pm$ 4.13 &
26 & \textbf{\scalebox{0.8}{SimplifiedTwoDesign-7}} & 26.66 $\pm$ 7.96 &
47 & \textbf{\scalebox{0.8}{RRQNN-5-1q}} & 43.73 $\pm$ 5.98 \\
6 & \textbf{\scalebox{0.8}{StronglyEntanglingLayers-60}} & 8.30 $\pm$ 5.07 &
27 & \textbf{\scalebox{0.8}{StronglyEntanglingLayers-6}} & 27.11 $\pm$ 7.08 &
48 & \textbf{\scalebox{0.8}{StronglyEntanglingLayers-3}} & 45.57 $\pm$ 6.51 \\
7 & \scalebox{0.8}{knn2} & 8.34 $\pm$ 9.03 & 
28 & \textbf{\scalebox{0.8}{BasicEntanglerLayers-9}} & 27.43 $\pm$ 4.86 &
49 & \textbf{\scalebox{0.8}{SimplifiedTwoDesign-2}} & 47.14 $\pm$ 6.79 \\
8 & \textbf{\scalebox{0.8}{StronglyEntanglingLayers-20}} & 9.50 $\pm$ 2.39 &
29 & \textbf{\scalebox{0.8}{RRQNN-40-2q}} & 27.48 $\pm$ 11.90 &
50 & \textbf{\scalebox{0.8}{SimplifiedTwoDesign-1}} & 48.82 $\pm$ 5.37 \\
9 & \textbf{\scalebox{0.8}{SimplifiedTwoDesign-20}} & 13.30 $\pm$ 10.43 &
30 & \scalebox{0.8}{SVR-RBF} & 29.27 $\pm$ 15.59 &
51 & \textbf{\scalebox{0.8}{BasicEntanglerLayers-1}} & 48.82 $\pm$ 5.37 \\
10 & \textbf{\scalebox{0.8}{RRQNN-120-2q}} & 15.48 $\pm$ 12.04 &
31 & \scalebox{0.8}{MLP500-500-ReLU} & 29.82 $\pm$ 20.01 &
52 & \textbf{\scalebox{0.8}{StronglyEntanglingLayers-2}} & 48.82 $\pm$ 5.37 \\
11 & \textbf{\scalebox{0.8}{StronglyEntanglingLayers-10}} & 16.14 $\pm$ 4.73 &
32 & \textbf{\scalebox{0.8}{SimplifiedTwoDesign-6}} & 30.11 $\pm$ 6.57 &
53 & \textbf{\scalebox{0.8}{BasicEntanglerLayers-2}} & 50.36 $\pm$ 4.97 \\
12 & \textbf{\scalebox{0.8}{RRQNN-120-1q}} & 17.20 $\pm$ 12.15 &
33 & \textbf{\scalebox{0.8}{RRQNN-20-1q}} & 30.25 $\pm$ 11.29 &
54 & \scalebox{0.8}{MLP500-ReLU} & 50.80 $\pm$ 6.53 \\
13 & \textbf{\scalebox{0.8}{BasicEntanglerLayers-20}} & 18.59 $\pm$ 6.09 &
34 & \textbf{\scalebox{0.8}{StronglyEntanglingLayers-5}} & 31.39 $\pm$ 7.87 &
55 & \scalebox{0.8}{SVR-sigmoid} & 52.18 $\pm$ 5.02 \\
14 & \textbf{\scalebox{0.8}{StronglyEntanglingLayers-9}} & 18.66 $\pm$ 5.34 &
35 & \textbf{\scalebox{0.8}{BasicEntanglerLayers-7}} & 31.57 $\pm$ 5.47 &
56 & \scalebox{0.8}{SVR-poly} & 54.25 $\pm$ 7.30 \\
15 & \textbf{\scalebox{0.8}{SimplifiedTwoDesign-10}} & 19.09 $\pm$ 8.38 &
36 & \textbf{\scalebox{0.8}{RRQNN-20-2q}} & 32.52 $\pm$ 7.65 &
57 & \scalebox{0.8}{MLP100-ReLU} & 57.43 $\pm$ 1.20 \\
16 & \textbf{\scalebox{0.8}{SimplifiedTwoDesign-40}} & 20.02 $\pm$ 12.71 &
37 & \textbf{\scalebox{0.8}{SimplifiedTwoDesign-5}} & 33.77 $\pm$ 6.66 &
58 & \textbf{\scalebox{0.8}{StronglyEntanglingLayers-1}} & 57.77 $\pm$ 2.33 \\
17 & \textbf{\scalebox{0.8}{BasicEntanglerLayers-40}} & 20.27 $\pm$ 7.51 &
38 & \textbf{\scalebox{0.8}{BasicEntanglerLayers-6}} & 33.89 $\pm$ 5.77 &
59 & \scalebox{0.8}{MLP100-Id} & 60.34 $\pm$ 1.52 \\
18 & \textbf{\scalebox{0.8}{BasicEntanglerLayers-60}} & 21.27 $\pm$ 11.10 &
39 & \textbf{\scalebox{0.8}{BasicEntanglerLayers-5}} & 36.52 $\pm$ 5.35 &
60 & \scalebox{0.8}{MLP500-500-Id} & 60.36 $\pm$ 1.91 \\
19 & \textbf{\scalebox{0.8}{SimplifiedTwoDesign-9}} & 21.82 $\pm$ 8.24 &
40 & \textbf{\scalebox{0.8}{StronglyEntanglingLayers-4}} & 38.20 $\pm$ 8.14 &
61 & \scalebox{0.8}{MLP100-100-Id} & 60.36 $\pm$ 1.78 \\
20 & \textbf{\scalebox{0.8}{StronglyEntanglingLayers-7}} & 22.20 $\pm$ 5.90 &
41 & \textbf{\scalebox{0.8}{RRQNN-10-1q}} & 38.23 $\pm$ 7.75 &
62 & \scalebox{0.8}{SVR-linear} & 60.45 $\pm$ 1.89 \\
21 & \textbf{\scalebox{0.8}{RRQNN-60-2q}} & 23.52 $\pm$ 13.78 &
42 & \textbf{\scalebox{0.8}{SimplifiedTwoDesign-4}} & 38.43 $\pm$ 6.75 &
63 & \scalebox{0.8}{MLP500-Id} & 60.64 $\pm$ 1.61 \\
\bottomrule
\end{tabular}
\label{tab:averageRankinALL}
\end{table}

The results of the 22 functions provide strong evidence of the potential of quantum machine learning models, especially RRQNNs, to offer high accuracy with far fewer parameters compared to classical models. Classical models, such as Random Forests and Decision Trees, remain dominant in functions with smooth, monotonic relationships. However, quantum models, particularly deep circuits and RRQNNs, excel in more complex, non-linear tasks and can surpass classical models when the function requires intricate entangling transformations. These findings suggest that quantum machine learning has a significant role to play in tackling complex regression problems, offering a promising future for their application in real-world, resource-constrained settings. It is possible to see in Figure~\ref{fig:resultsPlots} the plotting of the 22 functions for each of the best classical, RRQNN, and RQN models found during the experiment.

\subsubsection{Statistical Comparison Between Quantum and Classical Models}
\label{sec:statisticalComparison}

To deepen our understanding of the performance relationship between classical and quantum regressors, we applied the Wilcoxon signed-rank test to evaluate whether the predictive outputs of the two paradigms differ in a statistically meaningful way. The classical models included in this analysis whose average performance ranked above the quantum models in the global evaluation. This ensures that the statistical test focuses precisely on the most competitive classical baselines, i.e., those that empirically outperformed the quantum ansätze in the mean ranking. The Wilcoxon test then verifies whether such superior mean behavior is also reflected in statistically distinct predictive distributions.

Across all evaluated functions, Table~\ref{tab:wilcoxon_quantum_classical} shows that none of the resulting $p$-values reach statistical significance at the $\alpha = 0.05$ threshold. This indicates that, even when the comparison is restricted to the best-performing classical models, the quantum and classical regressors exhibit statistically indistinguishable predictive distributions over repeated runs.

For Function~1, both comparisons involving \texttt{StronglyEntanglingLayers-20} against \texttt{knn4} and \texttt{knn3} resulted in $p$-values of $0.375$ and $0.557$, respectively. Even the deeper \texttt{StronglyEntanglingLayers-40} circuit produced a $p$-value of $0.064$ when compared to \texttt{knn3}, which, although closer to the significance boundary, still fails to reject the null hypothesis. These results persist despite the fact that the selected classical models ranked above the quantum ones on average, highlighting the robustness of the statistical parity.

Function~2 includes a larger set of models, with depths up to 60 layers and multiple ansätze including \texttt{SimplifiedTwoDesign}. Yet the $p$-values remain comfortably above $0.05$, ranging from $0.064$ to $0.922$. This again underscores that higher classical mean performance does not necessarily imply statistically different behavior across paired predictions. The variability across runs remains similar for both paradigms.

Functions~4 and 5 follow a comparable pattern. All comparisons between quantum circuits and the classical models that outperformed them on average yielded non-significant $p$-values. For example, the comparison between \texttt{SimplifiedTwoDesign-40} and \texttt{knn4} in Function~5 resulted in a $p$-value of $0.160$, while comparisons with \texttt{knn2} yielded $p$-values as high as $0.625$ and $0.375$. These results extend to deeper circuits such as \texttt{StronglyEntanglingLayers-60}, again reinforcing the lack of evidence for any distributional shift.

Function~6 offers a particularly interesting case: in addition to variational quantum circuits, the hybrid recurrent quantum model \texttt{RRQNN-120-2q} is also included. Despite its distinct architecture, the statistical outcome aligns with the rest: $p$-values between $0.193$ and $0.770$ show no detectable difference relative to the top-performing classical baselines. This consistency suggests that, although quantum models often use drastically fewer trainable parameters, their predictive behavior remains statistically similar to that of the strongest classical competitors.

Function~11 provides one of the densest sets of comparisons, with quantum circuits of several depths evaluated against the classical decision tree, which ranked above all quantum models on average for this function. Yet even in this stringent scenario, all $p$-values remain above $0.10$, indicating that neither shallow nor deep quantum circuits produce distributions significantly different from the classical reference.

Finally, Function~16 continues to follow this global trend. The comparisons involving \texttt{SimplifiedTwoDesign-60} against \texttt{knn3} and \texttt{knn4} once again produce non-significant $p$-values ($0.064$ and $0.193$). Although the $p = 0.064$ case is relatively close to the significance threshold, it nonetheless aligns with the unified statistical picture emerging from all functions.

Overall, these findings allow us to draw two main conclusions. First, even when the analysis is restricted to \emph{the strongest classical models} --- those that achieved higher average performance than quantum models in the global ranking --- the Wilcoxon signed-rank test provides no evidence of statistically significant differences in predictive output distributions. Second, the empirical similarity across repeated runs suggests that quantum regressors, despite using considerably fewer trainable parameters and fundamentally different representational mechanisms, can match the classical models in terms of stability and distributional behavior. This reinforces the view that variational quantum models are not only expressive but also statistically reliable approximators in the regression tasks considered.

Across the 22 benchmark functions, the quantum models were competitive in 7 cases---exhibiting result distributions that were statistically indistinguishable from the best classical models---while in 1 function the quantum model outperformed all classical counterparts in terms of average performance.

\subsection{Analysis of Similarity between generated RRNQQ architectures}
\label{sec:similarityRRQNN_clustering}

The ten best RRQNN architectures produced by the genetic algorithm for each of the twenty-two nonlinear benchmark functions were analyzed in terms of structural similarity using unsupervised clustering methods. The underlying hypothesis was that architectures achieving high predictive accuracy---quantified by elevated \(R^{2}\) scores for a given target function---would exhibit convergent structural patterns and, consequently, would be assigned to the same cluster based on similarities in their operator sequences. If the genetic algorithm consistently discovered effective architectural pattern for a specific problem, these patterns should naturally form coherent clusters independent of the clustering algorithm used.

To evaluate this hypothesis, three clustering techniques were employed: KMeans, Birch, and Agglomerative Clustering, applied to the 220 architectures distributed across the twenty-two target functions. After clustering these architectures into twenty-two groups, the results revealed low values for the Silhouette coefficient, which is internal evaluation metrics that do not depend on the expected labels. Likewise, external clustering indices such as Adjusted Rand Index (ARI), Jaccard and Fowlkes--Mallows, which quantify how closely the cluster structure aligns with the underlying function for which each architecture was optimized, also indicated that the clusters did not reflect architectural performance.

Two complementary experiments were conducted. In the first experiment, each architecture was represented using the full operator-type encoding of the quantum circuit, as explained in Section~\ref{sec:mappingAG}, i.e. with integers \(1\!-\!13\) denoted, respectively, the identity operation (\(\mathrm{Id}\)), single-qubit rotations for input encoding (\(\mathrm{R}_{x}, \mathrm{R}_{y}, \mathrm{R}_{z}\)), single-qubit rotations for parameter encoding (\(\mathrm{R}_{x}, \mathrm{R}_{y}, \mathrm{R}_{z}\)), and their corresponding controlled operations (\(\mathrm{CRX}, \mathrm{CRY}, \mathrm{CRZ}, \mathrm{CRX}, \mathrm{CRY}, \mathrm{CRZ}\)). This representation preserved the full expressiveness of the architectural search space and allowed the clustering algorithms to explore fine-grained structural similarities. 

In the second experiment, each architecture was binarized: gates associated with input encoding were mapped to \(0\), while gates associated with parameter encoding were mapped to \(1\). This coarse representation was designed to test whether successful architectures share positional regularities---such as preferred locations for input loading or parameter insertion---regardless of the specific unitary type used.

Across all experiments, the conclusions were consistent: no significant similarity structure emerged among the architectures generated by the genetic algorithm. The results showed that even when the number of parameters defining each architecture was reduced (e.g., 20 or 40 operators), the clustering validity indices, although slightly higher, remained too low to support meaningful grouping. These observations suggest that effective RRQNN architectures do not converge to a stable or recurrent motif and that the architectural search landscape explored by the genetic algorithm remains highly irregular.

\subsubsection{Analysis of Clustering Metrics}
\label{sec:analysisClustering}

Tables~\ref{table:rrqnn1qCLUSTERING}, \ref{table:rrqnn1qCLUSTERINGBin}, \ref{table:rrqnn2qCLUSTERING}, and \ref{table:rrqnn2qCLUSTERINGBin} summarize the clustering metrics for the RRQNN architectures under four conditions: single-qubit (1q) architectures in their original operator encoding, single-qubit architectures after binarization, two-qubit (2q) architectures in their original encoding, and two-qubit architectures after binarization. These tables provide a comprehensive view of how architectural similarity behaves across different model complexities and representation schemes.

For the 1q architectures in their original encoding (Table~\ref{table:rrqnn1qCLUSTERING}), the Silhouette values are consistently low across all clustering algorithms, ranging from approximately \(0.004\) to \(0.214\). The Adjusted Rand Index (ARI), which measures alignment with the expected labels (i.e., the function each architecture was optimized for), remains extremely small, often close to zero or even negative. These results indicate that the architectural structures do not form coherent or natural clusters when the full operator information is used. Birch and AgglomerativeClustering occasionally produce slightly higher ARI values than KMeans, but the differences remain negligible in absolute terms.

When the 1q architectures are binarized (Table~\ref{table:rrqnn1qCLUSTERINGBin}), the clustering quality remains similarly low. The Silhouette values do not improve, and the ARI remains close to zero. The binarization was expected to highlight macro-level positional similarities between input-loading and parameter-loading operations, but the results suggest that no such positional regularities exist across the datasets. In particular, even for reduced architecture sizes (5 or 10 operators), where one might expect more structural regularity, the clustering indices remain weak.

The 2q architectures in their original encoding (Table~\ref{table:rrqnn2qCLUSTERING}) also exhibit low Silhouette and ARI scores. Interestingly, for very small architectures (20 operators), the silhouette values increase slightly (around \(0.051\)--\(0.054\)), and the ARI also increases modestly compared to larger architectures. However, these values remain far from those typically associated with meaningful clustering. For larger architectures (40, 60, 120 operators), the Silhouette and ARI values again approach zero, suggesting that architectural diversity increases with the number of operators and that the genetic algorithm does not converge toward a consistent structural pattern.

Finally, the 2q binarized architectures (Table~\ref{table:rrqnn2qCLUSTERINGBin}) show similar behavior: low Silhouette values, low ARI, and no evidence of meaningful structural regularity. The slight increase in Fowlkes--Mallows scores for small architectures (20 operators) again indicates that reducing architectural complexity can impose weak structural regularities, but these effects remain small and insufficient for any useful predictive interpretation.

Overall, the clustering metrics across all four tables demonstrate a consistent pattern: the RRQNN architectures exhibit high variability, and the genetic algorithm does not converge to a stable architectural template that can be detected through unsupervised clustering. Even architectures that achieve high predictive performance do not share enough structural similarity to form coherent clusters in either their full or binarized representations.

These findings underscore the inherent difficulty of designing RRQNN architectures capable of reliably solving nonlinear regression problems. The absence of consistent clustering structure suggests that high-performing architectures may be highly problem-specific and that generalized architectural motifs are unlikely to emerge from purely evolutionary search processes.

\subsection{Analysis of QNN Perfomance by size}
\label{sec:analysisQNNBySize}



\begin{figure}[ht]
\centering
  \begin{tabular}{@{}ccc@{}}
    \includegraphics[width=.32\textwidth]{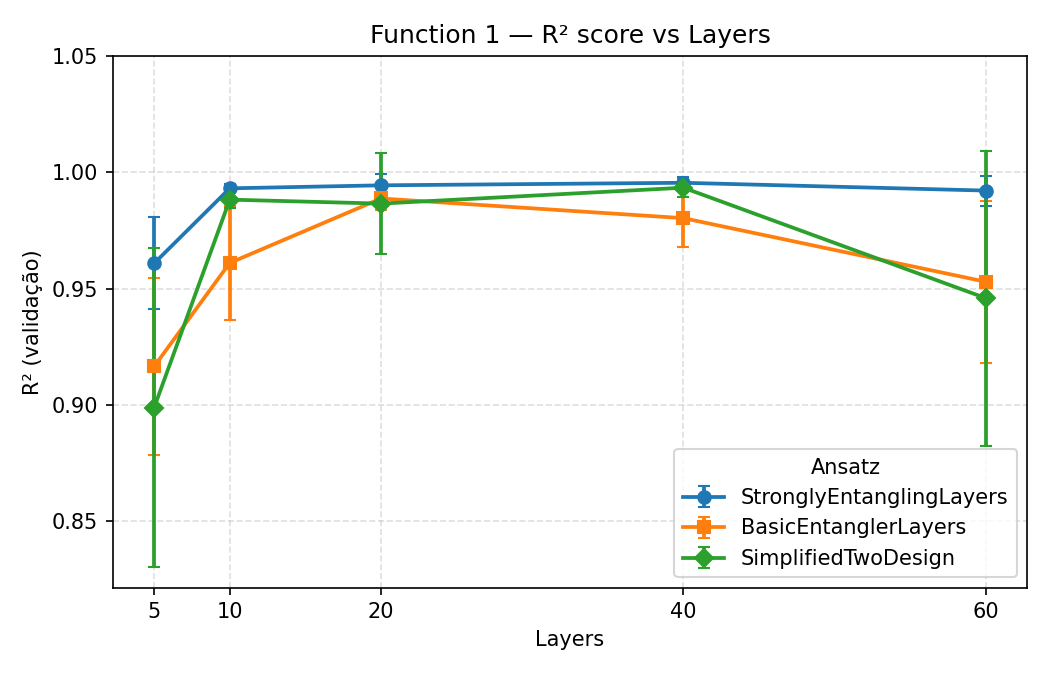} &
    \includegraphics[width=.32\textwidth]{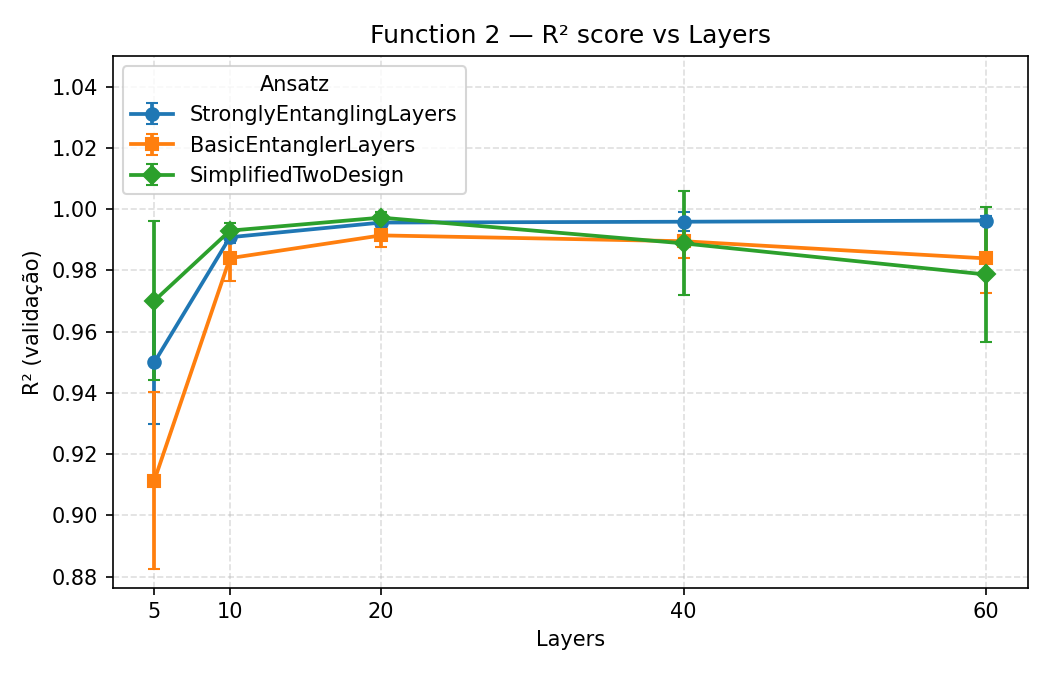} &
    \includegraphics[width=.32\textwidth]{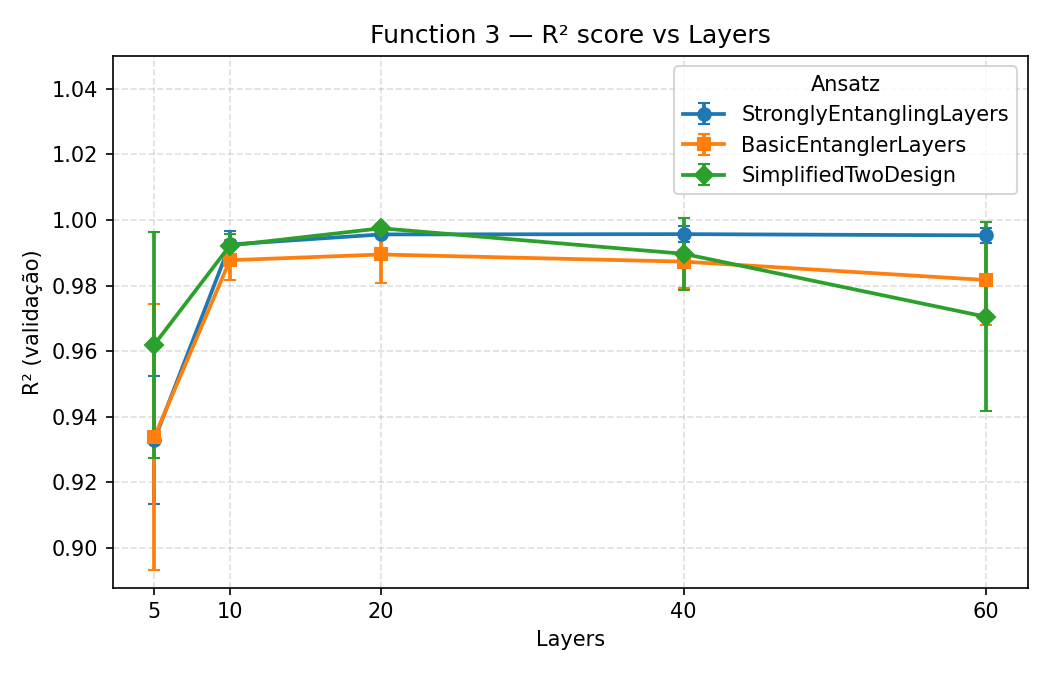} \\
    \includegraphics[width=.32\textwidth]{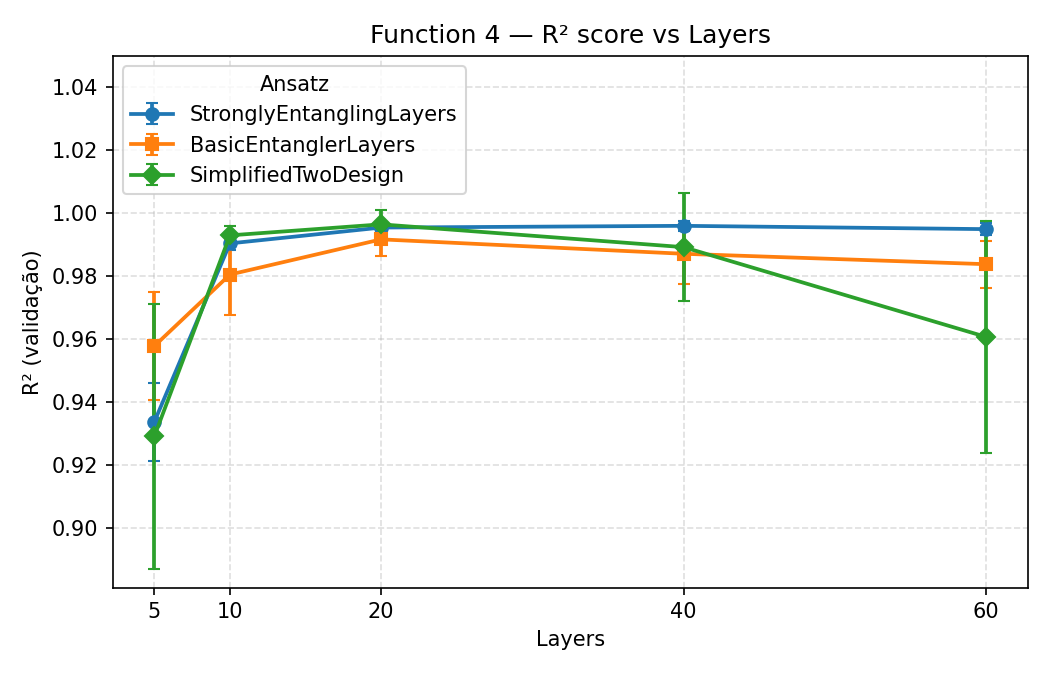} &
    \includegraphics[width=.32\textwidth]{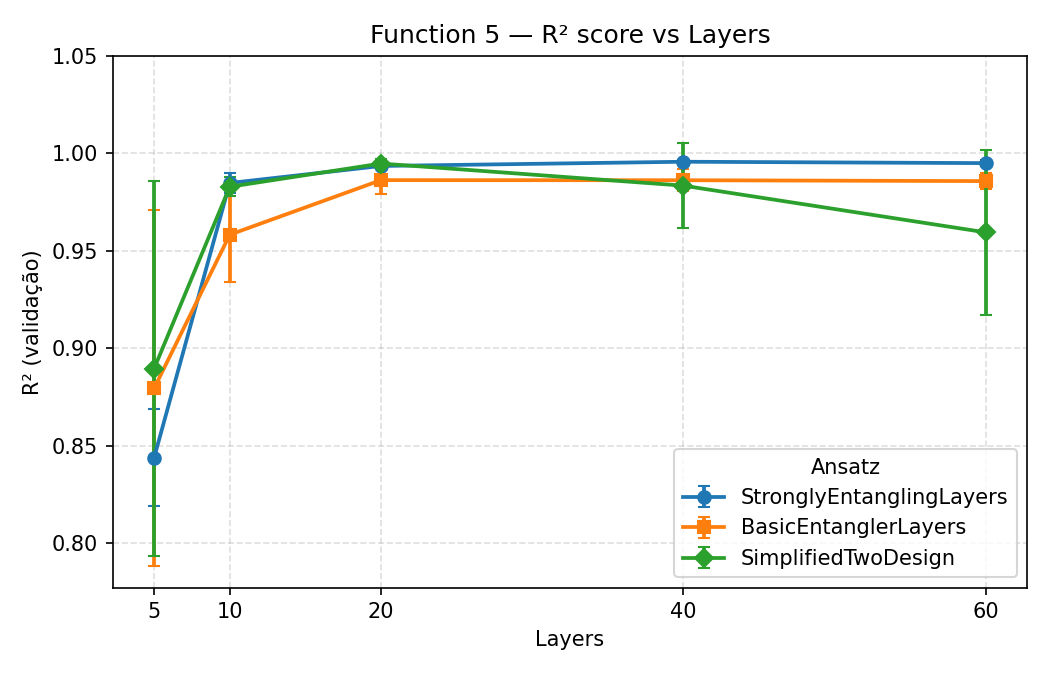} &
    \includegraphics[width=.32\textwidth]{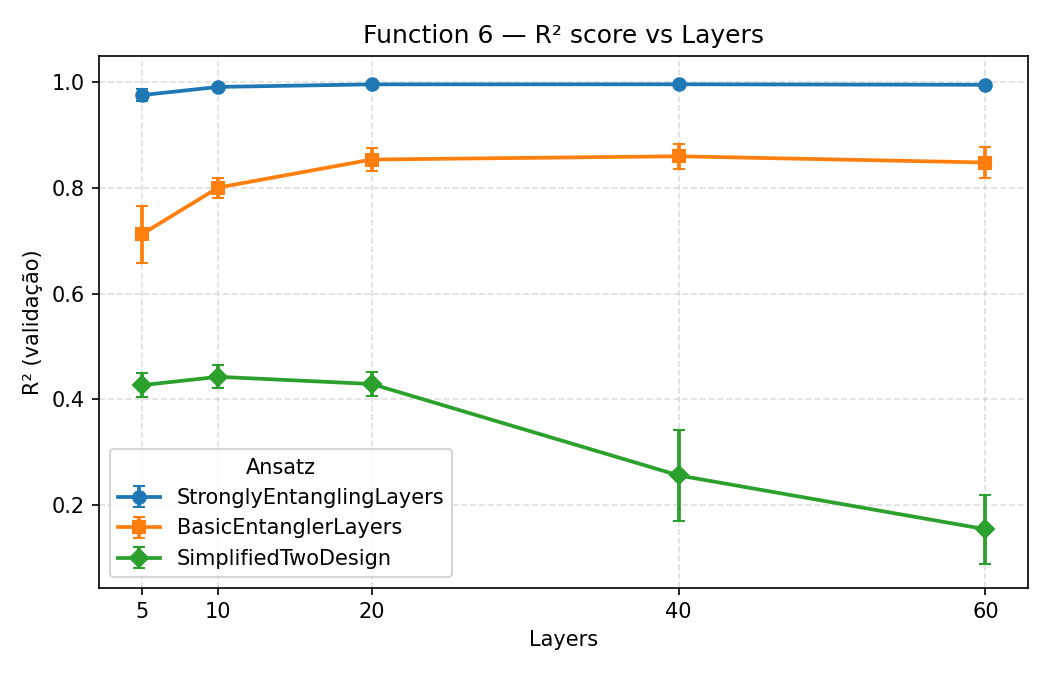} \\
    \includegraphics[width=.32\textwidth]{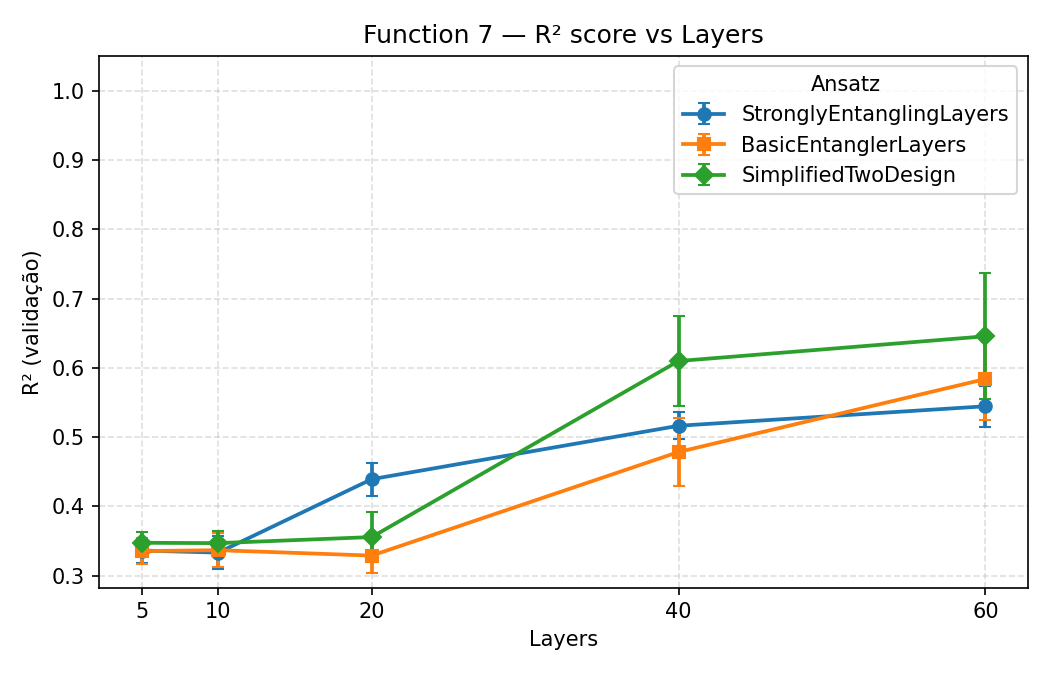} &
    \includegraphics[width=.32\textwidth]{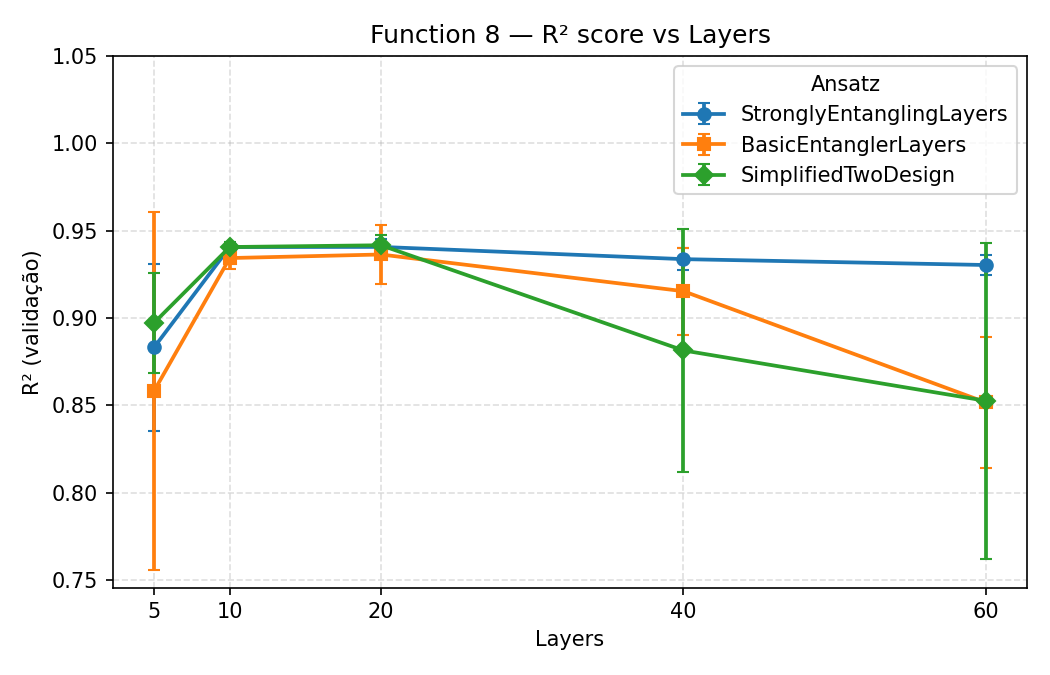} &
    \includegraphics[width=.32\textwidth]{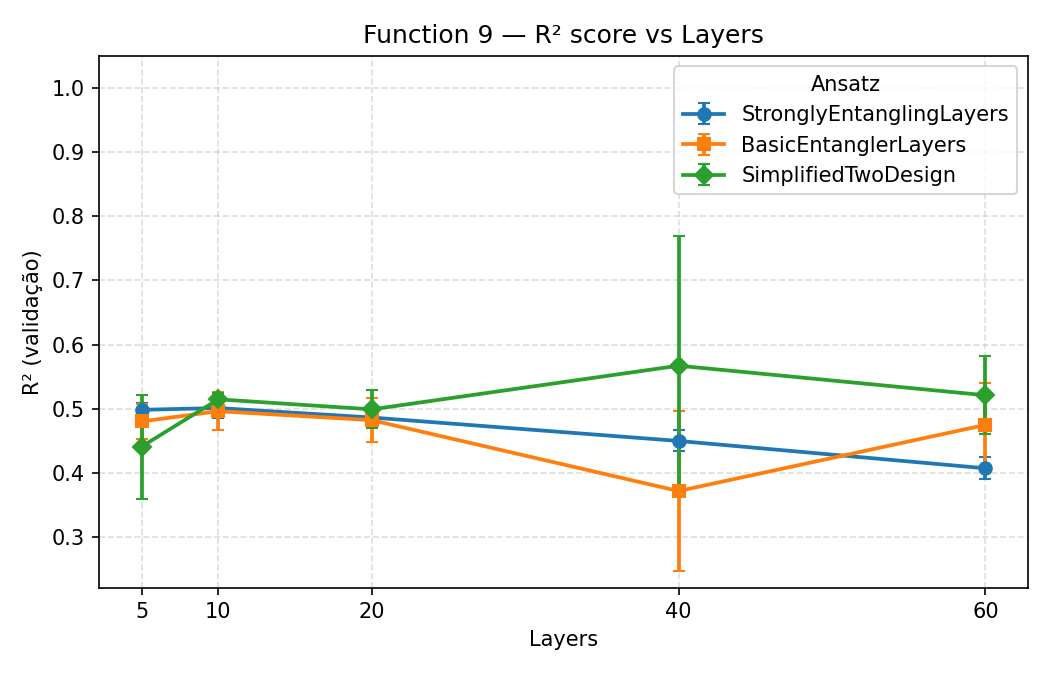} \\
  \end{tabular}\\[1ex] 
  \begin{tabular}{@{}cc@{}}
    \includegraphics[width=.32\textwidth]{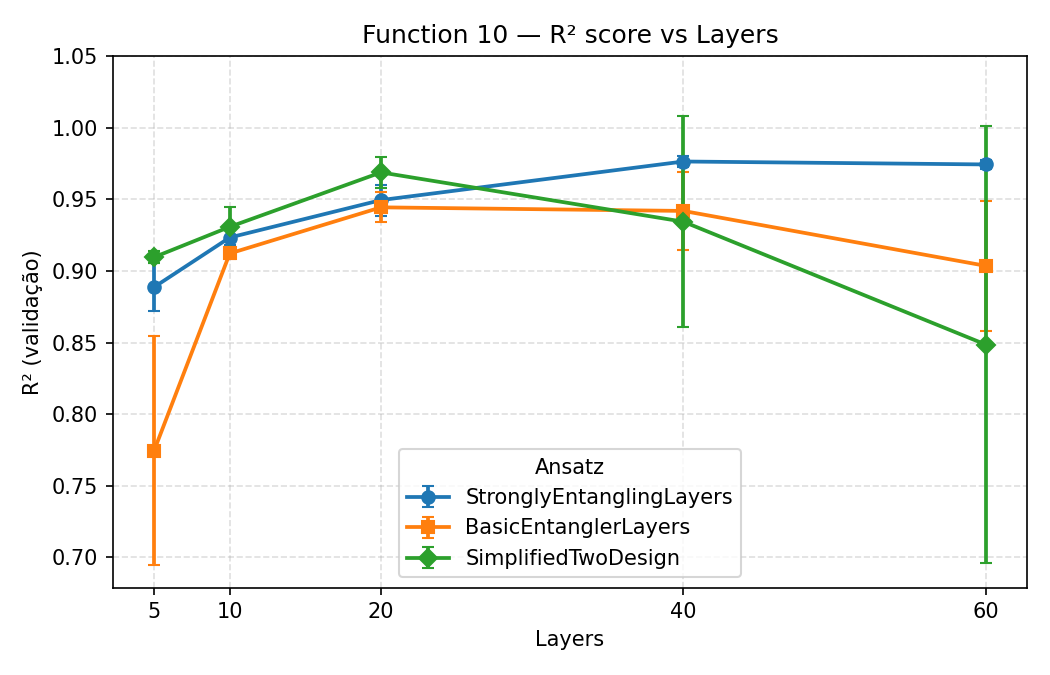} &
    \includegraphics[width=.32\textwidth]{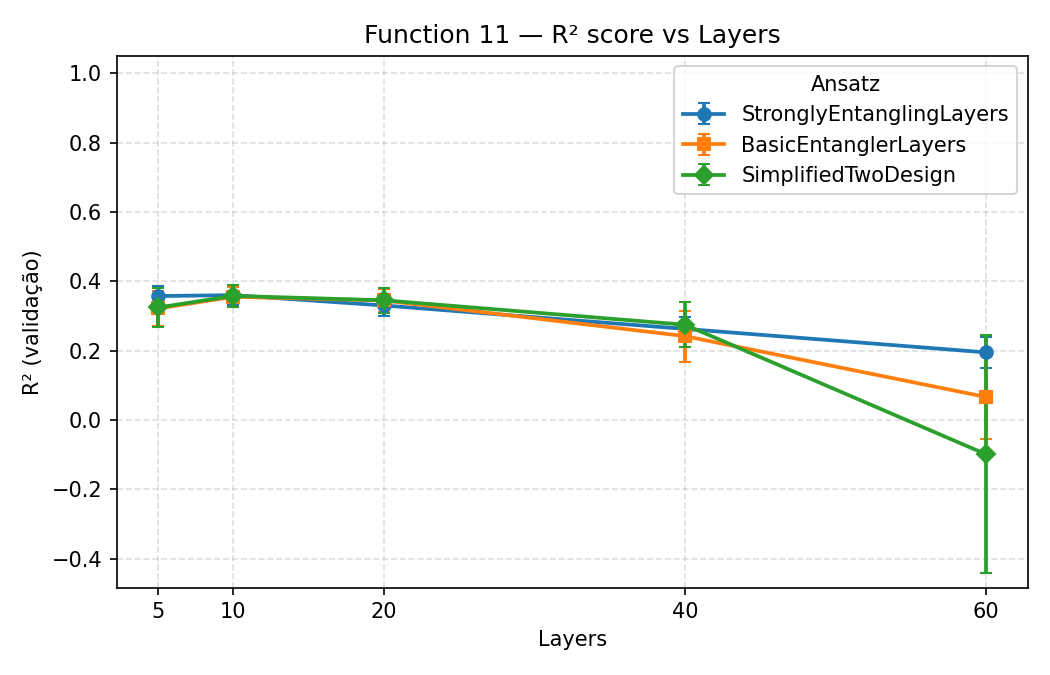} \\
  \end{tabular}
  \caption{Analysis of the $R^2$-Score for functions 1 to 11 (Mean and standard deviation across 10 runs).}
  \label{fig:functions_r2_vs_layers_part1}
\end{figure}

\begin{figure}[ht]
\centering
  \begin{tabular}{@{}ccc@{}}
    \includegraphics[width=.32\textwidth]{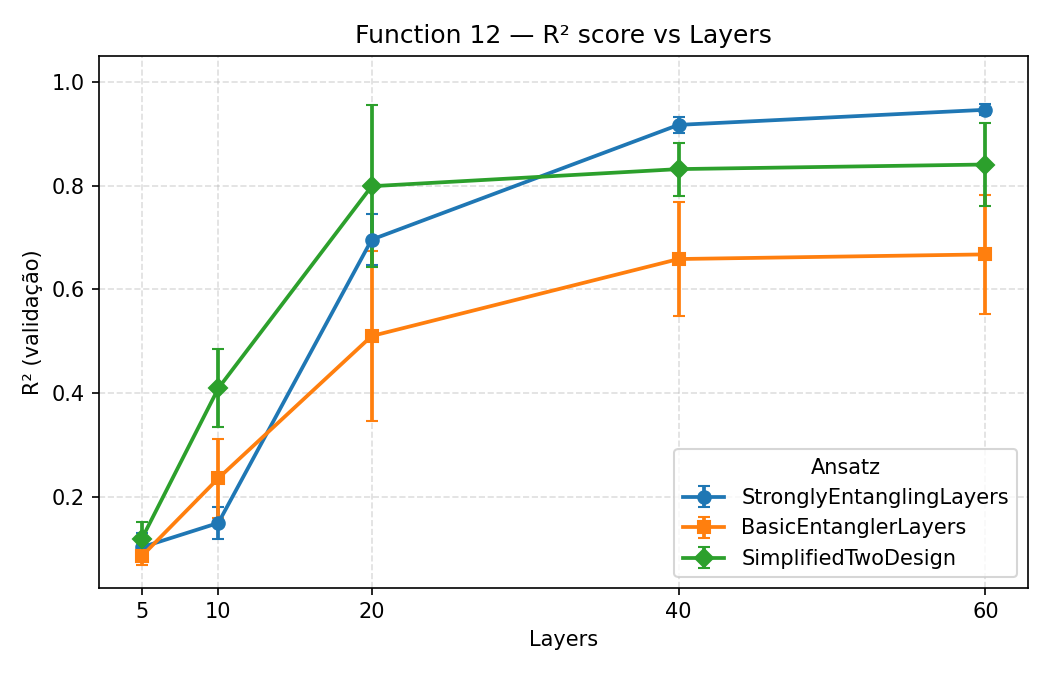} &
    \includegraphics[width=.32\textwidth]{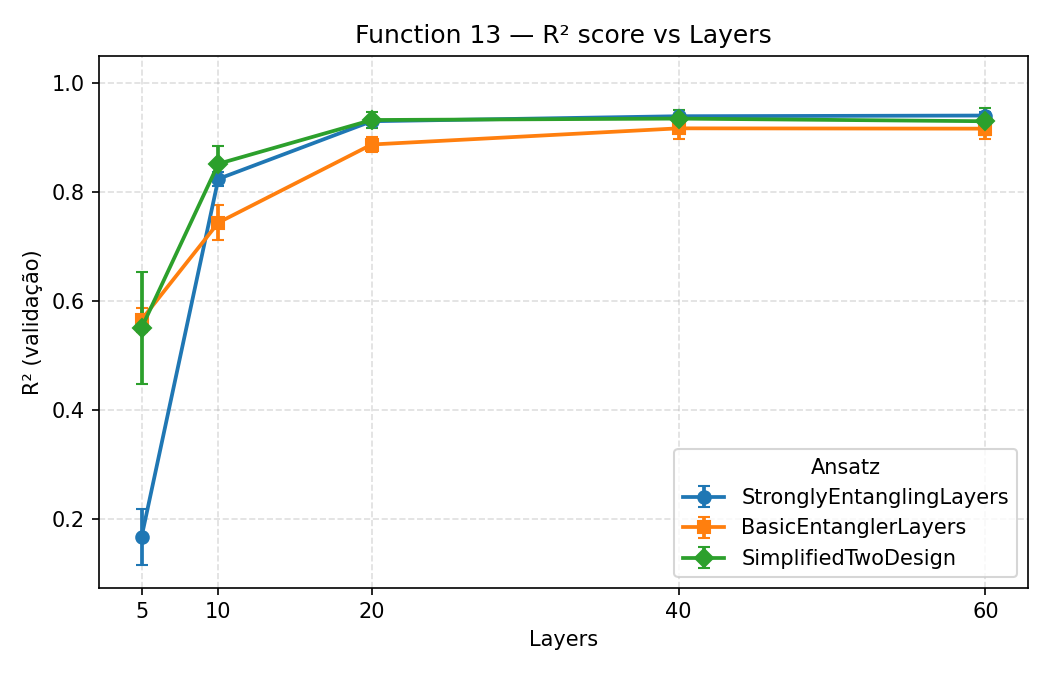} &
    \includegraphics[width=.32\textwidth]{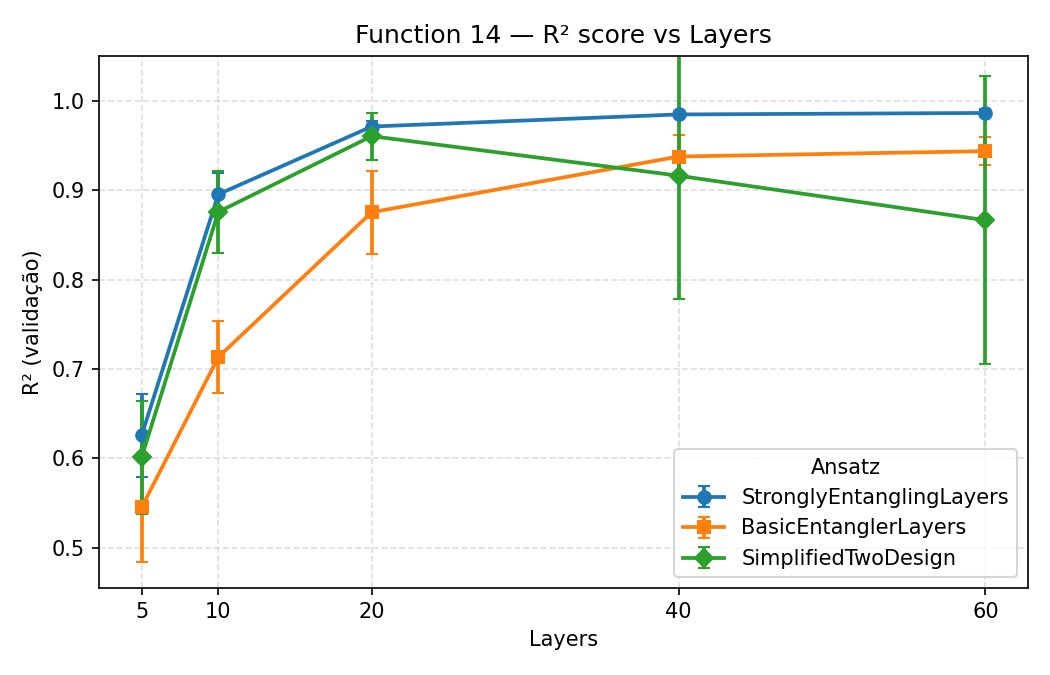} \\
    \includegraphics[width=.32\textwidth]{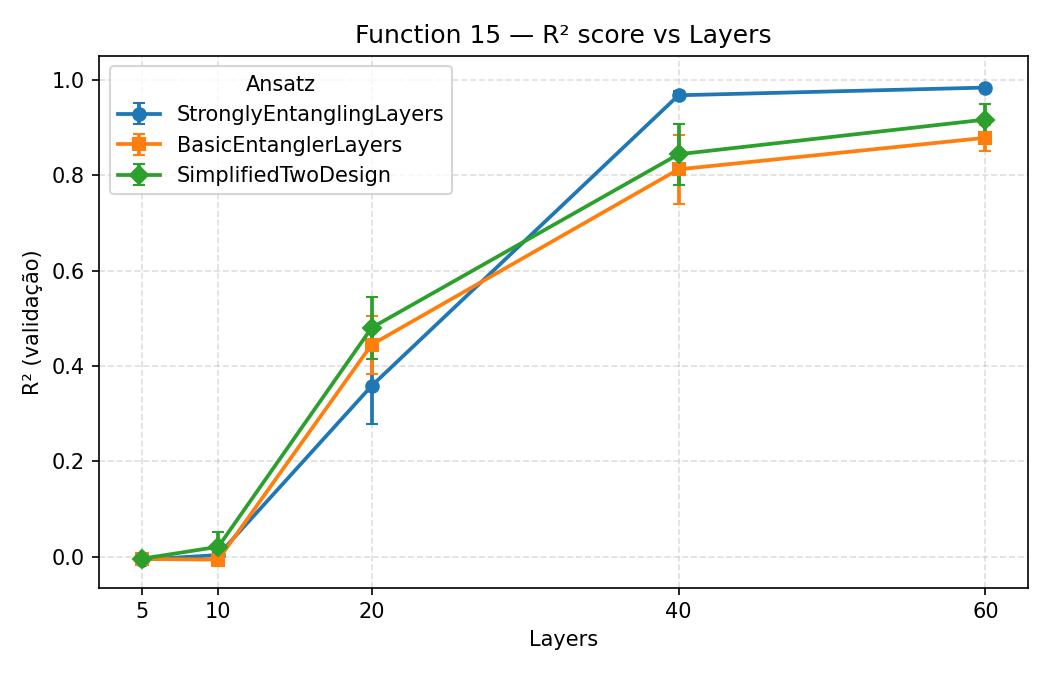} &
    \includegraphics[width=.32\textwidth]{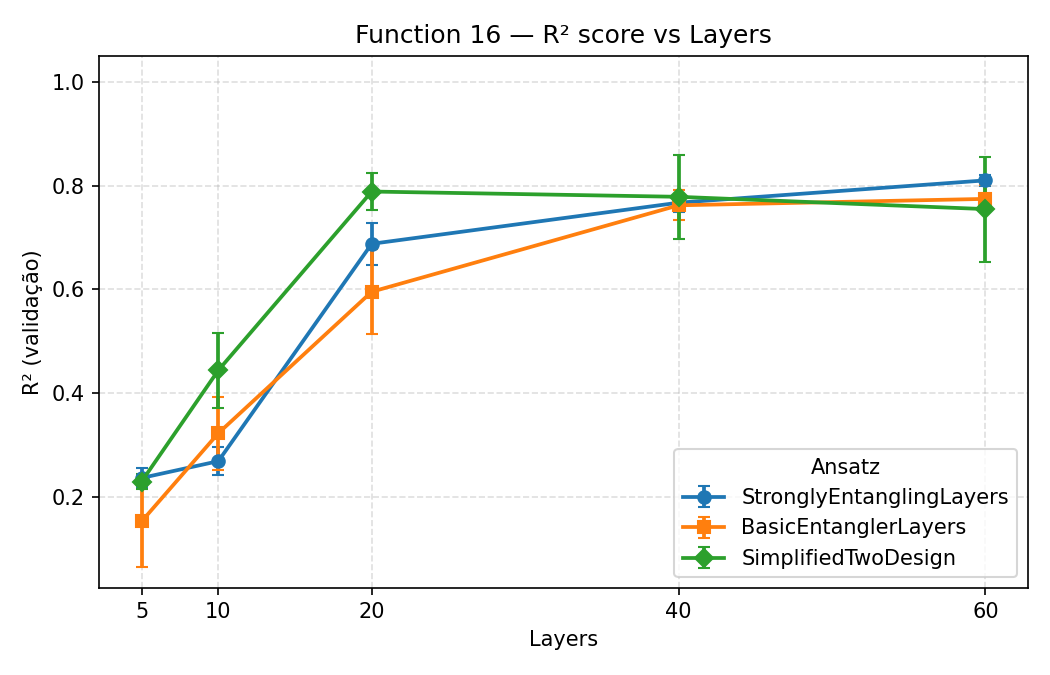} &
    \includegraphics[width=.32\textwidth]{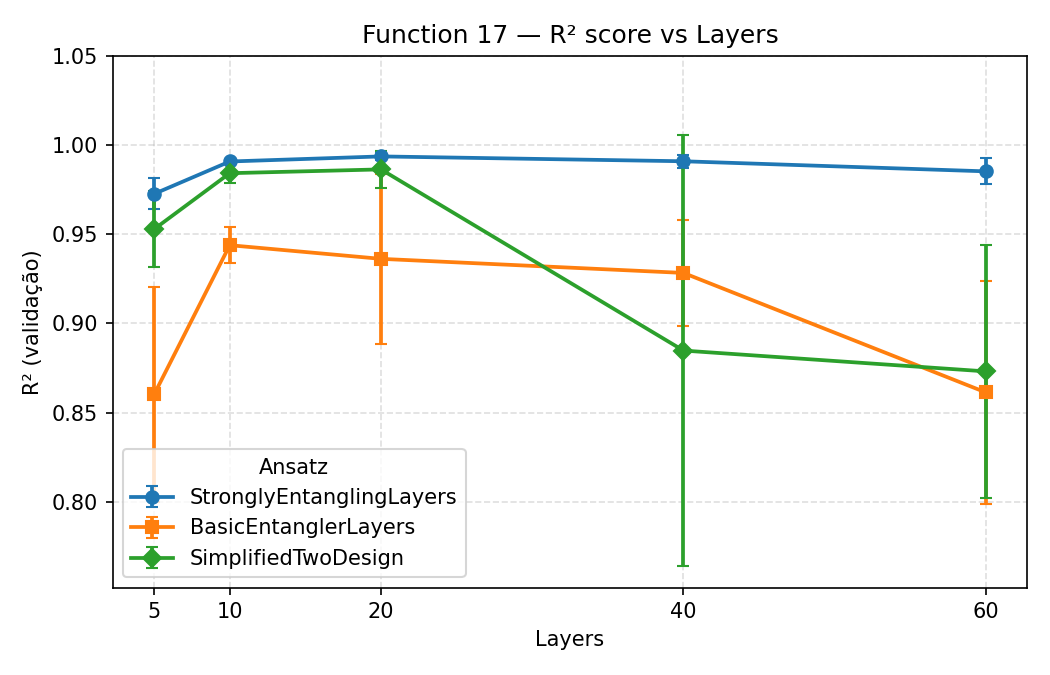} \\
    \includegraphics[width=.32\textwidth]{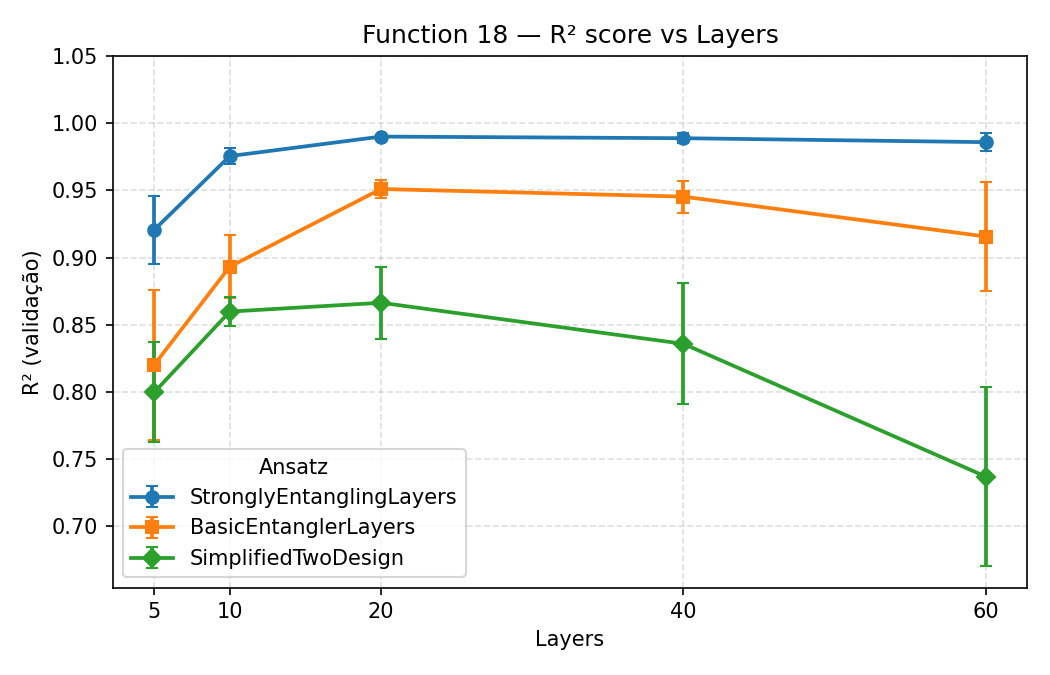} &
    \includegraphics[width=.32\textwidth]{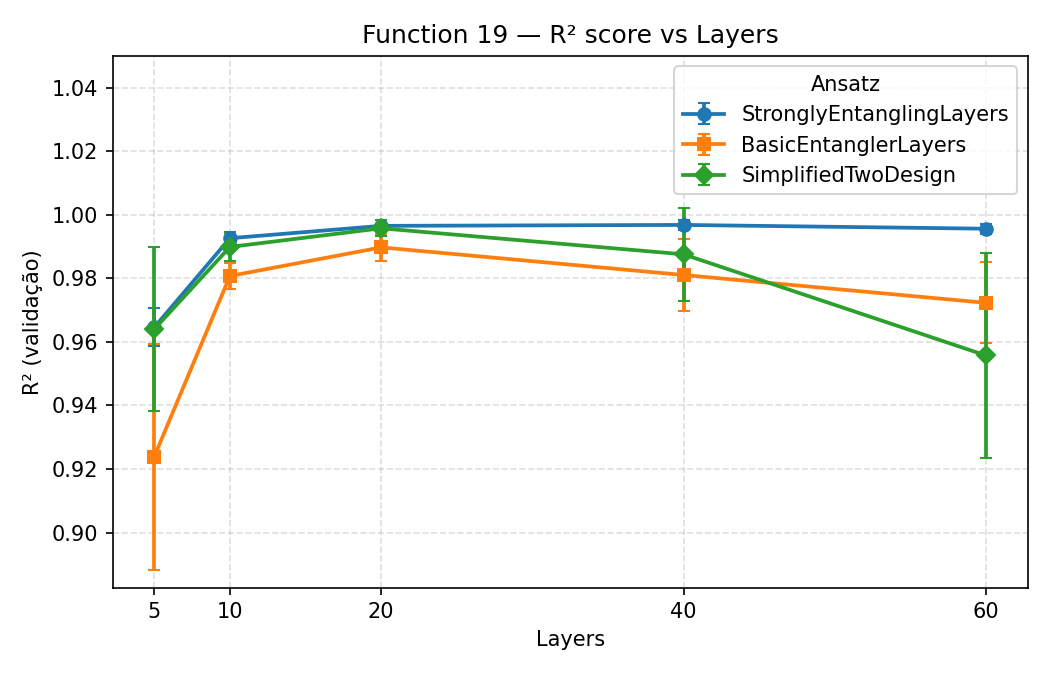} &
    \includegraphics[width=.32\textwidth]{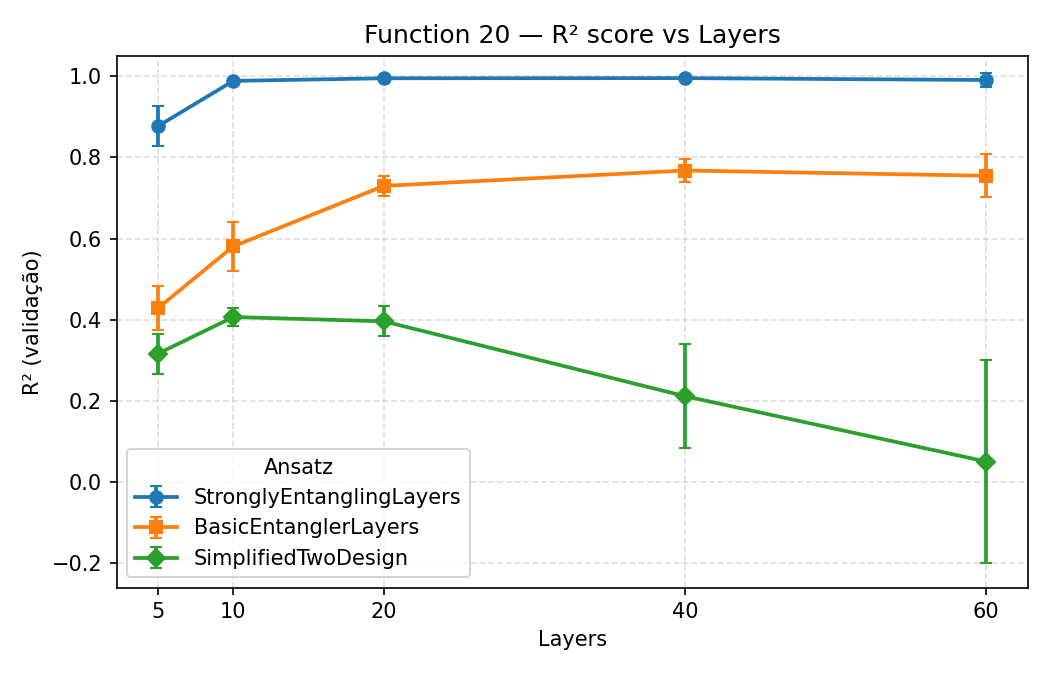} \\
  \end{tabular}\\[1ex] 
  \begin{tabular}{@{}cc@{}}
    \includegraphics[width=.32\textwidth]{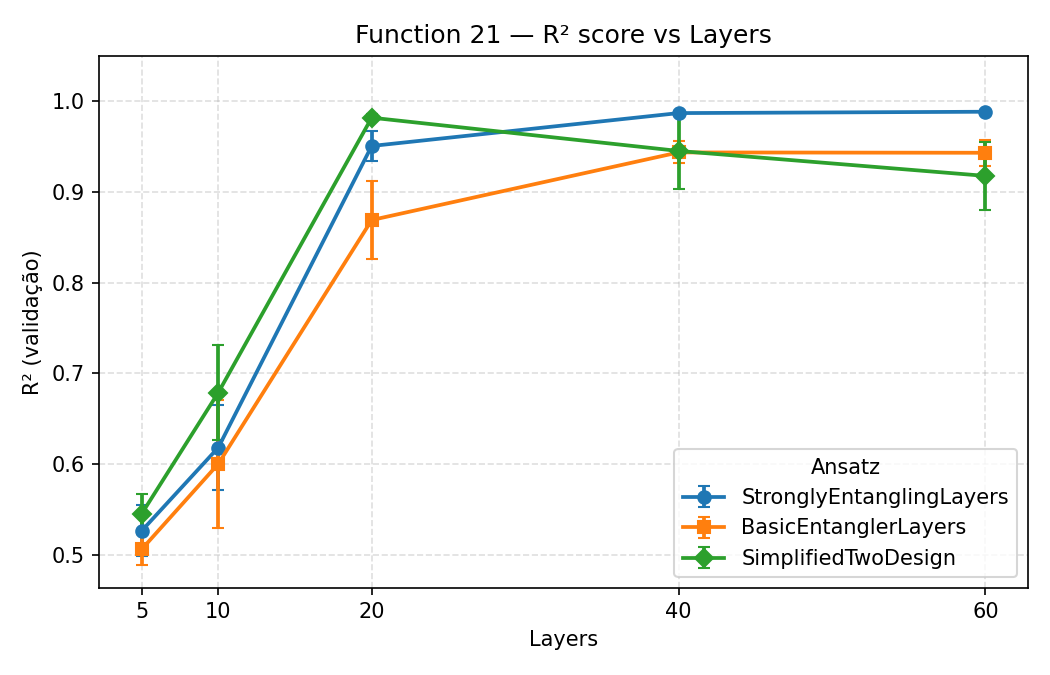} &
    \includegraphics[width=.32\textwidth]{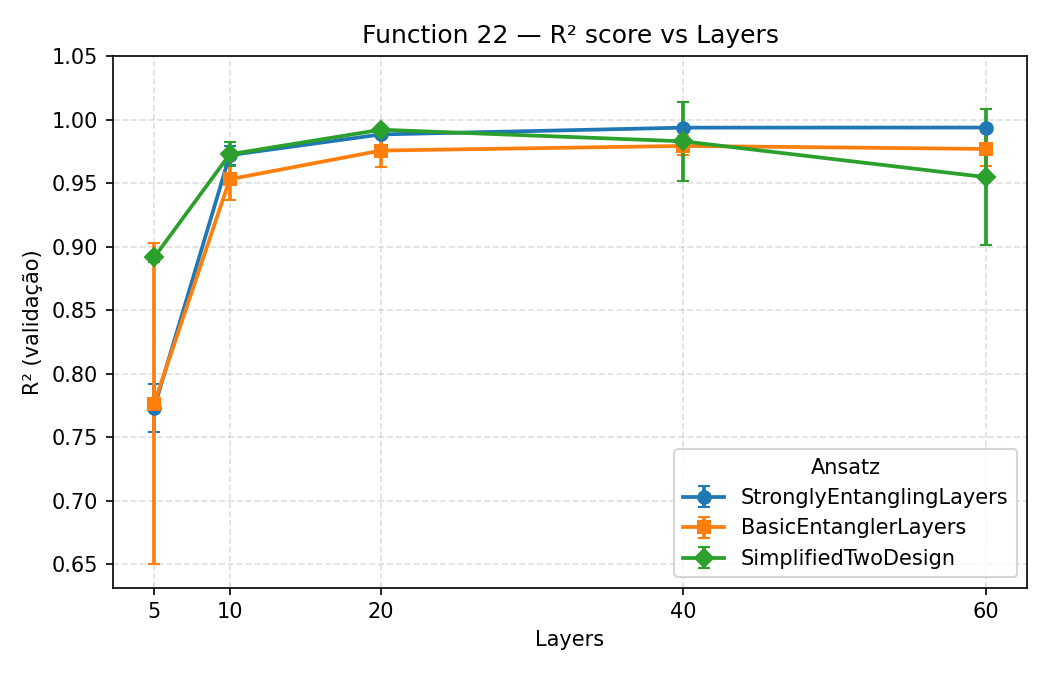} \\
  \end{tabular}
  \caption{Analysis of the $R^2$-Score for functions 12 to 22 (Mean and standard deviation across 10 runs).}
  \label{fig:functions_r2_vs_layers_part2}
\end{figure}

We checked the average behavior and its standard deviation of the 3 quantum neural network models used in Figures \ref{fig:functions_r2_vs_layers_part1} and \ref{fig:functions_r2_vs_layers_part2} for each of the 22 benchmark functions. It is possible to verify that there is an alternation of higher $R^2$-score between the SimplifiedTwoDesign and StronglyEntanglingLayers models, but the BasicEntanglingLayers model is almost always in second place in performance. The exceptions occur in function 8, when BasicEntanglingLayers ties with SimplifiedTwoDesign for second place, and in functions 12, 13, 15 and 17 when it is in third place. Therefore, a total of 18 times BasicEntanglingLayers is in second place.
It is also possible to see that sometimes increasing the number of layers does not mean increasing performance (increasing $R^2$ score) in the BasicEntanglingLayers and SimplifiedTwoDesign models.
The StronglyEntanglingLayers model generally benefits (increases its $R^2$ score) as the number of layers increases. It's even possible that the performance of function 12 could be improved if the number of layers in this model were increased, since the performance curve is still showing strong growth in the 60th repetition of layers (which doesn't seem to be the same behavior in function 13, where the performance of all 3 models seems to have converged).

An important point also analyzed in these graphs is that, despite the models not finding a high average value in some functions, within the 10 executions performed, it was still possible to find the maximum $R^2$-score, which suggests that the parametric initialization and the data used to train the models may prevent the data from converging to an optimal scenario. To see this situation, look at the standard deviation bar in the curves for functions 1, 5, 10, and 14.

\subsection{Defining best quantum architecture by dataset features}
\label{sec:RQQNN_metalearning}

In this section, we conducted a meta-learning analysis of five predictive scenarios, each designed to evaluate the ability of regression complexity measures to guide the automatic selection of quantum regression models. Across all scenarios, a classifier receives the same set of twelve meta-features introduced in Section~\ref{sec:complexityMetrics}, yet its prediction targets vary according to the architectural decision being automated. The objective is to determine whether the regression complexity descriptors contain sufficient information to discriminate between different quantum architectures, depths, and resource constraints. For each scenario, a Leave-One-Out Cross-Validation (LOOCV) protocol is used, and performance is compared against the accuracy of a naive majority-class predictor.
For each of the $\binom{12}{k}$ subsets of size $k$, a Random Forest classifier was trained on $21$ instances and tested on the remaining one. The average LOOCV accuracy was computed over the 22 splits and reported. 

\subsubsection{Scenario 1: Selecting the Best QNN Ansatz Among Two Leading Architectures}

The first scenario concerns the problem of selecting, for each dataset, the best-performing ansatz between the two highest-ranked quantum model families: the \textit{StronglyEntanglingLayers} and the \textit{SimplifiedTwoDesign} architectures. Across the 22 benchmark functions considered, the former achieved the best regression performance in 17 cases and the latter in 5, yielding a naive majority-class baseline of $77.27\%$. The meta-learning classifier, trained on the twelve complexity measures, surpasses this baseline when the appropriate subsets of features are used. Three combinations of complexity metrics achieve optimal LOOCV accuracy ($100\%$):
\begin{equation}
\text{(i)}\ \ (c1,\, c3,\, l1,\, s3,\, s4,\, t2),
\end{equation}
\begin{equation}
\text{(ii)}\ \ (c2,\, c3,\, l1,\, s3,\, s4,\, t2),
\end{equation}
\begin{equation}
\text{(iii)}\ (c1,\, c2,\, c3,\, s1,\, s3,\, l3,\, s4).
\end{equation}
The first two solutions share six metrics and differ only by substituting $c1$ with $c2$, whereas the third subset contains seven metrics but preserves the same structural components: global complexity indicators ($c1$, $c2$, $c3$), smoothness and local complexity descriptors ($s1$, $s3$, $s4$), and at least one landmarking-based nonlinearity measure ($l1$ or $l3$). These results indicate that optimal predictive accuracy arises from a stable configuration of complementary meta-features, rather than from arbitrary or isolated combinations.

\begin{figure}[ht]
    \centering
    \includegraphics[width=1.\linewidth]{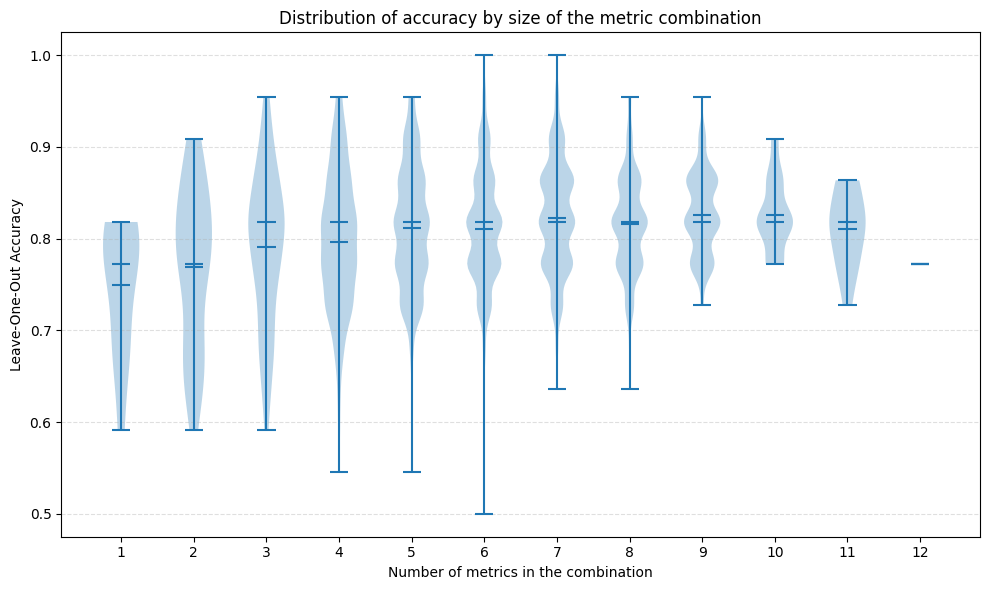}
    \caption{Violin plot of the number of metric combinations and their respective accuracy results when evaluated to determine which quantum circuit to use to solve the 22 nonlinear functions.}
    \label{fig:violinPlotMetalearning}
\end{figure}

The distribution of accuracies as a function of subset size is summarized in the violin plot shown in Figure~\ref{fig:violinPlotMetalearning}, which reveals substantial variability across subset sizes but also isolated regions of optimal predictive performance.

\subsubsection{Scenario 2: Selecting the Best RRQNN Model Under Operator and Qubit Constraints}

The second scenario evaluates the ability of the meta-learner to determine the best RRQNN variant among four competing configurations, each defined by its maximum number of quantum operators and number of qubits. Here is the frequency of each class:
\text{RRQNN-120-2q}: 11,\ \text{RRQNN-120-1q}: 7,\ 
\text{RRQNN-20-2q}: 3,\ \text{RRQNN-40-1q}: 1.
The majority-class baseline is $0.50$. Two feature subsets yield the best LOOCV accuracy ($0.8636$), far surpassing the baseline:
\begin{equation}
(c1,\, c3,\, l2,\, s1,\, l3,\, s4), \qquad
(c2,\, c3,\, l2,\, s1,\, l3,\, s4).
\end{equation}
Both optimal subsets share six complexity measures, differing only in the substitution of $c1$ and $c2$, mirroring the pattern observed in Scenario~1. Once again, the decisive role of $c3$, $s1$, $s4$, and higher-order linearity descriptors ($l2$, $l3$) suggests that the RRQNN architectural landscape is strongly governed by a mixture of global structure, local smoothness, and landmarking behavior.

\begin{figure}[ht]
    \centering
    \includegraphics[width=1.\linewidth]{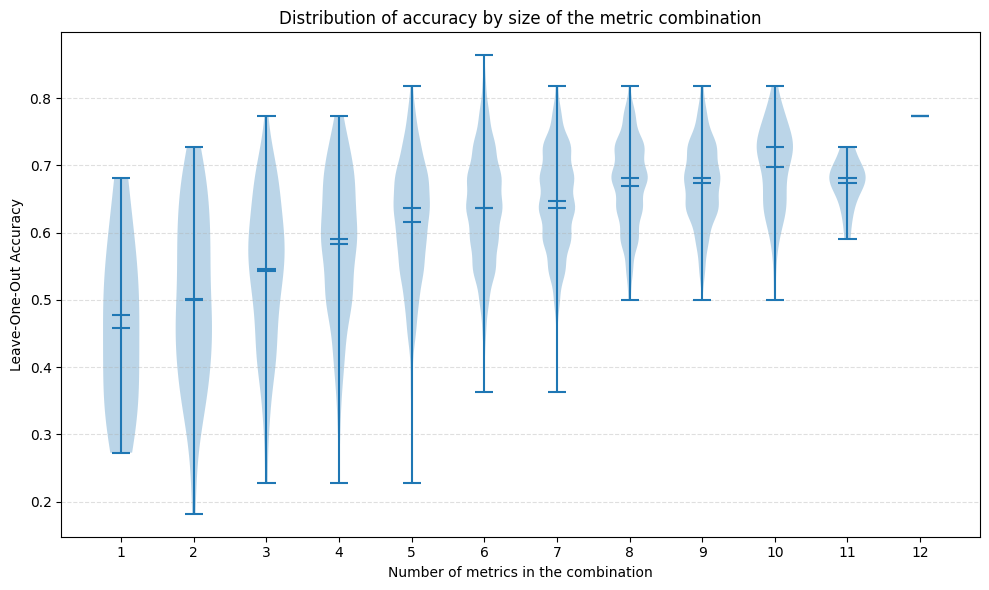}
    \caption{Violin plot of the number of metric combinations and their respective accuracy results when evaluated to determine which RRQNN architecture (considering maximum operators and qubit count) should be selected to solve the 22 nonlinear benchmark functions.}
    \label{fig:violinPlotMetalearning_RRQNN_operators}
\end{figure}

The distribution of accuracies as a function of subset size is summarized in the violin plot shown in Figure~\ref{fig:violinPlotMetalearning_RRQNN_operators}, which reveals substantial variability across subset sizes but also clear clusters of high-performing combinations capable of surpassing the majority-class baseline, thereby demonstrating that the complexity measures encode discriminative information regarding operator budget and qubit allocation in RRQNN models.

\subsubsection{Scenario 3: Determining the Optimal Number of Qubits in RRQNN Models}

In this scenario, the classifier must choose between one- and two-qubit RRQNN architectures. The empirical distribution of best models is
'2q': 14, and '1q': 8.
yielding a majority baseline of $0.6364$. Remarkably, three metric subsets achieve optimal LOOCV accuracy ($1.0$):
\begin{equation}
(c1,\, c2,\, c3,\, l2,\, l3,\, s4),
\end{equation}
\begin{equation}
(c1,\, c3,\, l2,\, l3,\, s4,\, t2),
\end{equation}
\begin{equation}
(c2,\, c3,\, l2,\, l3,\, s4,\, t2).
\end{equation}
These combinations reflect the same structural template observed in Scenarios~1 and~2: a core of global complexity indicators ($c1,c2,c3$), smoothness/nonlinearity descriptors ($s4$), and landmark metrics ($l2,l3$), with $t2$ acting as a stabilizing topological descriptor in two of the three optimal subsets. The consistent emergence of this motif across scenarios suggests that the number of qubits required for optimal regression behavior is highly dependent on the global complexity profile of the dataset.

\begin{figure}[ht]
    \centering
    \includegraphics[width=1.\linewidth]{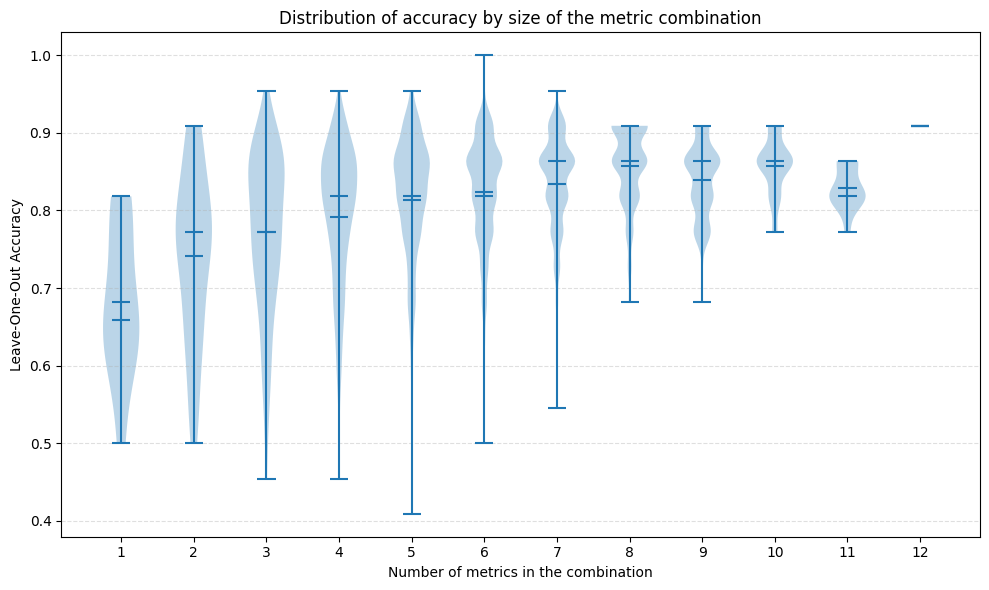}
    \caption{Violin plot of the number of metric combinations and their respective accuracy results when evaluated to determine whether a one-qubit or two-qubit RRQNN architecture should be used to solve the 22 nonlinear benchmark functions.}
    \label{fig:violinPlotMetalearning_RRQNN_qubits}
\end{figure}

Figure~\ref{fig:violinPlotMetalearning_RRQNN_qubits} shows that only subsets containing six meta-features were able to achieve $100\%$ accuracy, whereas all smaller subsets yielded lower performance. The figure further shows that as more meta-features are incorporated, the variability in accuracy decreases, reinforcing the role of richer complexity information in enabling reliable qubit-selection decisions.

\subsubsection{Scenario 4: Selecting the Best Overall Quantum Model Considering Architecture and Depth}

A more challenging classification task arises when the meta-learner must select, among all quantum models and depths tested, the best-performing configuration for each dataset. The empirical distribution of winners is:
\text{StronglyEntanglingLayers-40}: 5,\ 
\text{StronglyEntanglingLayers-60}: 5,\ 
\text{SimplifiedTwoDesign-20}: 4,\ 
\text{StronglyEntanglingLayers-20}: 4,\ 
\text{SimplifiedTwoDesign-40}: 3,\ 
\text{RRQNN-120-2q}: 1, yielding a naive baseline of $0.17$ (or $0.20$ if we ignore the lone RRQNN winner). The best LOOCV accuracy achieved by any subset of meta-features is $0.5455$, obtained by the following combinations:
\begin{equation}
(c1,\, c2,\, c3,\, s1), \quad
(c1,\, c3,\, s1,\, t2), \quad
(c2,\, c3,\, s1,\, t2),
\end{equation}
\begin{equation}
(c1,\, c2,\, c3,\, s1,\, s3), \quad
(c1,\, c2,\, c3,\, s1,\, t2), \quad
(c1,\, c2,\, c3,\, s1,\, s3,\, s4).
\end{equation}
Although lower than in the first three scenarios, the predictive accuracy remains significantly above the majority baseline and demonstrates that complexity measures can meaningfully guide model-depth selection, even when the decision space is substantially larger.

\begin{figure}[ht]
    \centering
    \includegraphics[width=1.\linewidth]{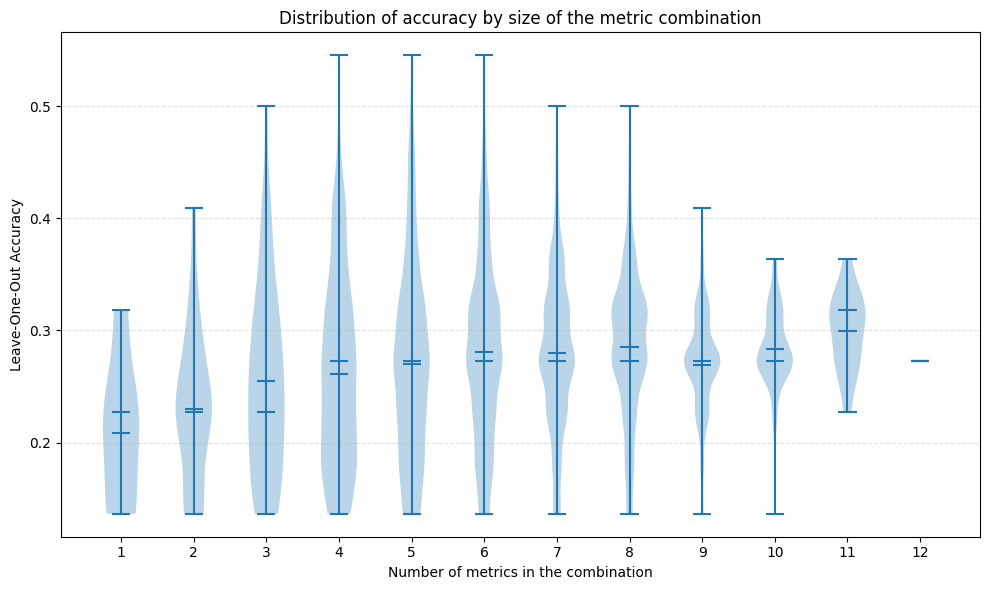}
    \caption{Violin plot of the number of metric combinations and their respective accuracy results when evaluated to determine which quantum model and layer configuration achieves the best performance across the 22 nonlinear benchmark functions.}
    \label{fig:violinPlotMetalearning_allModels_layers}
\end{figure}

Figure~\ref{fig:violinPlotMetalearning_allModels_layers}, which reveals substantial variability across subset sizes but also nontrivial pockets of above-baseline predictive accuracy. In particular, the highest accuracy levels are concentrated around combinations of 4, 5, and 6 meta-features, whereas nearly all subset sizes exhibit considerable dispersion in performance. This high degree of variability reflects the difficulty of this predictive scenario, where the classifier must discriminate among a broad and heterogeneous set of quantum architectures and depth configurations. Even under these challenging conditions, the meta-features still retain some discriminative power for identifying the architecture–depth pair.

\subsubsection{Scenario 5: Selecting the Best Quantum Model Without Considering Depth}

Finally, we collapse all layer-dependent variants into their respective architectural families, yielding the class distribution: \text{StronglyEntanglingLayers}: 14,\ 
\text{SimplifiedTwoDesign}: 6,\ 
\text{RRQNN-120-2q}: 2,
with a majority baseline of $0.6364$. The best LOOCV performance achieved by the meta-learner is $0.8182$, associated with two feature subsets:
\begin{equation}
(c3,\, s3), \qquad (c1,\, c2,\, c3,\, s1,\, s3).
\end{equation}
In contrast to previous scenarios, here a two-feature subset is already sufficient to surpass the naive baseline by a large margin, indicating that key distinctions across architectural families may be captured by a small set of highly informative regression complexity measures.

\begin{figure}[ht]
    \centering
    \includegraphics[width=1.\linewidth]{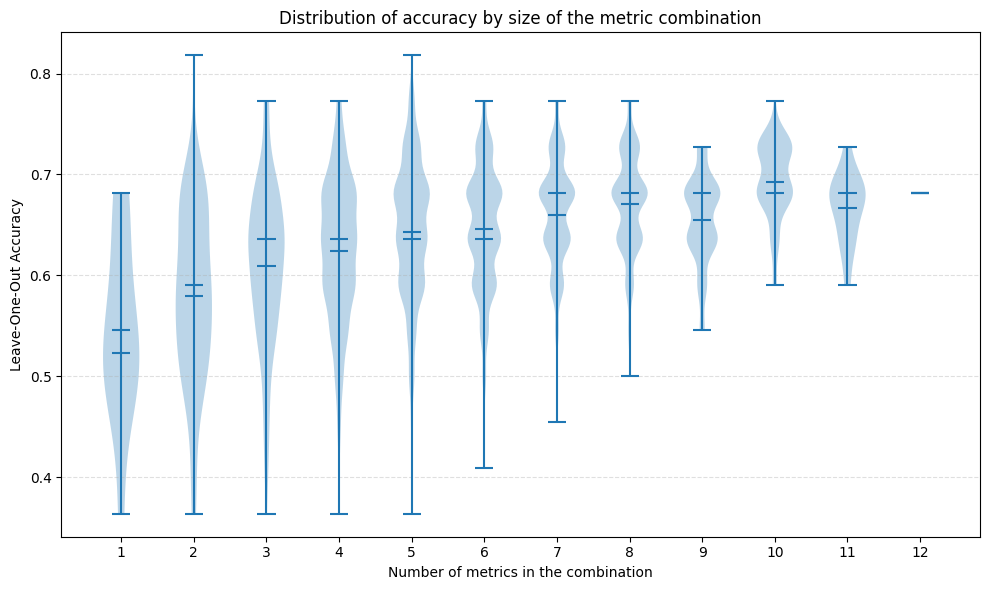}
    \caption{Violin plot of the number of metric combinations and their respective accuracy results when evaluated to determine which quantum model architecture—regardless of its number of layers—should be selected to solve the 22 nonlinear benchmark functions.}
    \label{fig:violinPlotMetalearning_allModels_noLayers}
\end{figure}

Figure~\ref{fig:violinPlotMetalearning_allModels_noLayers}, which reveals substantial variability across subset sizes but also regions of markedly enhanced predictive accuracy, thus demonstrating that the complexity metrics remain informative even when the task is reduced to distinguishing among architectural families independent of their layer configurations.

\subsubsection{Discussion}

Across all five scenarios, several consistent patterns emerge. First, the $C$-family metrics—particularly $C3$—appear essential for predicting which quantum architecture will succeed on a given dataset. Second, smoothness and nearest-neighbor–based descriptors from the $S$ family ($S1$, $S3$, $S4$) occur in nearly all high-performing subsets, highlighting the role of local geometric structure in governing quantum expressivity. Third, landmarking-based linearity indicators ($L1$, $L2$, $L3$) regularly appear in the optimal combinations, suggesting that the behavior of simple learners on the dataset provides powerful discriminative information. Finally, the ability of the meta-learner to exceed all majority-class baselines---often by a substantial margin---demonstrates that regression complexity measures contain highly actionable information for guiding quantum model selection, even in the presence of significant architectural variability and limited training examples.

The results reported in this section are obtained from a specific set of benchmark regression functions and therefore motivate further analysis on more diverse and generic datasets. Extending the evaluation to problems with different statistical properties, dimensionalities, and data-generation mechanisms will be important to better characterize the scope and generality of the proposed meta-learning approach.

At the same time, the observed performance trends indicate that structural descriptors of regression datasets capture information that is relevant for quantum architecture selection, suggesting that dataset-level characteristics can be exploited to inform the choice of suitable quantum ans\"atze in regression tasks.

\section{Conclusion}
\label{sec:conclusion}

In this work, it was possible to analyze the behavior of variational quantum neural networks applied to a benchmark of 22 non-linear function regression problems, with a variety of behaviors, including non-continuous, non-separable, and multimodal. An approach for defining reduced regression quantum neural networks (RRQNN) was proposed based on a search performed by Genetic Algorithms. The quantum models were compared with 17 classical baseline models. The results were extensively evaluated and compared, verifying the high impact of the number of parameters required for a classical model to perform the regression compared to the quantum model. 
Through statistical testing, it was possible to verify that although the average of the quantum models was not always greater than that of the classical models, there is statistical equivalence in 7 of the 22 functions.

It was verified that the RRQNN networks found by the Genetic Algorithm had good comparative results (resulting in the best quantum model in one of the scenarios). However, it was possible to assess that the circuits found did not have a significant pattern, based on the execution of 3 unsupervised clustering algorithms, and analyzing 4 metrics that evaluate the quality of the generated clusters.

It was also possible to verify that there is information (meta-features) in the databases that suggests identifying which quantum neural network layer can be used to maximize performance. We performed an exhaustive test combinatorially varying 12 metrics established in the literature, and verified that there are 3 combinations that achieve an optimal choice between the two best quantum models evaluated.

Based on all the analysis conducted, it is possible to conclude that quantum models are very promising in modeling nonlinear problems, being equivalent to or better than almost all classical models evaluated, performing the task, after training, with much smaller parameters. There are challenging factors in this area, among which we can point out as main ones the definition of layers (repetitions of their layers), premature convergence at barren plateaus due to sensitivity to training conditions, the lack of standardization of architectures that best solve the problems, and the training time due to the need to simulate the models in classical environments. This work paves the way for future work connected to the computability of variational quantum models and improved training.


As future work, several extensions of the present study are envisioned. First, the analysis can be expanded to regression problems with higher-dimensional input spaces and more complex functional structures, allowing a deeper assessment of the scalability of automatically synthesized quantum neural networks. In this context, an important research direction is the automation of meta-feature selection itself. The exhaustive searches conducted in this work indicate that specific subsets of regression complexity measures can optimally discriminate between competing quantum architectures. Future studies may therefore investigate principled mechanisms for automatically selecting or learning these meta-features, further strengthening the meta-learning framework for quantum model selection.

Moreover, the present study focuses on a restricted search space in terms of qubit count and circuit depth, and all experiments are performed under noiseless simulation. Extending the analysis to larger circuit architectures, higher-dimensional input spaces, and noisy quantum hardware constitutes a critical next step toward evaluating the practical viability of the proposed approach. In this direction, combining error-prediction strategies for noisy quantum processors \cite{saravanan2022data} with data-driven circuit-construction methodologies based on meta-features represents a promising avenue. Incorporating hardware-aware constraints, realistic noise models, and error-mitigation techniques directly into the metalearning loop may substantially improve the robustness and reliability of automatically synthesized quantum models.

Finally, although the comparison with classical regressors highlights the parameter efficiency of quantum neural networks, a tighter integration with classical metalearning pipelines remains an open challenge. Exploring connections with established techniques such as neural architecture search, ensemble-based model selection, and Bayesian optimization could provide deeper insights into when and why quantum models offer advantages over classical alternatives. Collectively, these directions point toward the development of principled, scalable, and hardware-adaptive metalearning frameworks for quantum regression, helping to bridge the gap between quantum machine learning and mature automated learning methodologies in the classical domain.

\backmatter


\bmhead{Acknowledgements}

Fernando M. de Paula Neto acknowledges financial support from the Brazilian National Council for Scientific and Technological Development (CNPq) under Grant No.~408499/2025-7. Fernando M. de Paula Neto is also a researcher affiliated with the National Institute of Science and Technology for Applied Quantum Computing, supported by CNPq under Grant No.~408884/2024-0. Fernando M. de Paula Neto and Paulo Mattos are part of the Brazilian National Institute for Artificial Intelligence, supported by CNPq under Grant No.~406417/2022-9. Paulo Mattos also acknowledges financial support from CNPq for an academic project under Grant No.~306934/2025-6.
Felipe Fanchini acknowledges financial support from the National Institute of Science and Technology for Applied Quantum Computing through CNPq process No. 408884/2024-0, and from the São Paulo Research Foundation (FAPESP), under grant numbers 2023/04987-6 and 2024/00998-6. Felipe Fanchini also acknowledges support from the Office of Naval Research Global (ONR Global) and the Air Force Office of Scientific Research (AFOSR), grant No. N62909-24-1-2012.


\section*{Declarations}

The authors declare no conflict of interest.

The data generated are detailed in the article.

FMPN conceived the idea, conducted the experiments, and wrote the article. LRS, PSGMN, and FFF contributed to the writing and revision of the manuscript, with LRS also participating in the experimental execution.


\begin{appendices}



\section{Results for 1D functions}

The detailed numerical results of the proof-of-concept experiments discussed in Section~\ref{sec:proofOfConcept} are in Table~\ref{tab:resultsConceptProof}.

\begin{table}[ht]
\tiny
\centering
\begin{tabular}{llrr}
\hline
Dataset      & Model          & $R^2$-score           & RMSE \\ \hline
$f_{1}^{1D}$ & RRQNN-25 gates & \textbf{0.999000} & 0.009191  \\
$f_{1}^{1D}$ & RRQNN-20 gates & 0.998200          & 0.012837  \\
$f_{1}^{1D}$ & RRQNN-15 gates & 0.997100          & 0.016551  \\
$f_{1}^{1D}$ & RF             & 0.995987          & 0.018765  \\
$f_{1}^{1D}$ & DT             & 0.994382          & 0.022859  \\
$f_{1}^{1D}$ & RRQNN-10 gates & 0.994000          & 0.022989  \\
$f_{1}^{1D}$ & KNN2           & 0.993750          & 0.023861  \\
\hline
\end{tabular}
\caption{Results of the Proof of Concept detailed in Section\ref{sec:proofOfConcept}}
\label{tab:resultsConceptProof}
\end{table}

\section{Results for 3D functions}

The detailed numerical results for each of the 22 benchmark regression datasets are reported in Tables~\ref{tab:resultCEC2020F1}--\ref{tab:resultCEC2020F22}. Each table corresponds to one benchmark function and summarizes the performance of all evaluated models under the experimental protocol described in this work.

\begin{table}[ht]
\centering
\tiny
\begin{tabular}{llll}
\toprule
Model & $R^2$-score $\pm$ std & RMSE $\pm$ std & \#Params $\pm$ std \\
\midrule
RF & 0.9989 $\pm$ 0.0001 & 0.0145 $\pm$ 0.0145 & 205285.8 $\pm$ 347.31 \\
knn4 & 0.9974 $\pm$ 0.0002 & 0.0221 $\pm$ 0.0221 & 0.0 $\pm$ 0.0 \\
knn3 & 0.9973 $\pm$ 0.0001 & 0.0227 $\pm$ 0.0227 & 0.0 $\pm$ 0.0 \\
RRQNN-120-2q & 0.9969 $\pm$ 0.0006 & 0.0249 $\pm$ 0.0249 & 84.4 $\pm$ 3.34 \\
DT & 0.9969 $\pm$ 0.0001 & 0.0245 $\pm$ 0.0245 & 3245.0 $\pm$ 9.29 \\
StronglyEntanglingLayers-40 & 0.9963 $\pm$ 0.0016 & 0.0262 $\pm$ 0.0262 & 240.0 $\pm$ 0.0 \\
knn2 & 0.9963 $\pm$ 0.0001 & 0.0266 $\pm$ 0.0266 & 0.0 $\pm$ 0.0 \\
\bottomrule
\end{tabular}

\caption{Results of the benchmark Function 1 }
 \label{tab:resultCEC2020F1}
 \end{table}

\begin{table}[ht]
\tiny
\centering
\begin{tabular}{llll}
\toprule
Model & $R^2$-score $\pm$ std & RMSE $\pm$ std & \#Params $\pm$ std \\
\midrule
RF & 1.0 $\pm$ 0.0 & 0.0 $\pm$ 0.0 & 173309.0 $\pm$ 1572.14 \\
DT & 1.0 $\pm$ 0.0 & 0.0 $\pm$ 0.0 & 2557.2 $\pm$ 30.63 \\
MLP500-500-ReLU & 0.9974 $\pm$ 0.0014 & 0.0306 $\pm$ 0.0306 & 252501.0 $\pm$ 0.0 \\
knn4 & 0.9974 $\pm$ 0.0002 & 0.031 $\pm$ 0.031 & 0.0 $\pm$ 0.0 \\
knn3 & 0.9973 $\pm$ 0.0002 & 0.0319 $\pm$ 0.0319 & 0.0 $\pm$ 0.0 \\
SimplifiedTwoDesign-20 & 0.9972 $\pm$ 0.0021 & 0.031 $\pm$ 0.031 & 42.0 $\pm$ 0.0 \\
StronglyEntanglingLayers-60 & 0.9968 $\pm$ 0.0014 & 0.0343 $\pm$ 0.0343 & 360.0 $\pm$ 0.0 \\
\bottomrule
\end{tabular}

\caption{Results of the benchmark Function 2 }
 \label{tab:resultCEC2020F2}
 \end{table}

\begin{table}[ht]
\tiny
\centering
\begin{tabular}{llll}
\toprule
Model & $R^2$-score $\pm$ std & RMSE $\pm$ std & \#Params $\pm$ std \\
\midrule
RF & 1.0 $\pm$ 0.0 & 0.0 $\pm$ 0.0 & 173309.0 $\pm$ 1572.14 \\
DT & 1.0 $\pm$ 0.0 & 0.0 $\pm$ 0.0 & 2557.2 $\pm$ 30.63 \\
SimplifiedTwoDesign-20 & 0.9977 $\pm$ 0.001 & 0.0296 $\pm$ 0.0296 & 42.0 $\pm$ 0.0 \\
MLP500-500-ReLU & 0.9974 $\pm$ 0.0014 & 0.0306 $\pm$ 0.0306 & 252501.0 $\pm$ 0.0 \\
knn4 & 0.9974 $\pm$ 0.0002 & 0.031 $\pm$ 0.031 & 0.0 $\pm$ 0.0 \\
knn3 & 0.9973 $\pm$ 0.0002 & 0.0319 $\pm$ 0.0319 & 0.0 $\pm$ 0.0 \\
knn2 & 0.9964 $\pm$ 0.0002 & 0.0372 $\pm$ 0.0372 & 0.0 $\pm$ 0.0 \\
\bottomrule
\end{tabular}

\caption{Results of the benchmark Function 3 }
 \label{tab:resultCEC2020F3}
 \end{table}

\begin{table}[ht]
\tiny
\centering
\begin{tabular}{llll}
\toprule
Model & $R^2$-score $\pm$ std & RMSE $\pm$ std & \#Params $\pm$ std \\
\midrule
RF & 1.0 $\pm$ 0.0 & 0.0 $\pm$ 0.0 & 173545.8 $\pm$ 1210.53 \\
DT & 1.0 $\pm$ 0.0 & 0.0 $\pm$ 0.0 & 2562.4 $\pm$ 22.84 \\
MLP500-500-ReLU & 0.9976 $\pm$ 0.0012 & 0.0297 $\pm$ 0.0297 & 252501.0 $\pm$ 0.0 \\
knn4 & 0.9975 $\pm$ 0.0001 & 0.0307 $\pm$ 0.0307 & 0.0 $\pm$ 0.0 \\
knn3 & 0.9973 $\pm$ 0.0001 & 0.0318 $\pm$ 0.0318 & 0.0 $\pm$ 0.0 \\
SimplifiedTwoDesign-20 & 0.9967 $\pm$ 0.0043 & 0.0319 $\pm$ 0.0319 & 42.0 $\pm$ 0.0 \\
StronglyEntanglingLayers-40 & 0.9966 $\pm$ 0.0015 & 0.0354 $\pm$ 0.0354 & 240.0 $\pm$ 0.0 \\
\bottomrule
\end{tabular}

\caption{Results of the benchmark Function 4 }
 \label{tab:resultCEC2020F4}
 \end{table}

\begin{table}[ht]
\tiny
\centering
\begin{tabular}{llll}
\toprule
Model & $R^2$-score $\pm$ std & RMSE $\pm$ std & \#Params $\pm$ std \\
\midrule
RF & 1.0 $\pm$ 0.0 & 0.0007 $\pm$ 0.0007 & 206018.8 $\pm$ 296.56 \\
DT & 1.0 $\pm$ 0.0 & 0.0008 $\pm$ 0.0008 & 3249.6 $\pm$ 7.89 \\
StronglyEntanglingLayers-40 & 0.9962 $\pm$ 0.0007 & 0.0361 $\pm$ 0.0361 & 240.0 $\pm$ 0.0 \\
StronglyEntanglingLayers-60 & 0.996 $\pm$ 0.0015 & 0.0369 $\pm$ 0.0369 & 360.0 $\pm$ 0.0 \\
knn3 & 0.9956 $\pm$ 0.0003 & 0.0387 $\pm$ 0.0387 & 0.0 $\pm$ 0.0 \\
knn4 & 0.9955 $\pm$ 0.0004 & 0.039 $\pm$ 0.039 & 0.0 $\pm$ 0.0 \\
SimplifiedTwoDesign-20 & 0.9951 $\pm$ 0.0021 & 0.0406 $\pm$ 0.0406 & 42.0 $\pm$ 0.0 \\
\bottomrule
\end{tabular}

\caption{Results of the benchmark Function 5 }
 \label{tab:resultCEC2020F5}
 \end{table}

\begin{table}[ht]
\tiny
\centering
\begin{tabular}{llll}
\toprule
Model & $R^2$-score $\pm$ std & RMSE $\pm$ std & \#Params $\pm$ std \\
\midrule
RF & 0.9992 $\pm$ 0.0001 & 0.0108 $\pm$ 0.0108 & 211158.0 $\pm$ 6.53 \\
knn4 & 0.9983 $\pm$ 0.0001 & 0.016 $\pm$ 0.016 & 0.0 $\pm$ 0.0 \\
knn3 & 0.9982 $\pm$ 0.0001 & 0.0166 $\pm$ 0.0166 & 0.0 $\pm$ 0.0 \\
DT & 0.9977 $\pm$ 0.0001 & 0.0188 $\pm$ 0.0188 & 3349.0 $\pm$ 0.0 \\
knn2 & 0.9975 $\pm$ 0.0001 & 0.0197 $\pm$ 0.0197 & 0.0 $\pm$ 0.0 \\
RRQNN-120-2q & 0.9971 $\pm$ 0.0007 & 0.0211 $\pm$ 0.0211 & 86.9 $\pm$ 4.28 \\
StronglyEntanglingLayers-40 & 0.9968 $\pm$ 0.0021 & 0.0214 $\pm$ 0.0214 & 240.0 $\pm$ 0.0 \\
\bottomrule
\end{tabular}

\caption{Results of the benchmark Function 6 }
 \label{tab:resultCEC2020F6}
 \end{table}

\begin{table}[ht]
\tiny
\centering
\begin{tabular}{llll}
\toprule
Model & $R^2$-score $\pm$ std & RMSE $\pm$ std & \#Params $\pm$ std \\
\midrule
RF & 0.8592 $\pm$ 0.0042 & 0.1275 $\pm$ 0.1275 & 206265.8 $\pm$ 319.62 \\
DT & 0.8532 $\pm$ 0.0103 & 0.1507 $\pm$ 0.1507 & 3250.6 $\pm$ 9.08 \\
knn2 & 0.8086 $\pm$ 0.0067 & 0.1583 $\pm$ 0.1583 & 0.0 $\pm$ 0.0 \\
knn3 & 0.8042 $\pm$ 0.0084 & 0.1508 $\pm$ 0.1508 & 0.0 $\pm$ 0.0 \\
knn4 & 0.771 $\pm$ 0.0112 & 0.1563 $\pm$ 0.1563 & 0.0 $\pm$ 0.0 \\
SimplifiedTwoDesign-60 & 0.7133 $\pm$ 0.0647 & 0.2106 $\pm$ 0.2106 & 122.0 $\pm$ 0.0 \\
BasicEntanglerLayers-60 & 0.656 $\pm$ 0.0312 & 0.2319 $\pm$ 0.2319 & 120.0 $\pm$ 0.0 \\
\bottomrule
\end{tabular}

\caption{Results of the benchmark Function 7 }
 \label{tab:resultCEC2020F7}
 \end{table}

\begin{table}[ht]
\tiny
\centering
\begin{tabular}{llll}
\toprule
Model & $R^2$-score $\pm$ std & RMSE $\pm$ std & \#Params $\pm$ std \\
\midrule
RF & 0.9927 $\pm$ 0.0005 & 0.03 $\pm$ 0.03 & 206403.6 $\pm$ 346.53 \\
DT & 0.9906 $\pm$ 0.0006 & 0.0343 $\pm$ 0.0343 & 3254.4 $\pm$ 8.11 \\
knn3 & 0.9854 $\pm$ 0.0007 & 0.0424 $\pm$ 0.0424 & 0.0 $\pm$ 0.0 \\
knn4 & 0.9845 $\pm$ 0.0009 & 0.0436 $\pm$ 0.0436 & 0.0 $\pm$ 0.0 \\
knn2 & 0.983 $\pm$ 0.0008 & 0.046 $\pm$ 0.046 & 0.0 $\pm$ 0.0 \\
StronglyEntanglingLayers-20 & 0.9431 $\pm$ 0.0044 & 0.0846 $\pm$ 0.0846 & 120.0 $\pm$ 0.0 \\
SimplifiedTwoDesign-20 & 0.943 $\pm$ 0.0039 & 0.0847 $\pm$ 0.0847 & 42.0 $\pm$ 0.0 \\
\bottomrule
\end{tabular}

\caption{Results of the benchmark Function 8 }
 \label{tab:resultCEC2020F8}
 \end{table}

\begin{table}[ht]
\tiny
\centering
\begin{tabular}{llll}
\toprule
Model & $R^2$-score $\pm$ std & RMSE $\pm$ std & \#Params $\pm$ std \\
\midrule
RF & 0.924 $\pm$ 0.0042 & 0.0922 $\pm$ 0.0922 & 206517.4 $\pm$ 283.51 \\
DT & 0.903 $\pm$ 0.0058 & 0.113 $\pm$ 0.113 & 3252.4 $\pm$ 6.8 \\
knn3 & 0.869 $\pm$ 0.006 & 0.1191 $\pm$ 0.1191 & 0.0 $\pm$ 0.0 \\
knn2 & 0.858 $\pm$ 0.0064 & 0.1292 $\pm$ 0.1292 & 0.0 $\pm$ 0.0 \\
knn4 & 0.8503 $\pm$ 0.0082 & 0.1237 $\pm$ 0.1237 & 0.0 $\pm$ 0.0 \\
SimplifiedTwoDesign-40 & 0.6354 $\pm$ 0.1613 & 0.2128 $\pm$ 0.2128 & 82.0 $\pm$ 0.0 \\
SimplifiedTwoDesign-60 & 0.6213 $\pm$ 0.0332 & 0.224 $\pm$ 0.224 & 122.0 $\pm$ 0.0 \\
\bottomrule
\end{tabular}

\caption{Results of the benchmark Function 9 }
 \label{tab:resultCEC2020F9}
 \end{table}

\begin{table}[ht]
\tiny
\centering
\begin{tabular}{llll}
\toprule
Model & $R^2$-score $\pm$ std & RMSE $\pm$ std & \#Params $\pm$ std \\
\midrule
RF & 0.9969 $\pm$ 0.0003 & 0.0212 $\pm$ 0.0212 & 210831.4 $\pm$ 41.85 \\
DT & 0.9949 $\pm$ 0.0004 & 0.0272 $\pm$ 0.0272 & 3346.6 $\pm$ 1.84 \\
knn3 & 0.9921 $\pm$ 0.0004 & 0.0337 $\pm$ 0.0337 & 0.0 $\pm$ 0.0 \\
knn4 & 0.9921 $\pm$ 0.0003 & 0.0336 $\pm$ 0.0336 & 0.0 $\pm$ 0.0 \\
knn2 & 0.9901 $\pm$ 0.0002 & 0.0378 $\pm$ 0.0378 & 0.0 $\pm$ 0.0 \\
StronglyEntanglingLayers-40 & 0.9789 $\pm$ 0.004 & 0.0552 $\pm$ 0.0552 & 240.0 $\pm$ 0.0 \\
StronglyEntanglingLayers-60 & 0.978 $\pm$ 0.0038 & 0.0565 $\pm$ 0.0565 & 360.0 $\pm$ 0.0 \\
\bottomrule
\end{tabular}

\caption{Results of the benchmark Function 10 }
 \label{tab:resultCEC2020F10}
 \end{table}

\begin{table}[ht]
\tiny
\centering
\begin{tabular}{llll}
\toprule
Model & $R^2$-score $\pm$ std & RMSE $\pm$ std & \#Params $\pm$ std \\
\midrule
DT & 0.3967 $\pm$ 0.044 & 0.2923 $\pm$ 0.2923 & 3264.4 $\pm$ 8.0 \\
StronglyEntanglingLayers-10 & 0.3796 $\pm$ 0.0019 & 0.2893 $\pm$ 0.2893 & 60.0 $\pm$ 0.0 \\
StronglyEntanglingLayers-9 & 0.3792 $\pm$ 0.0016 & 0.2894 $\pm$ 0.2894 & 54.0 $\pm$ 0.0 \\
SimplifiedTwoDesign-9 & 0.379 $\pm$ 0.0031 & 0.2895 $\pm$ 0.2895 & 20.0 $\pm$ 0.0 \\
StronglyEntanglingLayers-7 & 0.3775 $\pm$ 0.003 & 0.2898 $\pm$ 0.2898 & 42.0 $\pm$ 0.0 \\
SimplifiedTwoDesign-10 & 0.3745 $\pm$ 0.0117 & 0.2905 $\pm$ 0.2905 & 22.0 $\pm$ 0.0 \\
SimplifiedTwoDesign-7 & 0.3743 $\pm$ 0.0037 & 0.2906 $\pm$ 0.2906 & 16.0 $\pm$ 0.0 \\
\bottomrule
\end{tabular}

\caption{Results of the benchmark Function 11 }
 \label{tab:resultCEC2020F11}
 \end{table}

\begin{table}[ht]
\tiny
\centering
\begin{tabular}{llll}
\toprule
Model & $R^2$-score $\pm$ std & RMSE $\pm$ std & \#Params $\pm$ std \\
\midrule
RF & 0.9771 $\pm$ 0.0018 & 0.048 $\pm$ 0.048 & 210632.4 $\pm$ 91.3 \\
DT & 0.9694 $\pm$ 0.0021 & 0.0576 $\pm$ 0.0576 & 3345.8 $\pm$ 1.69 \\
StronglyEntanglingLayers-60 & 0.9564 $\pm$ 0.0087 & 0.0681 $\pm$ 0.0681 & 360.0 $\pm$ 0.0 \\
knn3 & 0.9479 $\pm$ 0.0031 & 0.0704 $\pm$ 0.0704 & 0.0 $\pm$ 0.0 \\
knn4 & 0.9448 $\pm$ 0.0041 & 0.0709 $\pm$ 0.0709 & 0.0 $\pm$ 0.0 \\
knn2 & 0.9387 $\pm$ 0.0027 & 0.0784 $\pm$ 0.0784 & 0.0 $\pm$ 0.0 \\
StronglyEntanglingLayers-40 & 0.9306 $\pm$ 0.0119 & 0.0861 $\pm$ 0.0861 & 240.0 $\pm$ 0.0 \\
\bottomrule
\end{tabular}

\caption{Results of the benchmark Function 12 }
 \label{tab:resultCEC2020F12}
 \end{table}

\begin{table}[ht]
\tiny
\centering
\begin{tabular}{llll}
\toprule
Model & $R^2$-score $\pm$ std & RMSE $\pm$ std & \#Params $\pm$ std \\
\midrule
RF & 0.9927 $\pm$ 0.0005 & 0.0457 $\pm$ 0.0457 & 207511.4 $\pm$ 171.93 \\
DT & 0.9905 $\pm$ 0.0009 & 0.0525 $\pm$ 0.0525 & 3260.8 $\pm$ 4.05 \\
knn3 & 0.974 $\pm$ 0.0013 & 0.084 $\pm$ 0.084 & 0.0 $\pm$ 0.0 \\
knn4 & 0.9725 $\pm$ 0.0014 & 0.0856 $\pm$ 0.0856 & 0.0 $\pm$ 0.0 \\
knn2 & 0.9684 $\pm$ 0.0009 & 0.0939 $\pm$ 0.0939 & 0.0 $\pm$ 0.0 \\
StronglyEntanglingLayers-60 & 0.9496 $\pm$ 0.0043 & 0.121 $\pm$ 0.121 & 360.0 $\pm$ 0.0 \\
StronglyEntanglingLayers-40 & 0.9451 $\pm$ 0.0025 & 0.1264 $\pm$ 0.1264 & 240.0 $\pm$ 0.0 \\
\bottomrule
\end{tabular}

\caption{Results of the benchmark Function 13 }
 \label{tab:resultCEC2020F13}
 \end{table}

\begin{table}[ht]
\tiny
\centering
\begin{tabular}{llll}
\toprule
Model & $R^2$-score $\pm$ std & RMSE $\pm$ std & \#Params $\pm$ std \\
\midrule
RF & 0.9964 $\pm$ 0.001 & 0.0205 $\pm$ 0.0205 & 206958.8 $\pm$ 254.69 \\
DT & 0.9955 $\pm$ 0.0008 & 0.0233 $\pm$ 0.0233 & 3274.0 $\pm$ 5.91 \\
StronglyEntanglingLayers-60 & 0.9906 $\pm$ 0.0019 & 0.0337 $\pm$ 0.0337 & 360.0 $\pm$ 0.0 \\
StronglyEntanglingLayers-40 & 0.9882 $\pm$ 0.0016 & 0.0379 $\pm$ 0.0379 & 240.0 $\pm$ 0.0 \\
knn3 & 0.9792 $\pm$ 0.0023 & 0.0482 $\pm$ 0.0482 & 0.0 $\pm$ 0.0 \\
knn2 & 0.9791 $\pm$ 0.002 & 0.049 $\pm$ 0.049 & 0.0 $\pm$ 0.0 \\
RRQNN-120-1q & 0.9784 $\pm$ 0.0028 & 0.0512 $\pm$ 0.0512 & 77.3 $\pm$ 4.88 \\
\bottomrule
\end{tabular}

\caption{Results of the benchmark Function 14 }
 \label{tab:resultCEC2020F14}
 \end{table}

\begin{table}[ht]
\tiny
\centering
\begin{tabular}{llll}
\toprule
Model & $R^2$-score $\pm$ std & RMSE $\pm$ std & \#Params $\pm$ std \\
\midrule
StronglyEntanglingLayers-60 & 0.9869 $\pm$ 0.0042 & 0.0703 $\pm$ 0.0703 & 360.0 $\pm$ 0.0 \\
StronglyEntanglingLayers-40 & 0.9722 $\pm$ 0.0066 & 0.103 $\pm$ 0.103 & 240.0 $\pm$ 0.0 \\
RF & 0.9348 $\pm$ 0.0028 & 0.1404 $\pm$ 0.1404 & 206129.6 $\pm$ 329.46 \\
SimplifiedTwoDesign-60 & 0.9283 $\pm$ 0.0291 & 0.1638 $\pm$ 0.1638 & 122.0 $\pm$ 0.0 \\
DT & 0.9153 $\pm$ 0.0063 & 0.1795 $\pm$ 0.1795 & 3250.2 $\pm$ 8.6 \\
RRQNN-120-1q & 0.8962 $\pm$ 0.0256 & 0.1992 $\pm$ 0.1992 & 74.3 $\pm$ 3.09 \\
knn3 & 0.8935 $\pm$ 0.0074 & 0.1802 $\pm$ 0.1802 & 0.0 $\pm$ 0.0 \\
\bottomrule
\end{tabular}

\caption{Results of the benchmark Function 15 }
 \label{tab:resultCEC2020F15}
 \end{table}

\begin{table}[ht]
\tiny
\centering
\begin{tabular}{llll}
\toprule
Model & $R^2$-score $\pm$ std & RMSE $\pm$ std & \#Params $\pm$ std \\
\midrule
RF & 0.9929 $\pm$ 0.0013 & 0.0341 $\pm$ 0.0341 & 40415.4 $\pm$ 1084.41 \\
DT & 0.9861 $\pm$ 0.0048 & 0.0479 $\pm$ 0.0479 & 545.2 $\pm$ 19.26 \\
knn3 & 0.8694 $\pm$ 0.0088 & 0.135 $\pm$ 0.135 & 0.0 $\pm$ 0.0 \\
knn4 & 0.8606 $\pm$ 0.0106 & 0.1377 $\pm$ 0.1377 & 0.0 $\pm$ 0.0 \\
StronglyEntanglingLayers-60 & 0.8476 $\pm$ 0.0048 & 0.1608 $\pm$ 0.1608 & 360.0 $\pm$ 0.0 \\
knn2 & 0.8257 $\pm$ 0.0066 & 0.1584 $\pm$ 0.1584 & 0.0 $\pm$ 0.0 \\
BasicEntanglerLayers-60 & 0.8083 $\pm$ 0.0268 & 0.18 $\pm$ 0.18 & 120.0 $\pm$ 0.0 \\
\bottomrule
\end{tabular}

\caption{Results of the benchmark Function 16 }
 \label{tab:resultCEC2020F16}
 \end{table}

\begin{table}[ht]
\tiny
\centering
\begin{tabular}{llll}
\toprule
Model & $R^2$-score $\pm$ std & RMSE $\pm$ std & \#Params $\pm$ std \\
\midrule
RF & 0.9992 $\pm$ 0.0001 & 0.0099 $\pm$ 0.0099 & 210822.0 $\pm$ 66.59 \\
knn4 & 0.9981 $\pm$ 0.0001 & 0.0152 $\pm$ 0.0152 & 0.0 $\pm$ 0.0 \\
knn3 & 0.998 $\pm$ 0.0001 & 0.0158 $\pm$ 0.0158 & 0.0 $\pm$ 0.0 \\
DT & 0.9979 $\pm$ 0.0001 & 0.0161 $\pm$ 0.0161 & 3346.8 $\pm$ 1.99 \\
knn2 & 0.9973 $\pm$ 0.0001 & 0.0185 $\pm$ 0.0185 & 0.0 $\pm$ 0.0 \\
RRQNN-120-2q & 0.9948 $\pm$ 0.0006 & 0.0259 $\pm$ 0.0259 & 86.6 $\pm$ 2.95 \\
StronglyEntanglingLayers-20 & 0.994 $\pm$ 0.0014 & 0.0272 $\pm$ 0.0272 & 120.0 $\pm$ 0.0 \\
\bottomrule
\end{tabular}

\caption{Results of the benchmark Function 17 }
 \label{tab:resultCEC2020F17}
 \end{table}

\begin{table}[ht]
\tiny
\centering
\begin{tabular}{llll}
\toprule
Model & $R^2$-score $\pm$ std & RMSE $\pm$ std & \#Params $\pm$ std \\
\midrule
RF & 0.9982 $\pm$ 0.0001 & 0.0159 $\pm$ 0.0159 & 210829.8 $\pm$ 52.5 \\
knn4 & 0.9962 $\pm$ 0.0002 & 0.0233 $\pm$ 0.0233 & 0.0 $\pm$ 0.0 \\
knn3 & 0.996 $\pm$ 0.0002 & 0.0238 $\pm$ 0.0238 & 0.0 $\pm$ 0.0 \\
DT & 0.9957 $\pm$ 0.0003 & 0.025 $\pm$ 0.025 & 3346.4 $\pm$ 2.5 \\
knn2 & 0.9948 $\pm$ 0.0001 & 0.0275 $\pm$ 0.0275 & 0.0 $\pm$ 0.0 \\
StronglyEntanglingLayers-20 & 0.9906 $\pm$ 0.0012 & 0.0371 $\pm$ 0.0371 & 120.0 $\pm$ 0.0 \\
StronglyEntanglingLayers-40 & 0.99 $\pm$ 0.0033 & 0.0378 $\pm$ 0.0378 & 240.0 $\pm$ 0.0 \\
\bottomrule
\end{tabular}

\caption{Results of the benchmark Function 18 }
 \label{tab:resultCEC2020F18}
 \end{table}

\begin{table}[ht]
\tiny
\centering
\begin{tabular}{llll}
\toprule
Model & $R^2$-score $\pm$ std & RMSE $\pm$ std & \#Params $\pm$ std \\
\midrule
RF & 0.9998 $\pm$ 0.0 & 0.0074 $\pm$ 0.0074 & 206308.8 $\pm$ 279.84 \\
DT & 0.9996 $\pm$ 0.0 & 0.0106 $\pm$ 0.0106 & 3250.8 $\pm$ 8.02 \\
StronglyEntanglingLayers-40 & 0.9975 $\pm$ 0.0008 & 0.026 $\pm$ 0.026 & 240.0 $\pm$ 0.0 \\
knn4 & 0.9975 $\pm$ 0.0002 & 0.0262 $\pm$ 0.0262 & 0.0 $\pm$ 0.0 \\
knn3 & 0.9973 $\pm$ 0.0001 & 0.0269 $\pm$ 0.0269 & 0.0 $\pm$ 0.0 \\
StronglyEntanglingLayers-20 & 0.9969 $\pm$ 0.0006 & 0.0292 $\pm$ 0.0292 & 120.0 $\pm$ 0.0 \\
StronglyEntanglingLayers-60 & 0.9966 $\pm$ 0.0009 & 0.0302 $\pm$ 0.0302 & 360.0 $\pm$ 0.0 \\
\bottomrule
\end{tabular}

\caption{Results of the benchmark Function 19 }
 \label{tab:resultCEC2020F19}
 \end{table}

\begin{table}[ht]
\tiny
\centering
\begin{tabular}{llll}
\toprule
Model & $R^2$-score $\pm$ std & RMSE $\pm$ std & \#Params $\pm$ std \\
\midrule
RF & 0.9981 $\pm$ 0.0004 & 0.0127 $\pm$ 0.0127 & 211067.8 $\pm$ 25.98 \\
StronglyEntanglingLayers-40 & 0.9965 $\pm$ 0.0015 & 0.0176 $\pm$ 0.0176 & 240.0 $\pm$ 0.0 \\
knn3 & 0.9963 $\pm$ 0.0006 & 0.0181 $\pm$ 0.0181 & 0.0 $\pm$ 0.0 \\
knn4 & 0.9961 $\pm$ 0.0008 & 0.0185 $\pm$ 0.0185 & 0.0 $\pm$ 0.0 \\
DT & 0.9959 $\pm$ 0.0003 & 0.0194 $\pm$ 0.0194 & 3348.2 $\pm$ 1.03 \\
StronglyEntanglingLayers-20 & 0.9956 $\pm$ 0.0012 & 0.0199 $\pm$ 0.0199 & 120.0 $\pm$ 0.0 \\
RRQNN-120-1q & 0.9954 $\pm$ 0.0023 & 0.0199 $\pm$ 0.0199 & 80.0 $\pm$ 3.56 \\
\bottomrule
\end{tabular}

\caption{Results of the benchmark Function 20 }
 \label{tab:resultCEC2020F20}
 \end{table}

\begin{table}[ht]
\tiny
\centering
\begin{tabular}{llll}
\toprule
Model & $R^2$-score $\pm$ std & RMSE $\pm$ std & \#Params $\pm$ std \\
\midrule
RF & 0.998 $\pm$ 0.0002 & 0.0209 $\pm$ 0.0209 & 211172.0 $\pm$ 0.0 \\
DT & 0.9979 $\pm$ 0.0002 & 0.0217 $\pm$ 0.0217 & 3349.0 $\pm$ 0.0 \\
StronglyEntanglingLayers-60 & 0.9912 $\pm$ 0.0013 & 0.0439 $\pm$ 0.0439 & 360.0 $\pm$ 0.0 \\
StronglyEntanglingLayers-40 & 0.9891 $\pm$ 0.0014 & 0.049 $\pm$ 0.049 & 240.0 $\pm$ 0.0 \\
SimplifiedTwoDesign-20 & 0.983 $\pm$ 0.0036 & 0.061 $\pm$ 0.061 & 42.0 $\pm$ 0.0 \\
knn4 & 0.9761 $\pm$ 0.0016 & 0.0698 $\pm$ 0.0698 & 0.0 $\pm$ 0.0 \\
knn3 & 0.9759 $\pm$ 0.0005 & 0.0708 $\pm$ 0.0708 & 0.0 $\pm$ 0.0 \\
\bottomrule
\end{tabular}

\caption{Results of the benchmark Function 21 }
 \label{tab:resultCEC2020F21}
 \end{table}

\begin{table}[ht]
\tiny
\centering
\begin{tabular}{llll}
\toprule
Model & $R^2$-score $\pm$ std & RMSE $\pm$ std & \#Params $\pm$ std \\
\midrule
RF & 1.0 $\pm$ 0.0 & 0.0 $\pm$ 0.0 & 207287.8 $\pm$ 559.86 \\
DT & 1.0 $\pm$ 0.0 & 0.0 $\pm$ 0.0 & 3273.2 $\pm$ 10.35 \\
StronglyEntanglingLayers-60 & 0.9948 $\pm$ 0.0011 & 0.0414 $\pm$ 0.0414 & 360.0 $\pm$ 0.0 \\
StronglyEntanglingLayers-40 & 0.9944 $\pm$ 0.0009 & 0.0431 $\pm$ 0.0431 & 240.0 $\pm$ 0.0 \\
knn3 & 0.9936 $\pm$ 0.0007 & 0.0452 $\pm$ 0.0452 & 0.0 $\pm$ 0.0 \\
knn4 & 0.9934 $\pm$ 0.0009 & 0.0457 $\pm$ 0.0457 & 0.0 $\pm$ 0.0 \\
SimplifiedTwoDesign-20 & 0.9924 $\pm$ 0.0029 & 0.0493 $\pm$ 0.0493 & 42.0 $\pm$ 0.0 \\
\bottomrule
\end{tabular}

\caption{Results of the benchmark Function 22 }
 \label{tab:resultCEC2020F22}
 \end{table}

\section{Similarity between architectures generated by genetic algorithms - Clustering Checking}
\label{sec:similarityArchsClustering}

Tables~\ref{table:rrqnn2qCLUSTERINGBin} and~\ref{table:rrqnn2qCLUSTERING} report the clustering metrics obtained for RRQNN architectures with two qubits, considering both binarized and non-binarized architectural representations across different operator counts and clustering algorithms. Analogously, Tables~\ref{table:rrqnn1qCLUSTERINGBin} and~\ref{table:rrqnn1qCLUSTERING} summarize the corresponding results for single-qubit RRQNNs, enabling a comparative analysis of how architectural representation, parameter count, and qubit configuration influence the clustering behavior of evolved quantum models.
A detailed discussion of these clustering results is provided in Section~\ref{sec:similarityRRQNN_clustering}.

\begin{table}[ht]
\centering
\begin{tabular}{cccccc}
\toprule
\# Param & Clustering Algorithm & ARI & Silhouette & Jaccard & Fowlkes--Mallows \\
\midrule
5  & KMeans(n\_clusters=22)                  & 0.027  & 0.214 & 0.028 & 0.069 \\
5  & Birch(n\_clusters=22)                   & 0.036  & 0.207 & 0.012 & 0.078 \\
5  & AgglomerativeClustering(n=22)          & 0.041  & 0.221 & 0.029 & 0.083 \\
\midrule
10 & KMeans(n\_clusters=22)                  & 0.025  & 0.107 & 0.017 & 0.067 \\
10 & Birch(n\_clusters=22)                   & 0.012  & 0.102 & 0.018 & 0.054 \\
10 & AgglomerativeClustering(n=22)          & 0.011  & 0.106 & 0.020 & 0.054 \\
\midrule
20 & KMeans(n\_clusters=22)                  & -0.002 & 0.042 & 0.012 & 0.043 \\
20 & Birch(n\_clusters=22)                   & -0.004 & 0.057 & 0.029 & 0.041 \\
20 & AgglomerativeClustering(n=22)          & -0.004 & 0.057 & 0.029 & 0.041 \\
\midrule
40 & KMeans(n\_clusters=22)                  & 0.006  & 0.015 & 0.024 & 0.049 \\
40 & Birch(n\_clusters=22)                   & 0.010  & 0.025 & 0.030 & 0.054 \\
40 & AgglomerativeClustering(n=22)          & 0.010  & 0.025 & 0.030 & 0.054 \\
\midrule
60 & KMeans(n\_clusters=22)                  & 0.011  & 0.004 & 0.020 & 0.054 \\
60 & Birch(n\_clusters=22)                   & -0.000 & 0.025 & 0.027 & 0.043 \\
60 & AgglomerativeClustering(n=22)          & -0.003 & 0.022 & 0.014 & 0.041 \\
\midrule
120 & KMeans(n\_clusters=22)                 & -0.001 & -0.006 & 0.025 & 0.047 \\
120 & Birch(n\_clusters=22)                  & -0.000 & 0.016  & 0.023 & 0.043 \\
120 & AgglomerativeClustering(n=22)         & -0.000 & 0.016  & 0.023 & 0.043 \\
\bottomrule
\end{tabular}
\caption{RRQNN - 1q - Clustering metrics for different RRQNN parameter counts and clustering algorithms.}
\label{table:rrqnn1qCLUSTERING}
\end{table}

\begin{table}[ht]
\centering
\begin{tabular}{cccccc}
\toprule
\# Param & Clustering Algorithm & ARI & Silhouette & Jaccard & Fowlkes--Mallows \\
\midrule
5  & KMeans(n\_clusters=22)                  & 0.027  & 0.214 & 0.028 & 0.069 \\
5  & Birch(n\_clusters=22)                   & 0.036  & 0.207 & 0.012 & 0.078 \\
5  & AgglomerativeClustering(n=22)          & 0.041  & 0.221 & 0.029 & 0.083 \\
\midrule
10 & KMeans(n\_clusters=22)                  & 0.025  & 0.107 & 0.017 & 0.067 \\
10 & Birch(n\_clusters=22)                   & 0.012  & 0.102 & 0.018 & 0.054 \\
10 & AgglomerativeClustering(n=22)          & 0.011  & 0.106 & 0.020 & 0.054 \\
\midrule
20 & KMeans(n\_clusters=22)                  & -0.002 & 0.042 & 0.012 & 0.043 \\
20 & Birch(n\_clusters=22)                   & -0.004 & 0.057 & 0.029 & 0.041 \\
20 & AgglomerativeClustering(n=22)          & -0.004 & 0.057 & 0.029 & 0.041 \\
\midrule
40 & KMeans(n\_clusters=22)                  & 0.006  & 0.015 & 0.024 & 0.049 \\
40 & Birch(n\_clusters=22)                   & 0.010  & 0.025 & 0.030 & 0.054 \\
40 & AgglomerativeClustering(n=22)          & 0.010  & 0.025 & 0.030 & 0.054 \\
\midrule
60 & KMeans(n\_clusters=22)                  & 0.011  & 0.004 & 0.020 & 0.054 \\
60 & Birch(n\_clusters=22)                   & -0.000 & 0.025 & 0.027 & 0.043 \\
60 & AgglomerativeClustering(n=22)          & -0.003 & 0.022 & 0.014 & 0.041 \\
\midrule
120 & KMeans(n\_clusters=22)                 & -0.001 & -0.006 & 0.025 & 0.047 \\
120 & Birch(n\_clusters=22)                  & -0.000 & 0.016  & 0.023 & 0.043 \\
120 & AgglomerativeClustering(n=22)         & -0.000 & 0.016  & 0.023 & 0.043 \\
\bottomrule
\end{tabular}
\caption{RRQNN - 1q - Clustering metrics for different operator counts and algorithms - considering binarized architecture vectors.}
\label{table:rrqnn1qCLUSTERINGBin}
\end{table}

\begin{table}[ht]
\centering
\begin{tabular}{cccccc}
\toprule
\# Param & Clustering Algorithm & ARI & Silhouette & Jaccard & Fowlkes--Mallows \\
\midrule
20 & KMeans(n\_clusters=22)                  & 0.000 & 0.040 & 0.031 & 0.047 \\
20 & Birch(n\_clusters=22)                   & 0.010 & 0.051 & 0.018 & 0.053 \\
20 & AgglomerativeClustering(n=22)          & 0.009 & 0.054 & 0.027 & 0.052 \\
\midrule
40 & KMeans(n\_clusters=22)                  & -0.008 & -0.003 & 0.008 & 0.040 \\
40 & Birch(n\_clusters=22)                   & 0.008  & 0.029  & 0.019 & 0.052 \\
40 & AgglomerativeClustering(n=22)          & 0.008  & 0.029  & 0.019 & 0.052 \\
\midrule
60 & KMeans(n\_clusters=22)                  & 0.009  & 0.004 & 0.028 & 0.052 \\
60 & Birch(n\_clusters=22)                   & 0.008  & 0.023 & 0.023 & 0.052 \\
60 & AgglomerativeClustering(n=22)          & 0.008  & 0.023 & 0.023 & 0.052 \\
\midrule
120 & KMeans(n\_clusters=22)                 & 0.004 & -0.009 & 0.025 & 0.051 \\
120 & Birch(n\_clusters=22)                  & 0.003 & 0.017  & 0.012 & 0.047 \\
120 & AgglomerativeClustering(n=22)         & 0.003 & 0.017  & 0.012 & 0.047 \\
\bottomrule
\end{tabular}
\caption{RRQNN - 2q - Clustering metrics for different operator counts and algorithms.}
\label{table:rrqnn2qCLUSTERING}
\end{table}

\begin{table}[ht]
\centering
\begin{tabular}{cccccc}
\toprule
\# Param & Clustering Algorithm & ARI & Silhouette & Jaccard & Fowlkes--Mallows \\
\midrule
20 & KMeans(n\_clusters=22)         & 0.001 & 0.042 & 0.015 & 0.046 \\
20 & Birch(n\_clusters=22)                         & 0.005 & 0.052 & 0.030 & 0.048 \\
20 & AgglomerativeClustering(n=22)                & 0.002 & 0.052 & 0.025 & 0.045 \\
\midrule
40 & KMeans(n\_clusters=22)         & -0.003 & 0.015 & 0.024 & 0.044 \\
40 & Birch(n\_clusters=22)                         & 0.008  & 0.026 & 0.023 & 0.053 \\
40 & AgglomerativeClustering(n=22)                & 0.008  & 0.026 & 0.023 & 0.053 \\
\midrule
60 & KMeans( n\_clusters=22)         & 0.016 & 0.002 & 0.010 & 0.065 \\
60 & Birch(n\_clusters=22)                         & 0.006 & 0.018 & 0.024 & 0.051 \\
60 & AgglomerativeClustering(n=22)                & 0.006 & 0.018 & 0.024 & 0.051 \\
\midrule
120 & KMeans(n\_clusters=22)        & -0.001 & -0.003 & 0.010 & 0.051 \\
120 & Birch(n\_clusters=22)                        & 0.003  & 0.011  & 0.019 & 0.047 \\
120 & AgglomerativeClustering(n=22)               & 0.008  & 0.015  & 0.019 & 0.052 \\
\bottomrule
\end{tabular}
\caption{RRQNN - 2q - Clustering metrics for different operator counts and algorithms - considering binarized architecture vectors.}
\label{table:rrqnn2qCLUSTERINGBin}
\end{table}

\section{Statistical comparing tests}

Table~\ref{tab:wilcoxon_quantum_classical} reports the results of the Wilcoxon statistical tests comparing quantum and classical regression models, which are discussed in detail in Section~\ref{sec:statisticalComparison}.

\begin{table}[ht]
\centering
\footnotesize
\begin{tabular}{llll}
\toprule
Function Number & Quantum Model & Classical Model & p-value \\
\midrule
1  & StronglyEntanglingLayers-20 & knn4              & 0.375 \\
1  & StronglyEntanglingLayers-20 & knn3              & 0.557 \\
1  & StronglyEntanglingLayers-40 & knn3              & 0.064 \\
2  & StronglyEntanglingLayers-20 & MLP500-500-ReLU   & 0.064 \\
2  & StronglyEntanglingLayers-60 & knn4              & 0.375 \\
2  & StronglyEntanglingLayers-60 & knn3              & 0.557 \\
2  & StronglyEntanglingLayers-60 & MLP500-500-ReLU   & 0.232 \\
2  & SimplifiedTwoDesign-40      & knn4              & 0.105 \\
2  & SimplifiedTwoDesign-40      & knn3              & 0.105 \\
2  & StronglyEntanglingLayers-40 & knn4              & 0.492 \\
2  & StronglyEntanglingLayers-40 & knn3              & 0.846 \\
2  & StronglyEntanglingLayers-40 & MLP500-500-ReLU   & 0.492 \\
2  & SimplifiedTwoDesign-20      & knn4              & 0.432 \\
2  & SimplifiedTwoDesign-20      & knn3              & 0.432 \\
2  & SimplifiedTwoDesign-20      & MLP500-500-ReLU   & 0.922 \\
4  & SimplifiedTwoDesign-40      & MLP500-500-ReLU   & 0.084 \\
4  & StronglyEntanglingLayers-40 & knn4              & 0.084 \\
4  & StronglyEntanglingLayers-40 & knn3              & 0.322 \\
4  & SimplifiedTwoDesign-20      & knn4              & 0.322 \\
4  & SimplifiedTwoDesign-20      & knn3              & 0.193 \\
4  & SimplifiedTwoDesign-20      & MLP500-500-ReLU   & 0.922 \\
5  & StronglyEntanglingLayers-20 & knn2              & 0.193 \\
5  & StronglyEntanglingLayers-60 & knn4              & 0.432 \\
5  & StronglyEntanglingLayers-60 & knn3              & 0.432 \\
5  & SimplifiedTwoDesign-40      & knn4              & 0.160 \\
5  & SimplifiedTwoDesign-40      & knn3              & 0.131 \\
5  & SimplifiedTwoDesign-40      & knn2              & 0.625 \\
5  & SimplifiedTwoDesign-20      & knn4              & 0.846 \\
5  & SimplifiedTwoDesign-20      & knn3              & 0.770 \\
5  & SimplifiedTwoDesign-20      & knn2              & 0.375 \\
6  & StronglyEntanglingLayers-60 & knn2              & 0.232 \\
6  & RRQNN-120-2q                & knn2              & 0.193 \\
6  & StronglyEntanglingLayers-40 & DT                & 0.232 \\
6  & StronglyEntanglingLayers-40 & knn2              & 0.770 \\
11 & StronglyEntanglingLayers-20 & DT                & 0.131 \\
11 & StronglyEntanglingLayers-9  & DT                & 0.322 \\
11 & StronglyEntanglingLayers-10 & DT                & 0.375 \\
11 & StronglyEntanglingLayers-7  & DT                & 0.322 \\
11 & SimplifiedTwoDesign-10      & DT                & 0.275 \\
11 & StronglyEntanglingLayers-5  & DT                & 0.105 \\
11 & SimplifiedTwoDesign-7       & DT                & 0.160 \\
11 & StronglyEntanglingLayers-6  & DT                & 0.160 \\
11 & SimplifiedTwoDesign-6       & DT                & 0.131 \\
11 & SimplifiedTwoDesign-9       & DT                & 0.232 \\
11 & SimplifiedTwoDesign-20      & DT                & 0.160 \\
16 & SimplifiedTwoDesign-60      & knn4              & 0.193 \\
16 & SimplifiedTwoDesign-60      & knn3              & 0.064 \\
\bottomrule
\end{tabular}
\caption{Wilcoxon signed-rank test p-values comparing quantum and classical models for selected CEC2020 functions.}
\label{tab:wilcoxon_quantum_classical}
\end{table}




\end{appendices}


\bibliography{bib}

\end{document}